\documentclass[10pt,aps,prd,showpacs,preprintnumbers,nofootinbib,amsmath,amssymb,fleqn]{revtex4-1}
\usepackage[dvips]{epsfig}
\usepackage{multirow}
\usepackage{color}
\usepackage{slashbox,pict2e}

\newcommand{\hm}{\hspace*{-0.6cm}}

\newcommand{\veck}{{\mathbf k}}

\newcommand{\kv}{{\mathbf k}}

\newcommand{\nn}{\nonumber}

\newcommand{\be}{\begin{equation}}
\newcommand{\ee}{\end{equation}}
\newcommand{\bea}{\begin{eqnarray}}
\newcommand{\eea}{\end{eqnarray}}
\newcommand{\bean}{\begin{eqnarray*}}
	\newcommand{\eean}{\end{eqnarray*}}

\newcommand{\bit}{\begin{itemize}}
\newcommand{\eit}{\end{itemize}}

\newcommand{\cO}{{\cal O}}
\newcommand{\Jpsi}{J/\psi}

\newcommand{\rec}{\rm rec}
\newcommand{\RGsub}{G^{\rm sub}/G^{\rm sub}_{\rm rec}}
\newcommand{\RGdiff}{G^{\rm diff}/G^{\rm diff}_{\rm rec}}

\newcommand{\MSbar}{\overline{\mathrm{MS}}}

\newcommand{\Tc}{\mathrm{T}_c}

\newcommand{\md}{\mathrm{d}}
\newcommand{\vecx}{{\mathbf x}}
\newcommand{\vecnull}{{\mathbf 0}}

\newcommand{\vecp}{{\mathbf p}}

\newcommand{\Dp}{D^{+}}
\newcommand{\Dm}{D^{-}}

\begin{document}


\title{Charmonium properties in hot quenched lattice QCD}
\author{H.-T. Ding$^{\rm1,2}$, A. Francis$^{\rm 1,3}$, O. Kaczmarek$^{\rm 1}$,  F. Karsch$^{\rm 1,2}$, \\ 
H. Satz$^{\rm 1}$ and  W. Soeldner$^{\rm 4}$}
\affiliation{
$^{\rm 1}$Fakult\"at f\"ur Physik, Universit\"at Bielefeld, D-33615 Bielefeld, Germany\\
$^{\rm 2}$Physics Department, Brookhaven National Laboratory,Upton, NY 11973, USA \\
$^{\rm 3}$Institut f\"ur Kernphysik, Johannes Gutenberg Universit\"at Mainz, D-55099 Mainz, Germany\\
$^{\rm 4}$Institut f\"ur Theoretische Physik, Universit\"at Regensburg, D-93040 Regensburg, Germany}

\preprint{BI-TP 2012/13}

\begin{abstract}
We study the properties of charmonium states at finite temperature in quenched QCD on large and fine isotropic lattices. We 
perform a detailed analysis of charmonium correlation and spectral functions both below and above $T_c$. 
Our analysis suggests that both S wave states ($\Jpsi$ and $\eta_c$) and P wave states 
( $\chi_{c0}$ and $\chi_{c1}$) disappear already at about $1.5~T_c$. 
The charm diffusion coefficient is estimated through the Kubo formula and 
found to be compatible with zero below $T_c$ and approximately $1/\pi T$ at $1.5~T_c\lesssim T\lesssim 3~T_c$. 

\end{abstract}


\maketitle

\newpage
\section{Introduction}
\label{intro}

The main goal of ongoing heavy ion programs at the RHIC and LHC is to 
study the properties of the hot and dense medium formed during the collision of two relativistic heavy nuclei.
It has been conjectured that at sufficiently high temperature the QCD medium will undergo a phase transition to 
a deconfined phase, in which the degrees of freedom are those of quarks and gluons. 
Unlike light mesons, the heavy mesons, e.g. $\Jpsi$, may survive in the hot medium up to certain temperatures before they get dissociated due to Debye screening~\cite{Matsui:1986dk}. Thus
due to the dissociation of heavy mesons, the suppression of their yield in nucleus-nucleus ($AA$) collisions compared to that in
proton-proton ($pp$) collisions can serve as a good probe for the properties of the medium. 
The experiments carried out at the SPS and LHC at CERN and the RHIC at BNL have indeed observed $\Jpsi$ suppression~\cite{Rapp08}. 
The interpretation of experimental data, however, is not as straightforward as the original idea proposed in Ref.~\cite{Matsui:1986dk}.
 The observed modification of $\Jpsi$ production in $AA$ collisions could be caused by two distinct classes of effects. On the one hand there are cold nuclear matter effects, which originate from the presence of cold nuclear matter in the target and projectile. On the other hand there are hot medium effects, which are of primary interest and reflect the properties of the medium we want to study. In order to disentangle these two effects, 
it is crucial to have a good understanding of the behavior of heavy quarks and quarkonia in the hot medium.

From the theoretical point of view, the meson spectral function at finite temperature~\cite{Kapusta06}, which contains all the
information on the hadron properties in the thermal medium, such as the presence, the location and the width of bound states (and thus about dissociation temperatures) 
as well as transport properties (e.g. heavy quark diffusion coefficients), is the key quantity to be investigated. As this is a difficult task, several theoretical approaches have been followed to determine the quarkonium properties at finite temperature.

 The most traditional approach is based on the analysis of nonrelativistic potential models. Here one assumes that the interaction between a heavy quark pair forming the quarkonium can be described by a potential~\cite{Eichten80}. Because of its success at zero temperature, the potential model approach has been used also at finite temperature~\cite{Mocsy08review}. The temperature dependent potential used in these calculations is based either on model calculations or on finite temperature lattice QCD results~\cite{Kaczmarek05}. It is used to solve a nonrelativistic Schr\"odinger equation. The resulting dissociation temperatures depend strongly on the potential used. Recently progress has been made in comparing directly heavy quark correlation functions calculated on the lattice with potential model results. This allows to eliminate certain ambiguities~\cite{Mocsy08review} and opens the possibility to 
 determine which potential is more appropriate for a description of the experimental data~\cite{Zhao10}. Nonetheless, the potential model approach at finite temperature is still under scrutiny.

Most recently a nonrelativistic effective theory approach at nonzero temperature, which requires the scales concerned to be in hierarchy, has been developed~\cite{Ghiglieri:2012rp}. By integrating out certain scales, one arrives at a complex real-time static potential, which includes effects of screening via its real part as well as the interaction with the medium via its imaginary part. The presence of an imaginary part in the heavy quark potential reduces the possibility for stable quarkonium states in the hot medium. This approach becomes more reliable as the quark mass increases and thus is more relevant for the analysis of bottomonium states. An approach to study charmonium spectral functions at finite temperature using QCD sum rules has also been developed recently~\cite{Gubler:2011ua}.

First principle calculations in lattice QCD are thus crucially needed to determine the nonperturbative behavior of heavy quarks and quarkonia in the hot medium. The investigations of charmonium states at finite temperature, which have been performed in both quenched and full lattice QCD, have led to the rather interesting result that $\Jpsi$ appears to survive up to temperatures well above $T_c$~\cite{Datta04, Asakawa04, Umeda05, Aarts07, Jakovac07, Iida06, Ohno11}. The most relevant quantities, meson spectral functions, however, cannot be obtained directly from lattice QCD calculations. Further input is needed to extract spectral functions from correlation functions calculated on the lattice. One of the commonly used methods is the Maximum Entropy Method (MEM). When using the Maximum Entropy Method, a very important issue is to get control over its input parameter (default model) dependence. The output spectral function from MEM can only be trusted if the default model dependence is eliminated or at least well understood. Additionally sufficient information on the Euclidean time dependence of the correlation function is crucially important in the MEM analysis. One economical way to increase the number of correlator data points in the temporal direction is to perform simulations on anisotropic lattices~\cite{Asakawa04, Umeda05, Aarts07, Jakovac07,Iida06,Ohno11}.  However, lattice cutoff effects are more significant on anisotropic lattices~\cite{Karsch03}. Thus in the present paper we use isotropic lattices and perform simulations on very large lattices. The finest lattices we performed simulations on are $128^{3}\times 96$, $128^3\times48$, $128^3\times32$ and $128^3\times24$ 
at $0.73~T_c$, $1.46~T_c$, $2.20~T_c$ and $2.93~T_c$, respectively. The number of data points in the temporal direction in the current study is doubled compared to our previous study in Ref.~\cite{Datta04} 
and  it is about 1.5 times larger than that in stuidies~\cite{Umeda05, Aarts07, Jakovac07,Iida06,Ohno11} and compatible with that used in Ref.\cite{Asakawa04}.
Based on the correlation functions calculated on these large lattices, we will report on a detailed
study of finite temperature charmonium correlators and perform a detailed MEM analysis of spectral functions, expanding on preliminary results reported in~\cite{Ding08,Ding09,Ding10,Ding11}. 
The signature obtained for the dissociation of charmonium states in the hot medium from the spectral function will be discussed.

Besides the properties of charmonium states in the medium, the behavior of a single charm quark in the hot medium is of great interest as well.
Experimentally a substantial elliptic flow of heavy quarks has been observed~\cite{Adare:2006nq}. The heavy quark diffusion $D$ can be connected to the energy loss of a heavy quark during its propagation in the medium and is also related to the ratio of shear viscosity to entropy density $\eta/s$~\cite{Riek:2010py,Moore:20004tg}. Various phenomenological model studies suggest the heavy quark diffusion coefficient $D\lesssim 1/T$ to accommodate data while various pQCD and T-Matrix calculations of the heavy quark diffusion coefficient differ significantly from each other~\cite{Moore:20004tg,CaronHuot:2007gq,Riek:2010fk,He:2011qa}. It is thus important to have a first principle calculation of the heavy quark diffusion coefficient. The heavy quark diffusion coefficient can be obtained from the vector spectral function at vanishing frequency through the Kubo formula. 
We will give here also an estimate for the value of the charm diffusion coefficient at different temperatures.

The rest of the paper is organized as follows. In Sec.~\ref{sec:intro} we discuss general features of quarkonium
correlators and spectral functions. In Sec.~\ref{sec:lattice_simulation} we give the lattice setup used in the calculation
of charmonium correlation functions. In Sec.~\ref{sec:rec_cor} we discuss information on the change of spectral functions from below to above $T_c$ that can be obtained from the analysis of correlation
functions only, i.e. on thermal modifications of charmonium states and also on the charm quark diffusion coefficient.  In Sec.~\ref{sec:spf_MEM} we will describe the Maximum Entropy Method used 
for the reconstruction of spectral functions and discuss the spectral functions below and above $T_c$ obtained from MEM.  Signatures for the dissociation of charmonium states and values of charm quark diffusion coefficients are
discussed. Finally we summarize in Sec.~\ref{sec:con}.
Some further details of our MEM analyses are given in an Appendix.

\section{Meson correlation and spectral functions}
\label{sec:intro}
In this section, we give the definition of the meson spectral function and its relation to the Euclidean correlation function, which can be calculated
directly on the lattice.

All information on quarkonium states is embedded in these spectral functions. The spectral function for a given 
meson channel $H$ in a system can be defined through the Fourier transform of the real-time two-point correlation functions
$\Dp$ and $\Dm$. The ensemble average of the commutator is
\be
D_H(t,\vecx) = -i\left<[J_H(t,\vecx),J_H(0,\vecnull)]\right> = \Dp_H(t,\vecx) - \Dm_H(t,\vecx),
\ee
and its spectral density $\rho(\omega,\vecp)$ can be expressed in terms of
the retarded correlator $D_{H}^{R}(\omega,\vecp)$~\cite{Blaizot02}
\bea
\rho_H(\omega,\vecp) = \Dp_H(\omega,\vecp) - \Dm_H(\omega,\vecp) = 2\,{\rm Im}D^R_H(\omega,\vecp),
                                         \label{spf_definition_p}
\eea
where 
\bea
&&D^{+(-)}_H(\omega,\vecp)= \int \md^4x\,e^{i\omega t -i\vecp\vecx} \,D_{H}^{+(-)}(t,\vecx), \\
&& i\Dp_H(t,\vecx) = \left\langle J_H(t,\vecx) J_H(0,\vecnull) \right\rangle, \\
&& i \Dm_H(t,\vecx) = \left\langle J_H(0,\vecnull) J_H(t,\vecx) \right\rangle.
\eea
The two-point correlation functions $D^{+(-)}_H$ satisfy the Kubo-Martin-Schwinger (KMS) relation
\bea
\Dp_H(t,\vecx) = \Dm_H( t+i\beta,\vecx),~~~~\Dp_H(\omega,\vecp)=e^{\beta\omega} \Dm_H(\omega,\vecp).
\label{KMS}
\eea
Inserting a complete set of states in Eq.~(\ref{spf_definition_p}) and using the KMS relation, one gets an explicit expression
for $\rho_H(\omega,\vecp)$
\bea
\rho_H(\omega,\vecp)    = \frac{2\pi}{Z}\sum_{n,m}e^{-\beta E_{n}\,} \Big( \delta\big(p + k_n - k_m\big) - \delta\big(p + k_m  - k_n\big) \Big)  \left | \left < n \left |J_H(0) \right | m \right > \right |^2,
\label{spf_energystates}
\eea
where $Z$ is the partition function, $p=(\omega,\vecp)$ and  $k_{n(m)}$ refer to the four-momenta of the state $|n(m)\rangle$. 
Given the above equation it is clear that the spectral function $\rho_H(\omega,\vecp)$ is an odd function of the frequency and momentum,
$\rho_H(-\omega,-\vecp)=-\rho_H(\omega,\vecp)$ and $\omega\rho_H(\omega,\vecp)\ge 0$. If the system is rotationally invariant, which means the state can have the same energy $\omega$ but opposite momentum $\vecp$, the spectral function $\rho_H(\omega,\vecp)$ would also be an odd function of $\omega$.

 The spectral function in the vector channel is related to the experimentally accessible differential cross section for thermal dilepton production~\cite{Braaten90},
\be
\frac{\md W}{\md\omega\,\md^3\vecp} = \frac{5\alpha^2}{54\pi^3}\frac{1}{\omega^2(e^{\omega/T} -1)}\,\rho_V(\omega,\vecp,T),
\ee
where $\alpha$ is the electromagnetic fine structure constant and $\rho_{V}$ is the spectral function in the vector channel. 
Additionally the spatial components of  the vector spectral function are related to the heavy quark diffusion constant $D$~\cite{Kapusta06}
\be
D = \frac{1}{6\chi^{00}}\lim_{\omega\rightarrow0}\sum_{i=1}^{3}\frac{\rho^{V}_{ii}(\omega,\vec{p}=0,T)}{\omega} ,
\label{eq:HQ_diffusion_formula}
\ee
where $\chi^{00}$ is the quark number susceptibility that is defined through the zeroth component of the temporal correlator 
in the vector channel.

In this work we consider local meson operators of the form 
 \be
 J_{H}(\tau,\vecx)=\bar{\psi}(\tau,\vecx)\Gamma_{H} \psi(\tau,\vecx) ,
 \label{local_current}
 \ee
  with $\Gamma_{H}=1,\gamma_5,\gamma_{\mu},\gamma_5\gamma_{\mu}$, for scalar ($SC$), pseudoscalar ($PS$), vector ($VC$) and axial-vector ($AV$) channels, respectively. The relation of these quantum numbers to different charmonium states  from the particle data book is summarized in Table \ref{table:meson_states}.

\begin{table}[htb]
\centering
\vspace{0.2cm}
\begin{tabular}{|c|c| c|c|c|c|}
\hline
${\rm Channel}$ &$\Gamma_{H}$ & $^{2S+1}L_{J}$  & $J^{PC}$  & $c\bar{c}$ & M($c\bar{c}$)[GeV] \\
\hline 
${\rm PS}$         & $\gamma_{5}$      &     $^{1}S_{0}$   &  $0^{-+}$ &   $\eta_{c}$ &2.980(1)\\
${\rm VC}$          & $\gamma_{\mu} $   &     $^{3}S_{1}$   &  $1^{--}$         &  $J/\psi$   &3.097(1)  \\
${\rm SC}$          & 1                 &     $^{3}P_{0}$   &  $0^{++}$   &  $\chi_{c0}$              & 3.415(1)     \\
${\rm AV}$         & $\gamma_{5}\gamma_{\mu}$     &     $^{3}P_{1}$   &  $1^{++}$ & $\chi_{c1}$  &3.510(1)\\ 
\hline
\end{tabular}
\caption{Charmonium states in different quantum number channels taken from the particle data book~\cite{Nakamura:2010zzi}.}
\label{table:meson_states}
\end{table}

The Euclidean temporal correlation function $G(\tau,\vecp)$ can then be defined as
\be
G_H(\tau,\vecp) =\int~\md^3\vecx~e^{-i\vecp\cdot\vecx}~\left <J_H(\tau,\vecx)J_{H}(0,\vecnull)\right > ,
\ee
where $G_{H}(\tau,\vecp)$ is the analytic continuation of $D^{+}(t,\vecp)$ from real to imaginary time
\be
G_H(\tau,\vecp)=D^{+}(-i\tau,\vecp).
\ee
By using the KMS relation and the above equation, one can easily relate the correlation function to the spectral function,
\be
G_H(\tau,\vecp) = \int_0^{\infty}\frac{\md\omega}{2\pi}\, \rho_H(\omega,\vecp) \,K(\omega,\tau),
\label{cor_spf_relation}
\ee
where the integration kernel $K(\omega,\tau)$ is
\be
K(\omega,\tau) = \frac{\cosh(\omega(\tau-1/2T))}{\sinh(\omega/2T)}.
\ee
Note that the kernel $K(\omega,\tau)$ is symmetric around $\tau=1/2T$.

Because of asymptotic freedom the spectral functions at very high energy are expected
to be described well by the propagation of a free quark antiquark pair. In this
noninteracting limit the spectral function is analytically given by~\cite{Karsch03,Aarts05}
 \bea
 \rho_H(\omega) =
\frac{N_c}{8\pi}  \Theta(\omega^2-4m^2)\, \omega^2\tanh\left(\frac{\omega}{4T}\right) \sqrt{1-\left(\frac{2m}{\omega}\right)^2}
&&\hm \nn \\
  \times\Bigg[\left( a_H^{(1)} -a_H^{(2)} \right) 
+\left(\frac{2m}{\omega}\right)^2\left( a_H^{(2)} -a_H^{(3)} \right) \Bigg]
 &&\hm \nn \\
+\,\,
N_c\Bigg[ \left(a_H^{(1)} + a_H^{(3)}\right) I_1 +
\left( a_H^{(2)} -a_H^{(3)}\right) I_2 \Bigg] \,\omega\,\delta(\omega), 
&&\hm
\label{eq:freespf_continuum_zerop}
\eea
with
\be
I_1 = -\int  \frac{\md^3 \veck}{2\pi^2}  \, \frac{\partial n_F(\omega_\kv)}{\partial \omega_\kv},~~~~~I_2 = -\int  \frac{\md^3 \veck}{2\pi^2} \,  \frac{\veck^2}{\omega_\kv^2 }\,\frac{\partial n_F(\omega_\kv)}{\partial \omega_\kv}.
\ee
The coefficients $a_H^{(1,2,3)}$ can be read off from Ref.~\cite{Aarts05}, $\omega_\kv^2=m^2+\veck^2$ and $n_F(\omega_\kv)$ is the Fermi distribution function.
 Note in the above expression that there is a term proportional to $\omega\delta(\omega)$, implying a $\tau$ independent contribution to the correlation function. This contribution
is also known as a zero mode contribution~\cite{Umeda07}. For correlators with massive quarks, the zero mode contribution vanishes only  in the $PS$ channel.

 \begin{figure}[!t]
  \begin{center}  
    \includegraphics[width=.45\textwidth]{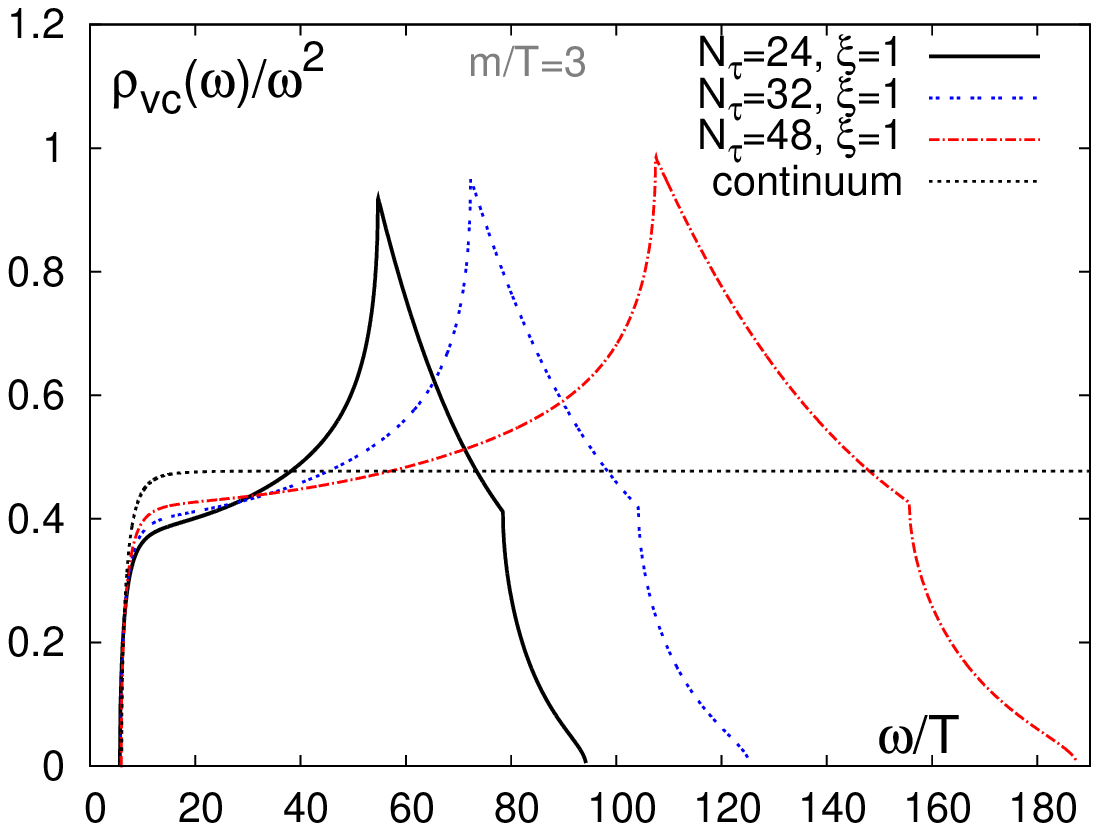}~    \includegraphics[width=.45\textwidth]{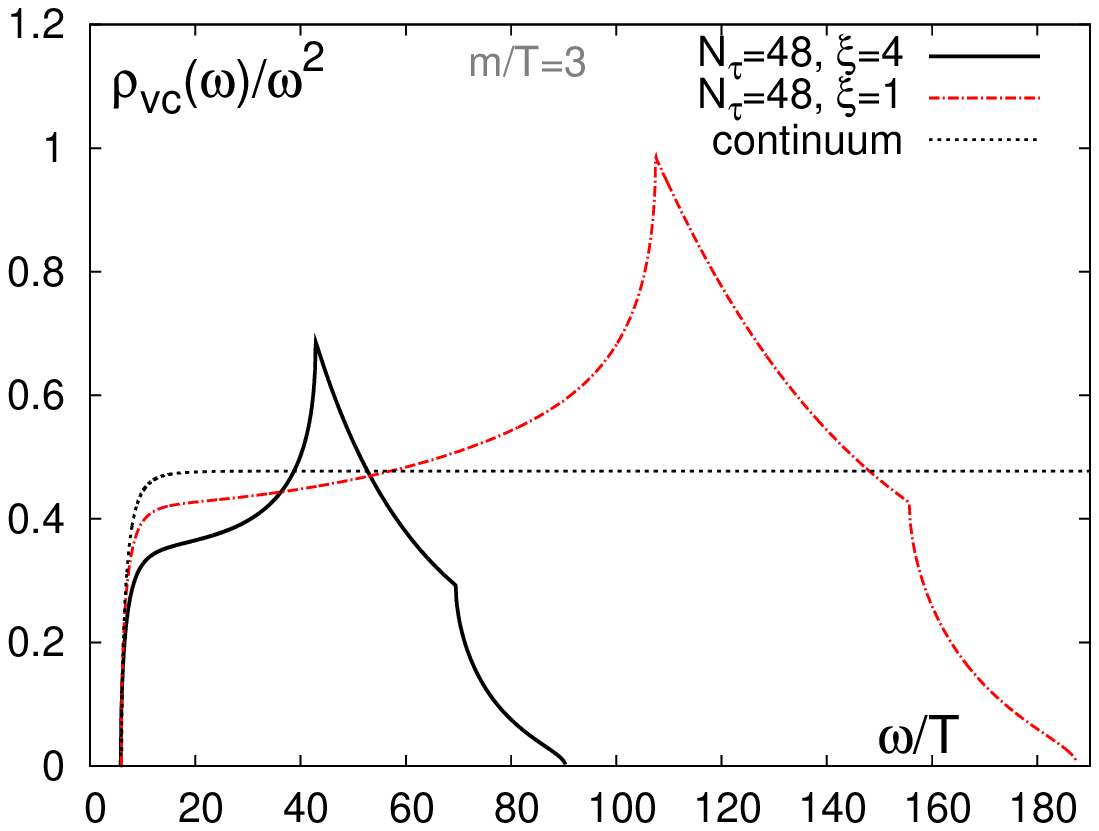}~\\
\caption{Left: free spectral functions on the isotropic lattice ($\xi=1$) versus free spectral functions in the continuum limit. The vector spectral functions $\rho_{\rm vc}(\omega)/\omega^2$ are plotted as function of $\omega/T$ for $N_{\tau}=24,$ 32, 48 at fixed $T$. Right: the free lattice spectral function on the isotropic lattice ($\xi=1$) versus the free spectral function on the anisotropic lattice ($\xi=4$). The spectral functions $\rho_{\rm vc}(\omega)/\omega^2$ are plotted as function of $\omega/T$ for $N_{\tau}=48$ at fixed $T$. In both plots the value of quark mass $m$ by $T$ is fixed to be 3.}
    \label{fig:spf_lat-cont}
  \end{center}
  \end{figure}

On lattices with finite temporal extent $N_\tau$ the spectral functions suffer from lattice cutoff effects. As shown in the left plot of Fig.~\ref{fig:spf_lat-cont}, the free lattice spectral functions for the Wilson fermion discretization, which is used in this work on isotropic lattices of temporal extent $N_{\tau}=24,$ 32, 48, strongly deviate from the free continuum spectral function in the large $\omega$ region. In contrast to the continuum case the lattice spectral function starts from $\omega_{min}/T=2N_{\tau}\log(1+ma_{\sigma}/\xi)$. It has two cusps at $2N_{\tau}\log(1+(2+ma_{\sigma})/\xi)$ and $2N_{\tau}\log(1+(4+ma_{\sigma})/\xi)$ and vanishes at $\omega_{max}/T=2N_{\tau}\log(1+(6+ma_{\sigma})/\xi)$~\cite{Karsch03}. Here $ma_{\sigma}$ is the value of the quark mass $m$ in units of spatial lattice spacing $a_{\sigma}$, $\xi$ is the anisotropic factor, i.e. the ratio of lattice spacing in the spatial direction over that in the temporal direction, $\xi=a_{\sigma}/a_{\tau}$. In Fig.~\ref{fig:spf_lat-cont} the value of quark mass $m/T=mN_{\tau}a_{\sigma}/\xi$=3. These lattice cutoff effects can be well separated from the region of physics interests when the number of points in the temporal direction $N_{\tau}$ is large, i.e. the lattice spacing at fixed temperature $T=1/N_{\tau}a_{\tau}$ becomes small. In this work we use isotropic lattices, i.e. $\xi=1$.  In order to increase $N_{\tau}$, which is very crucial 
in the spectral function analysis, an economic way is to perform simulations on anisotropic lattices. For instance, the anisotropic factor ratio $\xi=4$ or larger has typically been used in previous calculations~\cite{Umeda05,Asakawa04,Aarts07,Jakovac07}. However, the lattice spectral functions on anisotropic lattices ($\xi >1$) are much more distorted. As seen from the right plot of Fig.~\ref{fig:spf_lat-cont}, which shows the free lattice spectral function with $N_{\tau}=48$, 
the lattice spectral function on the $\xi=4$ anisotropic lattice vanishes at a much smaller energy, almost half of that on the isotropic lattice. As a consequence,
the two cusps move closer to the region of physics interests. Thus lattice spectral functions obtained on anisotropic lattices are more contaminated by lattice cutoff effects, e.g. lattice simulations with $N_{\tau}$ on the $\xi=4$ anisotropic lattices roughly correspond to those with $N_{\tau}/2$ on isotropic lattices. We therefore prefer to work on an isotropic lattice, although it is much more time consuming to generate gauge field configurations.

At finite temperature, i.e. in the interacting case,  due to the conservation of the vector current, the spectral function $\rho^{V}_{00}$ in the $\gamma^0$ channel 
will contribute a $\tau$ independent constant to the correlator
\bea
&&\rho^{V}_{00} = 2\pi\,\chi_{00}\, \omega\delta(\omega), \\
&&G^{V}_{00} = T\chi_{00}.
\eea
In the spectral function $\rho^{V}_{ii}$ in the $\gamma^{i}$ channel 
on the other hand the $\omega\delta(\omega)$ contribution present at infinite temperature changes into a smeared peak at finite temperature. From linear response theory the
shape of this peak is expected to be a Breit-Wigner like distribution~\cite{Petreczky05}
\bea
\rho^{V}_{ii}(\omega \ll T) = 2 \chi_{00}\frac{T}{M}\frac{\omega\eta}{\omega^2+\eta^2}\,, ~~~~~ \eta=\frac{T}{MD}.
\label{eq:spf_trans}
\eea  
Here $M$ is the mass of the heavy quark, $\eta$ is the drag coefficient and $D$ is the heavy quark transport coefficient defined in Eq.~(\ref{eq:HQ_diffusion_formula}). 
The contribution from Eq.~(\ref{eq:spf_trans}) to the correlation function is generally called the smeared zero mode contribution.

\section{Details of lattice simulations}
\label{sec:lattice_simulation}

In this work we present results based on quenched lattice QCD simulations performed on isotropic lattices using $\mathcal{O}(a)$-improved
Wilson (clover) fermions. The simulation parameters are shown in Table \ref{table:parameters}. As we are interested in temporal correlation functions there is a need for very fine lattices in order to have enough
data points in the temporal direction and reduce lattice cutoff effects. We thus performed simulations on lattices with lattice spacing ranging from $0.01$fm to $0.03$fm corresponding to the bare gauge couplings $\beta=6/g^2=7.793$, 7.457 and  6.872. At these $\beta$ values the lattice spacing has been determined from the string tension parameterization with $T_c/\sqrt{\sigma}=0.630(5)$ and $\sqrt{\sigma}=428~$MeV~{\cite{Edwards98,Beinlich:1997ia}}. The simulated temperatures range from about 0.75 $T_c$ to 3 $T_c$. Simulations have been performed at $T\approx0.75~T_c$ and $T\approx1.5~T_c$ with three different lattice spacings.
This allows an estimate of the magnitude of lattice cutoff effects. For the higher temperatures, $T\approx 2.2~T_c$ and $T\approx 2.9~T_c$, simulations are done only on the finest lattice. The number of correlator data points in the temporal direction is more than doubled compared to the finest lattice used in our previous study~\cite{Datta04}. To reduce the volume dependence on such fine lattices, we use a large spatial lattice $N_{\sigma}$=128. The spatial extent thus ranges from $1.3$ fm on the finest lattice to 3.9 fm on the coarsest lattice, which is in all cases significantly larger than the charmonium diameter. A subset of these lattices was used previously for the study of light meson spectral functions in Ref.~\cite{dilepton10}.

  \begin{table}[htdp]
\begin{center}
\begin{tabular}{ccccccccc}
\hline
\hline
$\beta$       & $a$[fm]  &   $a^{-1}$[GeV]     &  $L_{\sigma}$[fm]&$c_{\rm SW}$   &   $\kappa$   &      $N_{\sigma}^{3} \times N_\tau$  &      T/$\Tc$    & $N_{conf}$    \\
\hline
6.872           &   0.031    &     6.43                 & 3.93                           & 1.412488    &    0.13035    &        $128^{3} \times 32 $                &      0.74      &  126           \\
                      &                 &                               &                                     &                      &                      &           $128^{3} \times 16 $                &      1.49        &   198 \\
7.457           &  0.015     & 12.86                 & 1.96                             & 1.338927    &    0.13179    &          $128^{3} \times 64 $                &      0.74      &  179             \\
                      &                  &                              &                                      &                     &                      &          $128^{3} \times 32 $                &      1.49        &   250 \\
 7.793           &  0.010     & 18.97                & 1.33                             & 1.310381    &    0.13200    &         $128^{3} \times 96 $                &      0.73      &   234   \\
                      &                  &                              &                                      &                      &                      &          $128^{3} \times 48 $                &      1.46        &     461 \\
                       &                 &                              &                                     &                      &                      &           $128^{3} \times 32 $                &      2.20      &  105\\
                        &                &                              &                                      &                      &                      &          $128^{3} \times 24 $                &      2.93       & 81 \\
                        \hline
\hline
\end{tabular}
\end{center}
\caption{Lattice parameters and number of configurations used in the analysis with a clover improved Wilson fermion action. }
\label{table:parameters}
\end{table}

\begin{table}[ht]
\small
\begin{center}
\begin{tabular}{ccccccccc}
\hline
\hline
$\beta $  & $\kappa$       & $\kappa_c$ &   $am_b$    &  $T/\Tc$ & $N_\tau$ & $am_{\rm AWI}$ & $m_{\rm RGI}$[GeV] & $m_{\MSbar}(m)$[GeV]\\
\hline
6.872      & 0.13035     & 0.13497       &   0.13130    &   0.74     &    32           &   0.13305(2)         &  1.592(4)                             &1.255(2)     \\
                &                      &                       &                      &  1.49       & 16              & 0.13305(2)           &   1.592(4)                           & 1.255(2)      \\
7.457     &    0.13179    &   0.13398     &  0.06201 &      0.74     &  64             &  0.065430(6)        &  1.4742(3)                         & 1.1739(2)  \\
                &                      &                       &                      &  1.49        & 32              &  0.065352(4)       &    1.4734(8)                         &1.1733(6)\\
7.793     &    0.13200    &  0.13346     &  0.04143        & 0.73     & 96             & 0.044245(7)        &  1.358(3)                             &1.093(2)  \\
                &                      &                       &                      &  1.46        & 48             & 0.044222(2)         &    1.357(2)                          &1.094(1) \\
                &                      &                       &                      &  2.20      & 32             & 0.044280(6)        &   1.359(3)                               &1.096(2)\\ 
                &                      &                       &                      &  2.93           & 24             & 0.04420(1)     &      1.357(3)                            & 1.095(2)\\
                                            \hline
\hline
\end{tabular}
\end{center}
\caption[Quark masses on available lattices.]{Quark masses on available lattices. Here $m_b$ stands for the bare quark mass, $m_{\rm AWI}$ is obtained from the Axial Ward Identity at the scale of $\mu=1/a$ and $m_{\MSbar}(m)$ denotes the renormalized quark mass in the $\MSbar$ scheme at the scale of $\mu=m_{\MSbar}(\mu)$.}
\label{table:QuarkMass}
\end{table}

All gauge field configurations were generated using a heat bath algorithm combined with  5 over-relaxation steps, whereby neighboring 
configurations are separated by 500 sweeps.  For the fermion part the $\cO(a)$ nonperturbatively improved Sheikholeslami-Wohlert action~\cite{SW85} has been implemented in our simulation with nonperturbatively determined clover coefficients, $c_{\rm SW}$~\cite{Luescher96}, listed in Table~\ref{table:parameters}. The inversion of the Dirac matrix is carried out by using the 
 Conjugate Gradient (CG) algorithm. At the smallest lattice spacing, i.e. at $\beta=7.793$, we measured two-point correlation functions on lattices of size $128^3\times96$, $128^3\times48$, 
 $128^3\times32$ and $128^3\times24$ corresponding to temperatures $0.73~T_c$, $1.46~T_c$, $2.20~T_c$ and $2.93~T_c$, respectively. Since the temporal extent of the lattices is large and the exponentially decreasing correlation functions consequently become small at large distances, a rather stringent residue of $10^{-24}$ in the CG algorithm has been implemented in our simulations.

 The nonperturbatively improved clover action used in our calculations removes $\cO(a)$ discretization errors. However, in calculations with heavy quarks, discretization errors of order $am$ can also be large. We have estimated the quark mass values using the Axial Ward Identity (AWI) to compute the so called AWI current quark mass, $m_{\rm AWI}$, and the related Renormalization Group Invariant quark mass $m_{\rm RGI}$, for the different lattice data sets~\cite{deDivitiis:1997ka,Capitani:1998mq,Guagnelli:2000jw}. Here we used a nonperturbatively improved axial-vector current with coefficient $c_A$ taken from Ref.~\cite{Luescher96}. The commonly quoted quark mass for the heavy quark is the mass at its own scale. We therefore scaled $m_{\rm RGI}$ in the $\MSbar$ scheme to the scale $\mu=m$, where the evolution of $m_{\MSbar}(\mu)$ to $\mu$ is done using perturbative renormalization group functions known with four-loop accuracy~\cite{vanRitbergen97,Vermaseren97,Chetyrkin00}. The resulting quark masses are listed in Table~\ref{table:QuarkMass}. The $m_{\rm AWI}$ is independent of temperature, which consequently makes also the RGI quark mass $m_{\rm RGI}$ and running quark mass $m_{\MSbar}(m)$ temperature independent. On the finest lattice $am_{\MSbar}(\mu=m)$ is around 0.06 and thus the discretization errors proportional to the quark mass should be small. 
 After adjusting the quark mass parameters for our calculations it a posterior turned out that the $\Jpsi$ mass on the finest lattice is around $10\%$ larger than the physical $\Jpsi$ mass. In the other two cases our choice of parameters reproduces the $\Jpsi$ mass very well.

 As a local current, Eq.~(\ref{local_current}), is used in our calculations, it needs to be renormalized,
  \be
 J_{H}^{cont} = 2\kappa \,Z_{H}(a,m,\mu=1/a)\,J_{H}^{lat}\,a^{-3}~~.
 \ee
 The renormalization factors $Z_H(a,m,\mu=1/a)$ are estimated using one-loop tadpole improved perturbation theory
 \be
Z_{H}(am_q,g^2_{\overline{MS}}) = Z_{H}(am_q=0,g^2_{\overline{MS}},a\mu=1)\left(1\,+\,b_H(g^2_{\overline{MS}})am_q\right),
\label{relation:Massive_Z}
\ee
where $Z_{H}(am_q=0,g^2_{\overline{MS}},a\mu=1)$ are the renormalization constants for the massless quark case. $Z_{H}(am_q=0,g^2_{\overline{MS}},a\mu=1)$ has been determined perturbatively with two-loop accuracy for all the channels~\cite{2loop1,2loop2,tadpole2} and nonperturbatively for vector and axial-vector channels~\cite{Luescher97}. The coefficients $b_H(g^2_{\overline{MS}})$ can be expanded in powers of the gauge coupling,
\be
b_H(g^2_{\overline{MS}})=1\,+\,C_F\,b_H\,g^2_{\overline{MS}},
\ee
These coefficients have been calculated at one-loop level~\cite{Sint97,Sint98} and in particular, $b_H$ for the vector channel has been 
determined nonperturbatively~\cite{Luescher97}. The resulting renormalization factors used in our calculations are given in Table~\ref{table:Z_massive}.

\begin{table}[htdp]
\begin{center}
\begin{tabular}{|c|c|c|c|c|c|c|}
\hline        
$\beta$      &     $Z_{SC}$              &       $Z_{PS}$         &       $Z_{VC}$         & $Z_{AV}$\\
\hline
{6.872}   &   0.92         &        0.98         &       0.97       &  0.99\\         
           
                                           \hline
{7.457}    &  0.87         &        0.93       &       0.92       &  0.93\\
                                                                                 \hline
                                                                                 
{7.793}     & 0.87                &   0.92             &         0.91      & 0.92 \\

\hline 
\end{tabular} 
\end{center}
\caption{Renormalization constants of local operators for different channels.}
 \label{table:Z_massive}
\end{table}

 To check the magnitude of discretization errors, we analyzed the dispersion relation of the mesons. At nonzero ``momentum" ($\vecp_{\bot}\neq0$ or $\omega_n \neq0$), the exponential drop of the spatial correlator may be described by an energy $E_{sc}$
\be
G(z,\vecp_\bot,\omega_n)\sim \exp(-E_{sc}z),~~~~E^2_{sc}=\vecp_{\bot}^2+\frac{\omega_n^2}{A^2} + m_{sc}^2,
\label{def:scr_mass_dispersion_relation}
\ee 
where $\omega_{n}=2\pi nT$ are the Matsubara frequencies, $\vecp_{\bot}$ is the transverse momentum, and $m_{sc}$ is the screening mass which can differ from the pole mass if $A(T)\neq 1$. It is worth noting that the above ansatz is based on the dispersion relation in the continuum limit.

  \begin{figure}[!t]
      \begin{center}
    \includegraphics[width=.6\textwidth]{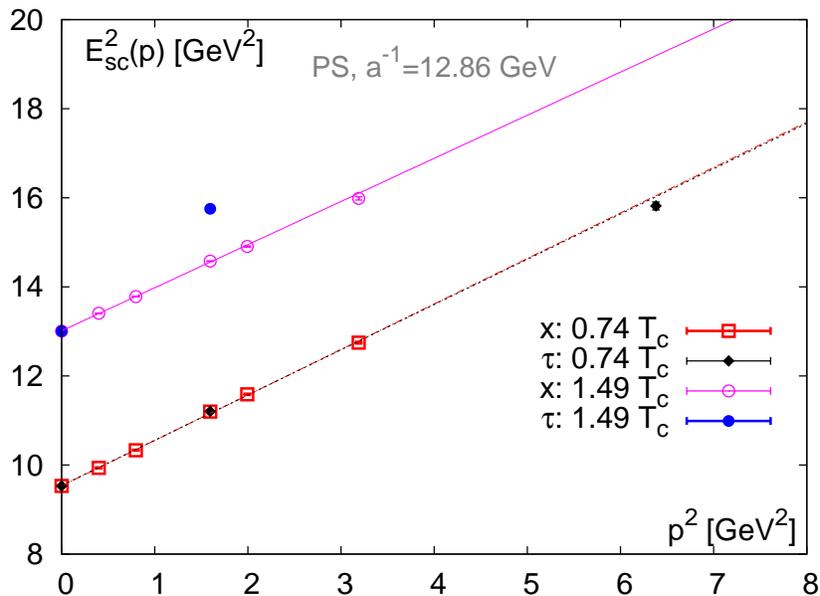}  
    \caption{The dispersion relation of the screening mass in the $PS$ channel obtained from lattices with $\beta=7.457$. Labels in the figure indicate whether spatial (``x") or temporal components (``$\tau$") of $(p_x, p_y, p_\tau)$ were chosen to be nonzero. The lines denote the dispersion relation obtained by fitting with the form of Eq.~(\ref{def:scr_mass_dispersion_relation}).}
    \label{fig:dispersion_relation_PS_beta7p457}
  \end{center}
  \end{figure}
  
  We show the dispersion relation of the screening mass in the $PS$ channel in Fig.~\ref{fig:dispersion_relation_PS_beta7p457}. The results are obtained from calculations performed on $128^3\times N_{\tau}$ lattices at 0.74 $T_c~(N_\tau=64)$ and at 1.49 $T_c~(N_\tau=32)$ with $a^{-1}=12.86~$GeV. Labels in the figure indicate whether spatial (``x") or temporal components (``$\tau$") of $(p_x, p_y, p_\tau)$ were chosen to be nonzero. The lines denote the dispersion relation obtained by fitting with an $Ansatz$ of $E_{sc}^2(p)=ap^2+b$. At 0.74 $T_c$, for the results from the spatial directions, we have a good fit with parameters $a$=1.02$\pm$0.01 and $b$=9.530$\pm$0.013. The applicability of the $Ansatz$ $ap^2+b$ indicates that our lattice is very close to the continuum limit; for the results from the temporal direction, even though we only have 3 data points, at this temperature, the data points have the same behavior as those from the spatial direction. We also performed a $\chi^2$ fit and obtained $a$=1.01$\pm$0.03 and $b$=9.539$\pm$0.033. The slope parameter $a$ here equals $A^{-2}$ in Eq.~(\ref{def:scr_mass_dispersion_relation}). The proximity of $a$ to 1 confirms that at $0.74~T_c$ the screening mass is a good approximation for the pole mass. The meson masses obtained from the spatial correlation functions at $T<T_c$ are shown in Table~\ref{table:Meson_Mass}. When going to the higher temperature of $1.49~T_c$, the data points from the temporal direction differ strongly from the fitting line for the results from spatial directions. Thus, the temporal direction is distinguished from the spatial direction and the breaking of Lorentz symmetry is clearly observed at this temperature. We also note that the screening mass at $1.49~T_c$ is about 10\% larger than the mass determined in the confined phase.
 
  \begin{table}[!t]
\begin{center}
\begin{tabular}{ccccc}
\hline
\hline
                   &                  &         Mass in GeV       &                      & \\
$\beta$     &  $\Jpsi$  &$\eta_c$ & $\chi_{c1}$ & $\chi_{c0}$ \\
\hline
6.872         &  3.1127(6)      & 3.048(2) & 3.624(36) & 3.540(25) \\
7.457         &  3.147(1)(25)  & 3.082(2)(21) & 3.574(8) & 3.486(4) \\
7.793         &  3.472(2)(114)  & 3.341(2)(104) & 4.02(2)(23) & 4.52(2)(37) \\
\hline
\hline
\end{tabular}
\end{center}
\caption{Meson masses (in GeV) for different charmonium states. The errors in the first bracket are statistical errors and the errors in the second bracket are systematic errors from effects of the physical distance.}
\label{table:Meson_Mass}
\end{table}

In the following sections, we restrict ourselves to the case of vanishing momentum and suppress the $\vecp$ indices.

   \section{Euclidean Correlators above $T_c$}
   \label{sec:rec_cor}

Following Ref.~\cite{Datta04} we introduce the ``reconstructed" correlator at temperature $T$ from a spectral function determined at temperature $T^{\prime}$
\be
G_{\rec}(\tau,T;T^{\prime})= \int_{0}^{\infty} \mathrm{d}\omega ~K(\tau,T,\omega)~\rho(\omega,T^{\prime}).
\ee
$G_{\rec}(\tau,T;T^{\prime})$ at the temperature $T$ is computed from a spectral function at the temperature $T^{\prime}$ and an integral kernel $K$ at the temperature $T$. 
 In the following subsections, we will study the Euclidean correlation functions at $T>T_c$ with respect to a reconstructed correlation function that uses a spectral function obtained at $T^{\prime}<T_c$.

\subsection{Remarks on the reconstructed Euclidean correlation function}
We want to compare  correlators calculated at $T>T_c$ to those at $T<T_c$, i.e. we consider the ratio of 
the measured correlator to the reconstructed correlator~\cite{Datta04}, 
 \be
\frac{G(\tau,T)}{G_{\rec}(\tau,T;T^{\prime})} = \frac{\int_{0}^{\infty} \mathrm{d}\omega ~K(\tau,T,\omega)~\rho(\omega,T)}{\int_{0}^{\infty} \mathrm{d}\omega ~K(\tau,T,\omega)~\rho(\omega,T^{\prime})}.
\label{rec_cor}
 \ee
This reduces the influence of the trivial temperature dependence of the kernel $K(\tau,T,\omega)$ in the correlation function. If the ratio is equal to unity at all distances it would suggest that the spectral function does not vary with temperature. 
 In fact, in order to obtain the reconstructed correlator at temperature $T$ from a
spectral function at $T^{\prime}$ one does not require any knowledge of the spectral function
at that temperature. It suffices to know the correlator at $T^{\prime}$.

To arrive at the desired correlation function at temperature $T$, we first exploit the following relation ~\cite{Ding10} which is a generalization of the relation derived in Ref.~\cite{Meyer10},

\be
 \frac{\cosh[\omega(\tilde{\tau}-N_{\tau}/2)]}{\sinh (\omega N_{\tau}/2)}~~\equiv~\sum_{\tilde{\tau}^{\prime}=\tilde{\tau};~\Delta\tilde{\tau}^{\prime}=N_{\tau}}^{N_{\tau}^{\prime}-N_{\tau}+\tilde{\tau}} \frac{\cosh[\omega(\tilde{\tau}^{\prime}-N_{\tau}^{\prime}/2)]}{\sinh (\omega N_{\tau}^{\prime}/2)},
\label{kernel_rules}
\ee
where $T^{\prime}=(a N_{\tau}^{\prime})^{-1},~~T=(aN_{\tau})^{-1},~~\tilde{\tau}^{\prime}=(\tau^{\prime}/a)\in[0,~N_{\tau}^{\prime}-1],~~\tilde{\tau}=(\tau/a)\in[0,~N_{\tau}-1],~~N_{\tau}^{\prime}=m~ N_{\tau},~~m\in\mathbb{Z}^{+}$. $N_{\tau}$ and $N_{\tau}^{\prime}$ are the number of time slices in the temporal direction at temperature $T$ and $T^{\prime}$, respectively; $\tilde{\tau}$ denotes the time slice of the correlation function at temperature $T$ while $\tilde{\tau}^{\prime}$ denotes the time slice of the correlation function at 
temperature $T^{\prime}$. The sum over $\tilde{\tau}^{\prime}$ on the right hand side of  Eq.~(\ref{kernel_rules}) starts from $\tilde{\tau}^{\prime}=\tilde{\tau}$ with a step length of $\Delta\tilde{\tau}^{\prime}=N_{\tau}$ and ends at the upper limit $N_{\tau}^{\prime}-N_{\tau}+\tilde{\tau}$. After multiplying both sides of Eq.~(\ref{kernel_rules}) with $\rho(\omega,T^{\prime})$  and performing the integration over $\omega$, one immediately arrives at
\be
G_{\rec}(\tilde{\tau},T;T^{\prime}) = \sum_{\tilde{\tau}^{\prime}=\tilde{\tau};~\Delta\tilde{\tau}^{\prime}=N_{\tau}}^{N_{\tau}^{\prime}-N_{\tau}+\tilde{\tau}}  G(\tilde{\tau}^{\prime},T^{\prime}),
\label{eq:Grec_data}
\ee
which shows that $G_{\rec}(\tau, T; T^{\prime})$ is obtained directly by using the correlator $G(\tau^{\prime}, T^{\prime})$ at $T^{\prime}$. Using relation~(\ref{eq:Grec_data}) we can calculate $G_{\rec}(\tau ,T;T^{\prime})$ directly from the correlator data at temperature $T^{\prime}$. 
An immediate consequence clearly is that one has a better control over systematic errors in the calculation of ratios used in Eq.~(\ref{rec_cor}). In the following subsections, we will implement Eq.~(\ref{eq:Grec_data}) to calculate the reconstructed correlators and compare them with the measured correlation functions. We will discuss what can be learned about the modification of spectral functions  from the analysis of correlation functions. In the following sections we will suppress the index $T^{\prime}$ in the $G_{\rec}$.

\subsection{Ratios of $G(\tau, T)$ to $G_{\rec}(\tau ,T)$ }

\begin{figure}[htbl]
 \begin{center}
\includegraphics[width=.45\textwidth]{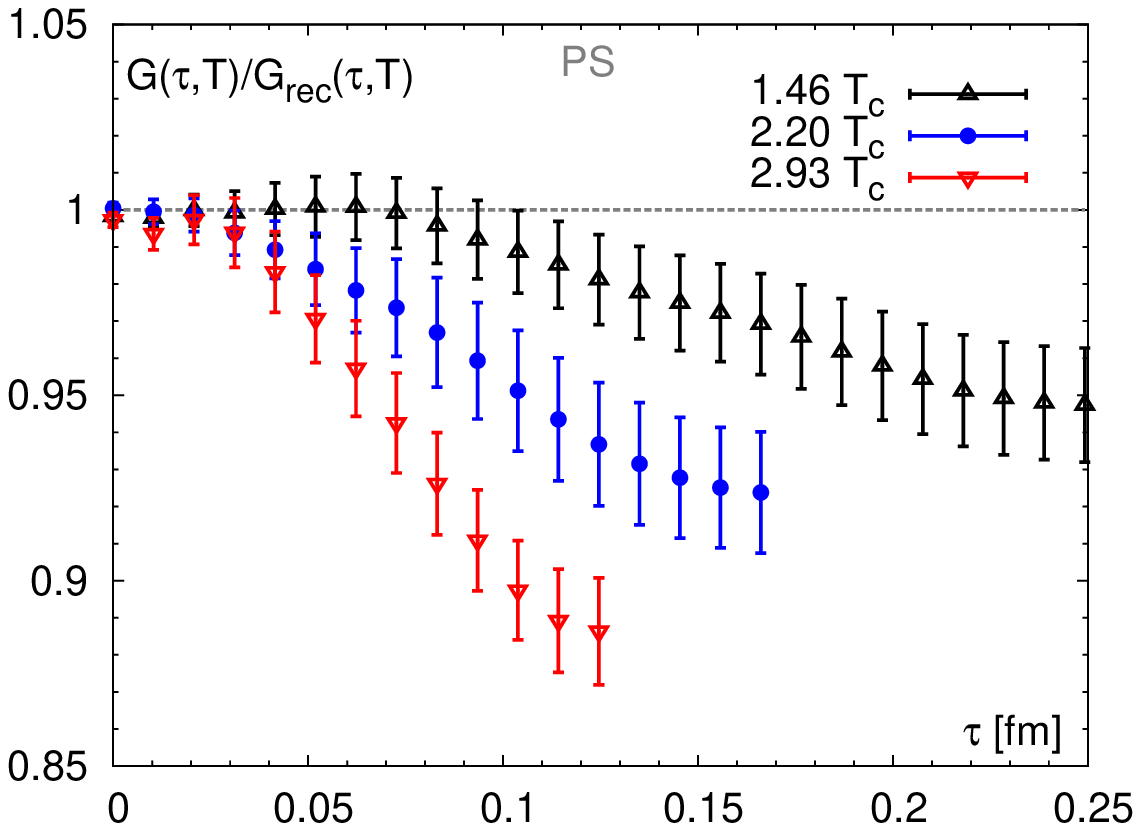}~
~~\includegraphics[width=.45\textwidth]{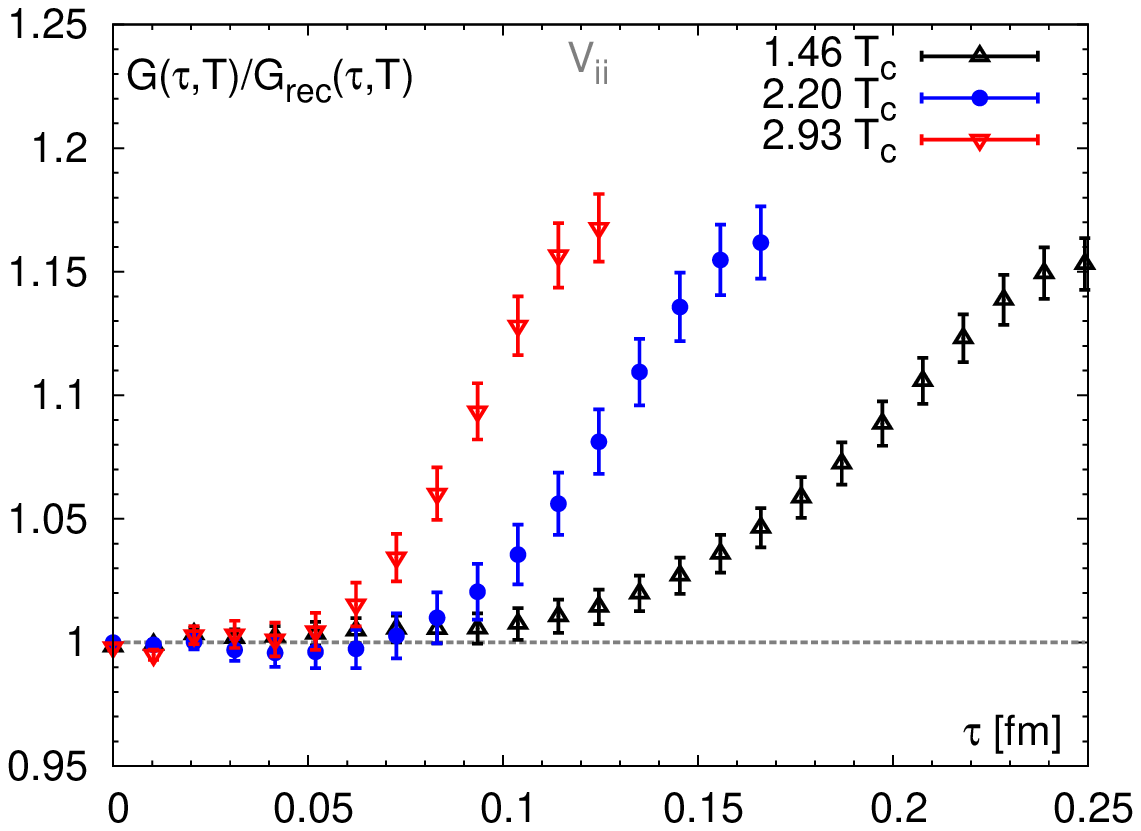}
\caption{The ratio $G(\tau,T)/G_{\rm rec}(\tau,T)$ for $PS$ (left) and $V_{ii}$ (right) channels as a function of the Euclidean distance $\tau$ on our finest lattice with $\beta=7.793$ ($a=0.01$ fm)
at $T=1.46$, 2.20 and 2.93$~T_c$. The reconstructed correlator $G_{\rec}$ is obtained directly from correlator data at $0.73~T_c$.}
 \label{fig:GoverGrec_Swave_beta7p793}
 \end{center}
\end{figure}

We first investigate the temperature dependence of the pseudoscalar correlators.  We show the numerical results for $G(\tau,T)/G_{\rm rec}(\tau,T)$ at 1.46$~T_c$, 2.20$~T_c$ and $2.93~T_c$ on our finest lattice in the left plot of Fig.~\ref{fig:GoverGrec_Swave_beta7p793}. $G_{\rec}(\tau,T)$ are evaluated from the correlator data at  $T^{\prime}=0.73~T_c$ using Eq.~(\ref{eq:Grec_data}). Note that the error bars shown in the plots 
are statistical errors obtained from a Jackknife analysis. As seen from the left plot of Fig.~\ref{fig:GoverGrec_Swave_beta7p793}, the ratio $G(\tau,T)/G_{\rm rec}(\tau,T)$ approaches unity at small distances and starts to deviate from unity at larger distances. Temperature effects start to set in at about $0.06$ fm at $1.46~T_c$ and make the ratio smaller than unity. Deviations are about $5\%$ at the largest distance. The small temperature dependence of the pseudoscalar correlator might indicate that the corresponding spectral function is subject to only small thermal modifications. When going to the higher temperature, $2.20~T_c$, the temperature effects set in at a smaller distance ($\approx 0.03$ fm). The ratio rapidly drops and the deviation from unity ($\approx 8\%$) becomes larger at the largest distance $\tau=1/2T$. Turning to the highest available temperature, i.e. $2.93~T_c$, the temperature effects also set in at around 0.03 fm and the deviation of the ratio from unity increases to about 12\% at the largest distance. This may suggest considerable modifications of the lowest state in the $PS$ channel at this temperature.

The ratio $G(\tau,T)/G_{\rm rec}(\tau,T)$ for the vector correlator $V_{ii}$ (summing over spatial components only) on our finest lattice is shown in the right plot
of Fig.~\ref{fig:GoverGrec_Swave_beta7p793}. Clearly this plot shows that the temperature dependence of $G(\tau,T)/G_{\rm rec}(\tau,T)$ is quite different from that in the $PS$ channel. At all temperatures the ratios are larger than unity. This is already an indication that  different temperature dependent contributions arise in the vector channel at large distances, related to the low frequency region in the spectral function. The temperature effects set in at larger distances compared to pseudoscalar correlators: around 0.1, 0.08 and 0.06 fm at 1.46, 2.20 and 2.93 $T_c$, respectively. A unique feature seen in the $V_{ii}$ channel is that the magnitude of $G(\tau,T)/G_{\rm rec}(\tau,T)$ at the largest distance $\tau=1/2T$ does not vary with temperature. All ratios deviate from unity by about
 16\%. In fact, the ratios seem to be to a good approximation a function of $\tau T$ only. 
 However, one has to be careful with the interpretation of this result in terms of bound state modifications 
as their effect may be compensated by possible positive diffusion contributions in the $V_{ii}$ channel at temperatures above $T_c$. We will examine this
in more detail in the next two subsections.

 \begin{figure}[htbl]
\begin{center}
  \includegraphics[width=.45\textwidth]{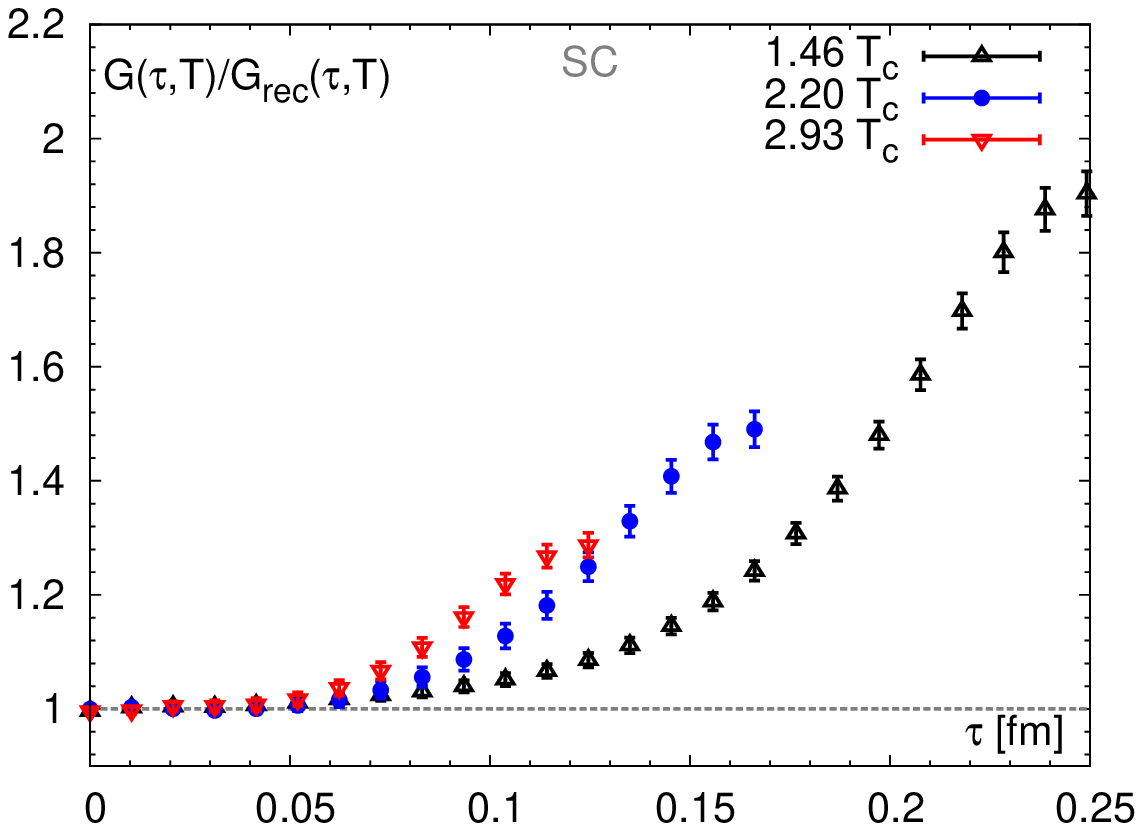}~\includegraphics[width=.45\textwidth]{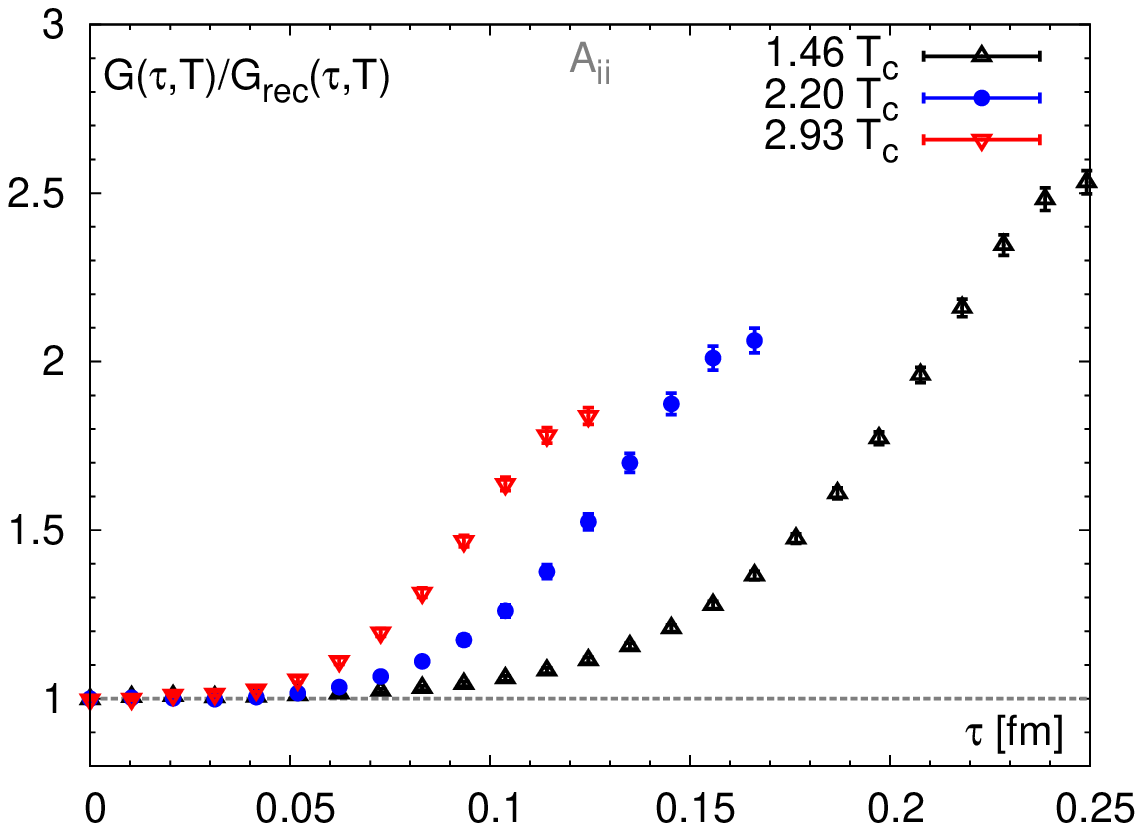}
  \caption{Same as Fig.~\ref{fig:GoverGrec_Swave_beta7p793} but for P wave states. The left plot is for the $SC$ channel and the right one is for the $A_{ii}$ channel.}
 \label{fig:GoverGrec_Pwave_beta7p793}
\end{center}
\end{figure}

The numerical results for the ratio $G/G_{\rec}$ for P wave states obtained on our finest lattice are shown in Fig.~\ref{fig:GoverGrec_Pwave_beta7p793}. The left plot of Fig.~\ref{fig:GoverGrec_Pwave_beta7p793} is for the scalar channel  while the right plot is for the axial-vector channel. 
They show similar features as we have seen in S wave correlators: at short distance the ratio is close to unity while at the large distances the ratio
deviates from unity and the deviations start at shorter distance at higher temperatures.  We find a significant deviation of $G/G_{\rec}$ from unity in both channels already at $1.46~T_c$: at the largest distance $\tau=1/2T$, $G/G_{\rec}$ reaches about 1.9 in the SC channel and about 2.5 in the $A_{ii}$ channel. This deviation at the largest distance is much larger compared to the case of  S wave correlators and the magnitude of this deviation at the largest distance decreases with increasing temperatures. In order to connect these features with the thermal modification
of bound states, one needs to separate the contribution of the smeared zero mode at low frequency, which is present in $SC$ and $A_{ii}$ channels.

We also compared the ratios $G(\tau , T)/G_{\rec}(\tau , T)$ obtained from the finest lattice with those from the two coarser lattices, $\beta=6.872~(a=0.031~\mathrm{fm})$ and $\beta=7.457~(a=0.015~\mathrm{fm}$). We found that the lattice cutoff effects are small in these ratios. The ratios at the largest distance obtained on the finest lattices are about 7\% larger  than those on the coarser lattices. 
 This is mainly due to our choice of
quark mass parameters, which lead to somewhat larger charmonia masses
on our finest lattice. In fact, the exponential decrease of the ratio
of correlation functions, $G/G_{\rec}\sim \exp(-\Delta M\, \tau)$, is
controlled by the difference of effective meson masses $M_{\rm eff}$ below and above $T_c$, i.e.
$\Delta M\simeq M_{\rm eff}(T>T_c) - M_{\rm eff}(T<T_c)$. As the mass
$M_{\rm eff}(T\simeq 0.75~T_c)$ is larger on our finest lattice, the high temperature
mass $M_{\rm eff}(T\simeq1.5~T_c)$ differs less from the low temperature
value and thus leads to a smaller value of $\Delta M$. This in turn
leads to a smaller decrease in $G/G_{\rec}$ relative to the result on
the coarser lattice and explains the somewhat larger values for
$G/G_{\rm rec}$ on our finest lattice.

\subsection{Smeared zero mode contributions}
\label{sec:zeromode}

As discussed in Section~\ref{sec:intro}, there are zero mode contributions in $A_{ii}$, $SC$ and $V_{ii}$
channels in the infinite temperature limit. In the study of spectral functions at temperatures below $T_c$, which will be shown in Section~\ref{sec:spf_finite_T}, we found no zero mode contributions
in $SC$, $PS$ and $V_{ii}$ channels at $0.73~T_c$. Thus smeared zero mode contributions (or smeared $\delta$ functions) are expected to arise in  $A_{ii}$, $SC$ and $V_{ii}$ channels only at temperatures above $T_c$. In $A_{ii}$, $SC$ and $V_{ii}$ channels, the information on bound states and smeared zero modes are thus entangled in the low frequency region and may partly compensate each other. 
In order to retrieve reliable information on possible bound states it is 
therefore necessary to filter out or separate the smeared zero mode contribution. To emphasize this point we note that previous studies suggested that the temperature dependence of $G/G_{rec}$ is in fact mainly due to zero mode contributions~\cite{Umeda07,Datta:2008dq,Petreczky:2008px}. On the other hand, the smeared zero mode contribution is interesting in its own. 
For example in the vector channel it is related to the diffusion (process) of
a single quark in the medium. In this subsection we will discuss the evidence we have for smeared zero mode contributions and thermal modifications of bound states at the correlation function level in more detail.

To get a better understanding of the $\tau$ dependence of the ratios defined in Eq.~(\ref{rec_cor}), 
we perform a Taylor expansion of the correlators at the largest distance accessible at finite temperature
\bea
\label{eq:TaylorExpansion}
G(\tau ,T) &&=\int_0^{\infty}\frac{\mathrm{d}\omega}{2\pi}~\rho(\omega)~\frac{\cosh(\omega(\tau-1/2T))}{\sinh(\omega/2T)}\\
                &&=\int_0^{\infty}\frac{\mathrm{d}\omega}{2\pi}~\frac{\rho(\omega)}{\sinh(\omega/2T)}\Big[1+\frac{1}{2!}\left(\frac{\omega}{T}\right)^2(\tau T-\frac{1}{2})^2 + \frac{1}{4!}\left(\frac{\omega}{T}\right)^4(\tau T-\frac{1}{2})^4+\cdots \Big] \nonumber.
                \eea
This allows us to explore the properties of the low frequency behavior of the spectral function.
Here we define the Taylor expansion coefficients, i.e. the time derivatives of the Euclidean correlation functions,
\be
G^{(n)}=\frac{1}{n!}~\frac{\mathrm{d}^nG(\tau,T)}{\mathrm{d}(\tau T)^n}\Bigg|_{\tau T=1/2}~ =~ \frac{1}{n!}\int_0^{\infty}\frac{\mathrm{d}\omega}{2\pi}~\left(\frac{ \omega}{T}\right)^n~ \frac{\rho(\omega)}{\sinh(\omega/2T)},
\label{eq:def_moments}
\ee
as thermal~moments~\cite{dilepton10}. By going to higher order  thermal moments, one probes higher frequency region in the spectral function. In particular the value of the zeroth order thermal moment $G^{(0)}$ is the same as the value of the correlator at the  symmetry point, $G(\tau T=1/2)$. We have extracted the zeroth and second order thermal moments ($G^{(0)}$ and $G^{(2)}$) from correlation functions on the finest lattices. The results are shown in Table~\ref{table:moments}.

We rewrite the Taylor expansion of the Euclidean correlators $G(\tau , T)$ (Eq.~(\ref{eq:TaylorExpansion})) as 
\be
G(\tau , T)=G^{(0)}~\sum_{n=0}^{\infty}\,R_{2n,0}\,\left(\tau T-\frac{1}{2}\right)^{2n},~~R_{n,m}\equiv \frac{G^{(n)}}{G^{(m)}}.
\ee
Thus the ratio of measured correlator to the reconstructed correlator can be expanded as
\bea
\frac{G(\tau ,T)}{G_{\rec}(\tau ,T)} = \frac{G^{(0)}}{G^{(0)}_{\rec}} \left( 1 + \big(R^{2,0}-R^{2,0}_{\rec}\big)\,\left(\tau T-\frac{1}{2}\right)^2+\cdots\right),
\label{eq:GoverGrec_expansion}
\eea
which shows that the sign of $R^{2,0}-R^{2,0}_{\rec}$ determines whether $G(\tau ,T)/G_{\rec}(\tau ,T)$ is decreasing or increasing with $\tau T$
at large distances. Take the S wave states for example. From Table~\ref{table:moments} we find that $R^{2,0}-R^{2,0}_{\rec}$ is negative in the $V_{ii}$
channel and positive in the $PS$ channel at all three temperatures above $T_c$. It indicates that $G(\tau ,T)/G_{\rec}(\tau ,T)$ increases with $\tau T$ in the $V_{ii}$ channel and deceases with $\tau T$ in the $PS$ channel at large distances at $T>T_c$, which is consistent with Fig.~\ref{fig:GoverGrec_Swave_beta7p793}.

\begin{table}[ht]
\begin{center}
\begin{tabular}{c  c c c c c c c c}
\hline
\hline
channel   &  $T/\Tc$ & $G^{(0)}/T^3$ & $G^{(0)}_{\rec}/T^3$ &$G^{(2)}/T^3 $ & $G^{(2)}_{\rec}/T^3$  &$ \Delta G^{(0)}/T^3$&$\Delta G^{(2)}/T^3$ & $\Delta G^{(4)}/T^3$\\
\hline

$V_{ii}$         &  1.46       & 0.955(5)   &  0.829(6)  & 46.39(4) & 46.43(7)   &   0.126(8)  &  -0.04(8)  \\
                        &  2.20      &  1.81(2)     & 1.561(9)  & 57.3(2)    & 59.2(1)      &  0.25(2)    &  -1.9(2)\\  
                        &  2.93       & 2.33(2)     & 1.99(1)   & 59.6(3)       & 62.6(3)     &  0.34(2)     &  -3.0(4)\\
$PS$             &  1.46       & 0.858(8)    &  0.91(1)    & 44.73(3) & 45.74(7)    &  -0.05(1)   & -1.01(7)\\
                        &  2.20      &  1.44(2)     & 1.56(2)  &  52.6(1) & 54.2(1)       &  -0.12(3)    & -1.6(1)\\  
                        &  2.93       & 1.68(2)      & 1.90(2)  & 51.2(1)  & 54.1(2)       &  -0.22(3)   & -2.9(2)\\
$A_{ii}$          &  1.46       & 0.708(6)  &  0.280(3)  & 23.80(4) & 22.71(2)  & 0.428(7) & 1.13(2)  & 45(4) \\
                        &  2.20      &   1.57(3)    & 0.761(6)   &  39.9(2)  & 40.0(1)    & 0.81(3)   & -0.1(2)\\  
                        &  2.93       &2.18(3)      &1.186(7)     & 47.9(2)   & 48.8(3)    & 0.99(3)   & -0.9(3)\\ 
$SC$              &  1.46       & 0.493(5)  &  0.259(5)  & 21.32(2) & 20.15(3) & 0.234(7) &  1.12(1)  & 33(2) \\
                        &  2.20      &   0.99(2)    & 0.665(7)   & 32.3(1)  & 32.21(9)   &  0.33(2)  &  -0.1(1)\\  
                        &  2.93       & 1.26(2)     & 0.980(9)    & 35.2(1)  & 36.2(2)     &  0.28(2)  &    -1.0(2)\\

                                            \hline
\hline
\end{tabular}
\end{center}
\caption{Thermal moments extracted from correlator data on the finest lattice ($\beta=7.793$). $\Delta G^{(n)}$ is the difference between $G^{(n)}$ and $G^{(n)}_{\rec}$ as defined in Eq.~(\ref{eq:deltaGn}).}
\label{table:moments}
\end{table}

As mentioned before the contribution from (smeared) zero modes and bound states to the correlation function is difficult to disentangle by investigating 
the ratios $G(\tau ,T)/G_{\rm rec}(\tau ,T)$. 
To investigate the modification of bound states, one has to separate the zero mode contribution. Since the $\omega\delta(\omega)$ term only contributes to $G^{(0)}$, one can construct  quantities that only include the higher order thermal moments. One possibility is to look 
at the ratio of  differences of the correlators at neighboring Euclidean time slices to the difference of the corresponding reconstructed correlators,
\be
\frac{G^{\rm diff}(\tau,T)}{G^{\rm diff}_{\rm rec}(\tau,T)} \equiv \frac{G(\tilde{\tau},T) -  G(\tilde{\tau}+1,T)}{G_{\rm rec}(\tilde{\tau},T) -  G_{\rm rec}(\tilde{\tau}+1,T)}.
\label{eq:GoverGrec_diff}
\ee
This approximates the ratio of the time derivative of the measured correlators to the time derivative of the reconstructed correlators at $\tilde{\tau}+1/2$. Here $\tilde{\tau}=\tau/a$ and the difference of correlators $\tilde{G}(\tau, T)=G(\tau, T)-G(\tilde{\tau} +1,T)$ can be expanded as
\be
\tilde{G}(\tau, T)=\tilde{G}^{(1)}~\sum_{n=0}^{\infty}\,\tilde{R}_{2n+1,1}\,\left(\tau T-\frac{1}{2}+\frac{1}{2N_\tau}\right)^{2n+1},
~~~\tilde{R}_{n,m}\equiv \frac{\tilde{G}^{(n)}}{\tilde{G}^{(m)}}.
\ee
where 
\be
\tilde{G}^{(n)}=\frac{1}{n!}~\frac{\mathrm{d}^n\tilde{G}(\tau,T)}{\mathrm{d}(\tau T)^n}\Bigg|_{\tau T=1/2-1/2N_\tau}~ =~ -\frac{2}{n!}\int_0^{\infty}\frac{\mathrm{d}\omega}{2\pi}~\left(\frac{ \omega}{T}\right)^n~\rho(\omega)~\frac{\sinh(\omega/2N_{\tau}T)}{\sinh(\omega/2T)} ,
\label{eq:def_moments2}
\ee
Thus Eq.~(\ref{eq:GoverGrec_diff}) can be rewritten as 
\be
\frac{G^{\rm diff}(\tau,T)}{G^{\rm diff}_{\rm rec}(\tau,T)} =  \frac{\tilde{G}^{(1)}}{\tilde{G}^{(1)}_{\rec}} \left( 1 + \big(\tilde{R}^{3,1}-\tilde{R}^{3,1}_{\rec}\big)\,\left(\tau T-\frac{1}{2}+\frac{1}{2N_\tau}\right)+\cdots\right).
\label{eq:GoverGrec_diff_exp}
\ee
 Alternatively  one can  consider the ratio of midpoint subtracted correlators
\be
\frac{G^{\rm sub}(\tau,T)}{G^{\rm sub}_{\rm rec}(\tau,T)} \equiv \frac{G(\tau,T) -  G(N_{\tau}/2,T)}{G_{\rm rec}(\tau,T) -  G_{\rm rec}(N_{\tau}/2,T)} = \frac{G^{(2)}}{G^{(2)}_{\rec}} \left( 1 + \big(R^{4,2}-R^{4,2}_{\rec}\big)\,\left(\tau T-\frac{1}{2}\right)^2+\cdots\right). 
\label{eq:GoverGrec_midpoint-subtracted}
\ee

As seen from Eq.~(\ref{eq:GoverGrec_diff_exp}) and Eq.~({\ref{eq:GoverGrec_midpoint-subtracted}) the zeroth order thermal moment $G^{(0)}$ drops out in ratios $G^{\rm diff}(\tau,T)/G^{\rm diff}_{\rm rec}(\tau,T)$ and $G^{\rm sub}(\tau,T)/G^{\rm sub}_{\rm rec}(\tau,T)$.
Since a $\omega\delta(\omega)$ term in the spectral function only contributes to the
zeroth order thermal moment and,  moreover,  its contribution vanishes in the higher order moments $\tilde{G}^{(n\geq1)}$ and $G^{(n\geq 2)}$, it is thus possible
to completely remove the zero mode contribution in the two ratios $G^{\rm diff}(\tau,T)/G^{\rm diff}_{\rm rec}(\tau,T)$ and $G^{\rm sub}(\tau,T)/G^{\rm sub}_{\rm rec}(\tau,T)$. However, at finite temperature above $T_c$, the $\omega\delta(\omega)$ term is likely to be smeared out as a Breit-Wigner like distribution (Eq.~(\ref{eq:spf_trans})). This Breit-Wigner like distribution in the very low frequency region of the spectral function does not lead to a $\tau$ independent constant, and it contributes to the thermal moments at all orders. Thus the smeared zero mode contributions cannot be completely removed from the above two ratios. 
 However, these contributions, which are located only in the frequency region $\omega\lesssim T$ in the spectral function,  are suppressed at higher orders of the thermal moments due to the presence of a factor $(\frac{\omega}{T})^n$  in Eq.~(\ref{eq:def_moments}) and Eq.~(\ref{eq:def_moments2}).

 \begin{figure}[htbl]
 \centering
 \includegraphics[width=.45\textwidth]{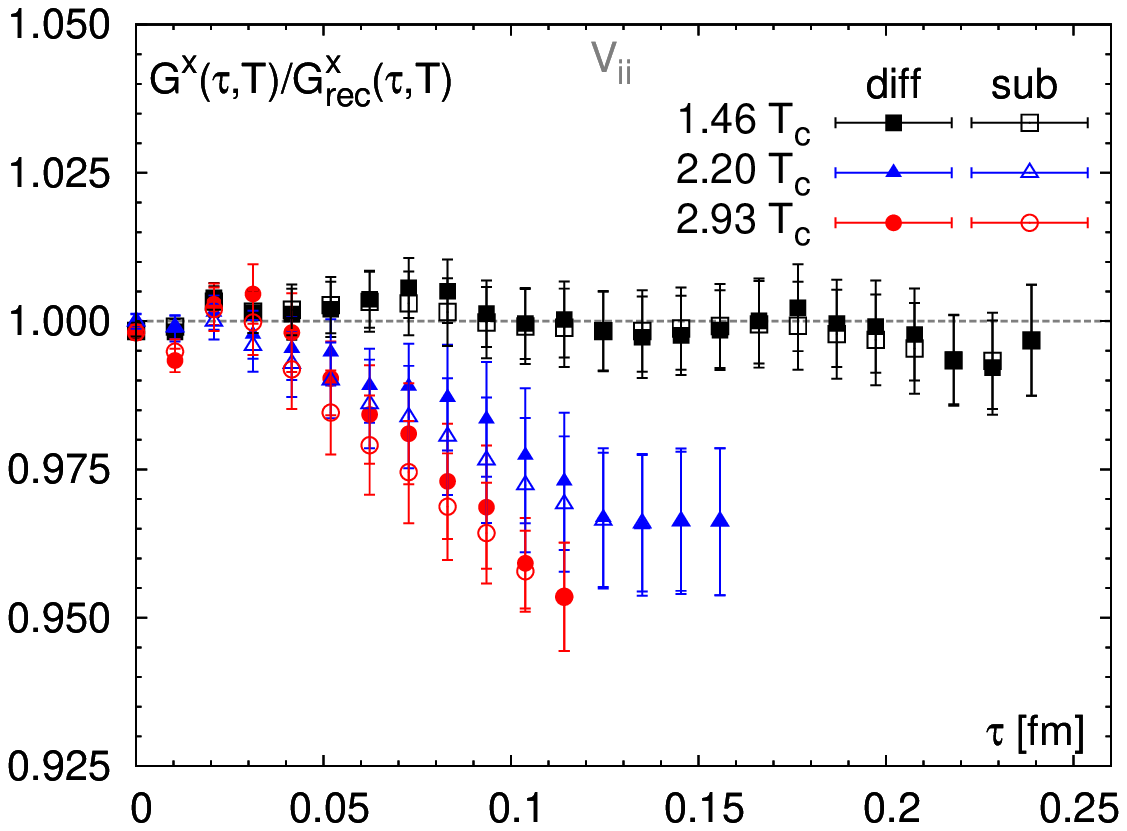}~\includegraphics[width=.45\textwidth]{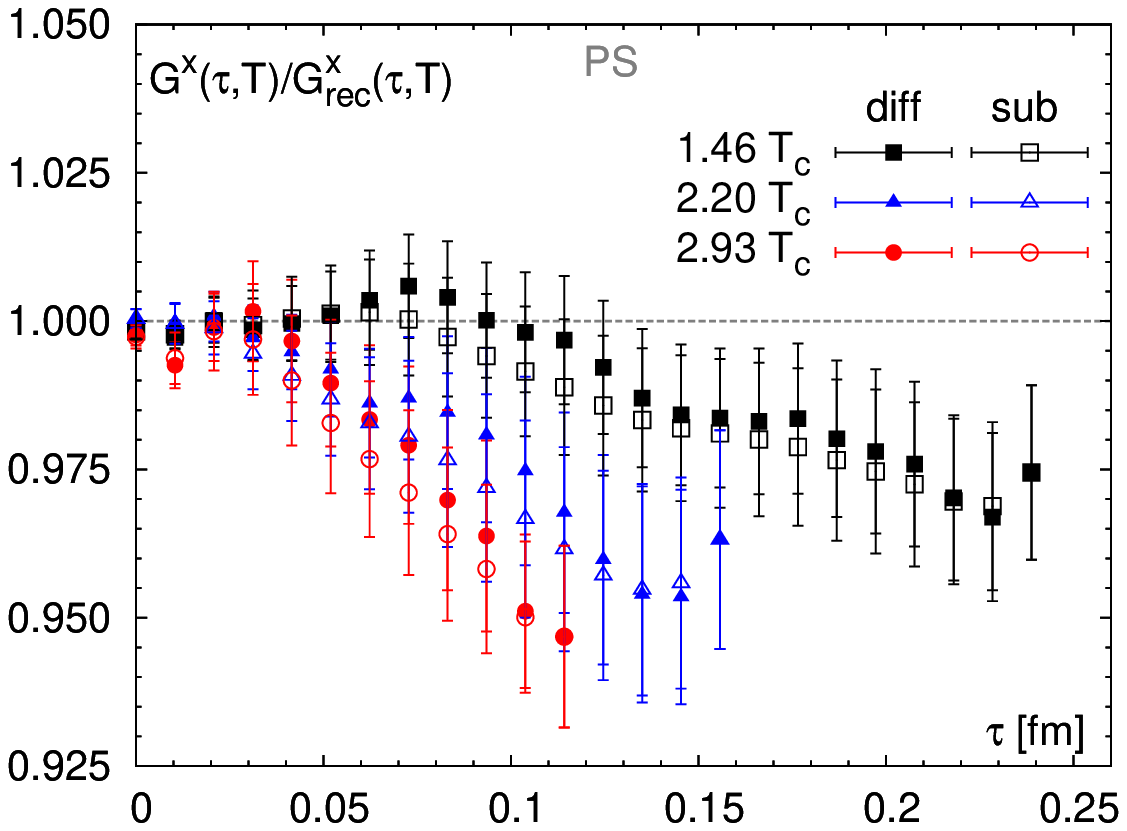}~
\caption{The ratio $G^{\rm diff}(\tau,T)/G^{\rm diff}_{\rec}(\tau,T)$ ($G^{\rm sub}(\tau,T)/G^{\rm sub}_{\rec}(\tau,T)$) of S wave sates as a function of the Euclidean distance $\tau$ on our finest lattice with $\beta=7.793$ at $T=1.46$, 2.20 and 2.93$~T_c$. The superscript ``x" denotes either ``diff" or ``sub". The left plot is for the $V_{ii}$ channel and the right one is for the $PS$ channel.}
 \label{fig:GoverGrec_Swave_diff_sub_beta7p793}
\end{figure}

 In Fig.~\ref{fig:GoverGrec_Swave_diff_sub_beta7p793} we show  results for~$\RGsub $ and $\RGdiff $ in the $V_{ii}$ (left) and also in the $PS$ channel (right).
The open symbols denote the ratio $\RGsub $ while filled symbols label the ratio $\RGdiff $. The ratios $\RGsub$ and $\RGdiff$ give similar results at all distances. 
In the $V_{ii}$ channel we observe that values of $\RGsub $ and $\RGdiff $ are much smaller than those of $G/G_{\rm rec}$ at large distances.
The values at the largest distance are reduced by almost 15\%. In the $PS$ channels deviations of $\RGsub $ and $\RGdiff $ from unity are also reduced compared to those of  $G/G_{\rm rec}$ at large distances.
However, the change is not as large as in the $V_{ii}$ channel. The values at the largest distance are increased only by about 3\% at both $1.46~T_c$ and $2.20~T_c$ and about 6\% at $2.93~T_c$.
The larger changes that occur in the $V_{ii}$ channel correlators
relative to those in the $PS$ channel may be understood
in terms of large smeared zero mode contributions that contribute
in the vector channel and get almost completely eliminated in subtracted
correlation functions. Such contributions do not seem to be present in the
$PS$ correlator and the resulting changes are thus smaller. We will confirm this interpretation through
the explicit construction of the spectral functions in the next section.
We also note that in the subtracted correlators there are
significant differences in the pseudoscalar and vector channels. In fact,
in the subtracted correlators the situation now seems to be reversed
compared to the unsubtracted correlators. At $T=1.46 T_c$ the subtracted
vector correlator now stays close to unity at all distances $\tau T$, while
we observe a clear drop in the pseudoscalar correlator. 
Of course, one has to keep in mind that the subtracted correlators also modify the spectral contributions in the
bound state region and thus may also suppress contributions that result from modifications of the bound states.
The weak temperature dependence seen in Fig.~\ref{fig:GoverGrec_Swave_diff_sub_beta7p793} thus does not necessarily mean that the bound states suffer negligible thermal modifications.
We will look into this issue more closely in the next subsection. At the two higher temperatures, $T=2.20 T_c$ and $2.93 T_c$, the subtracted correlators
in both channels show almost identical behavior.

 \begin{figure}[htbl]
 \centering
 \includegraphics[width=.45\textwidth]{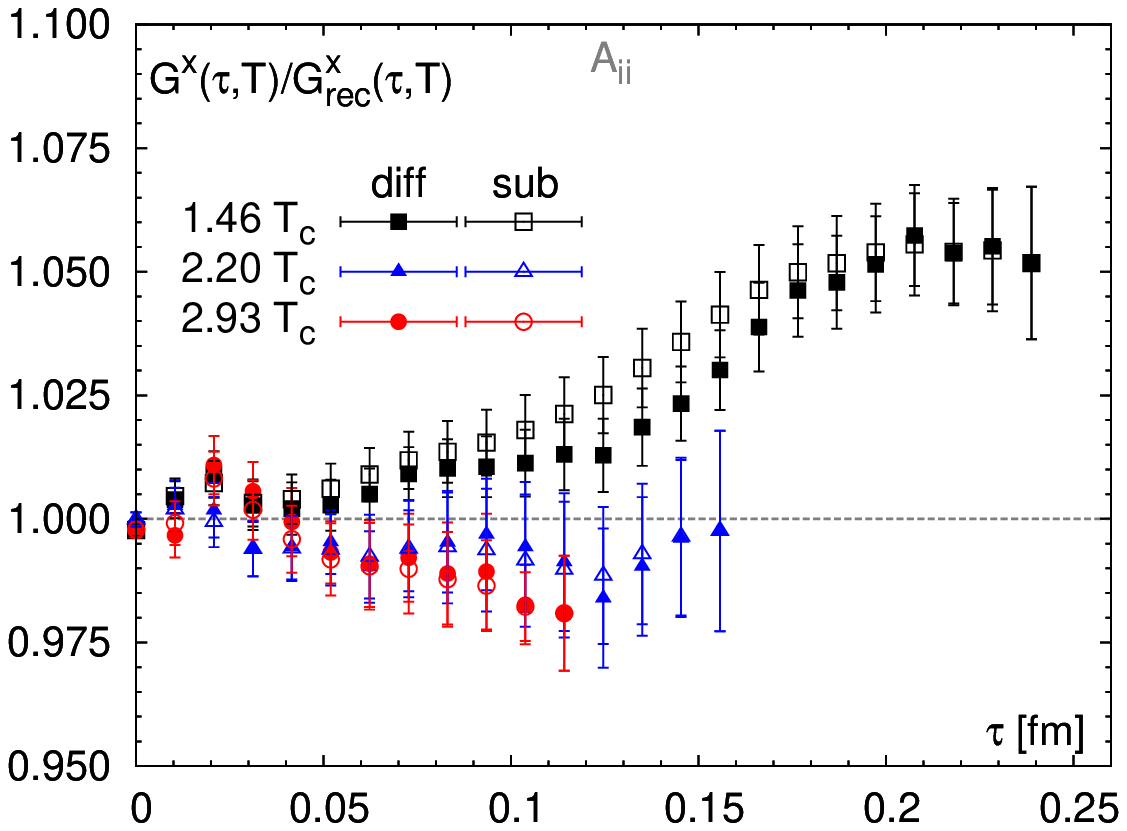}~\includegraphics[width=.45\textwidth]{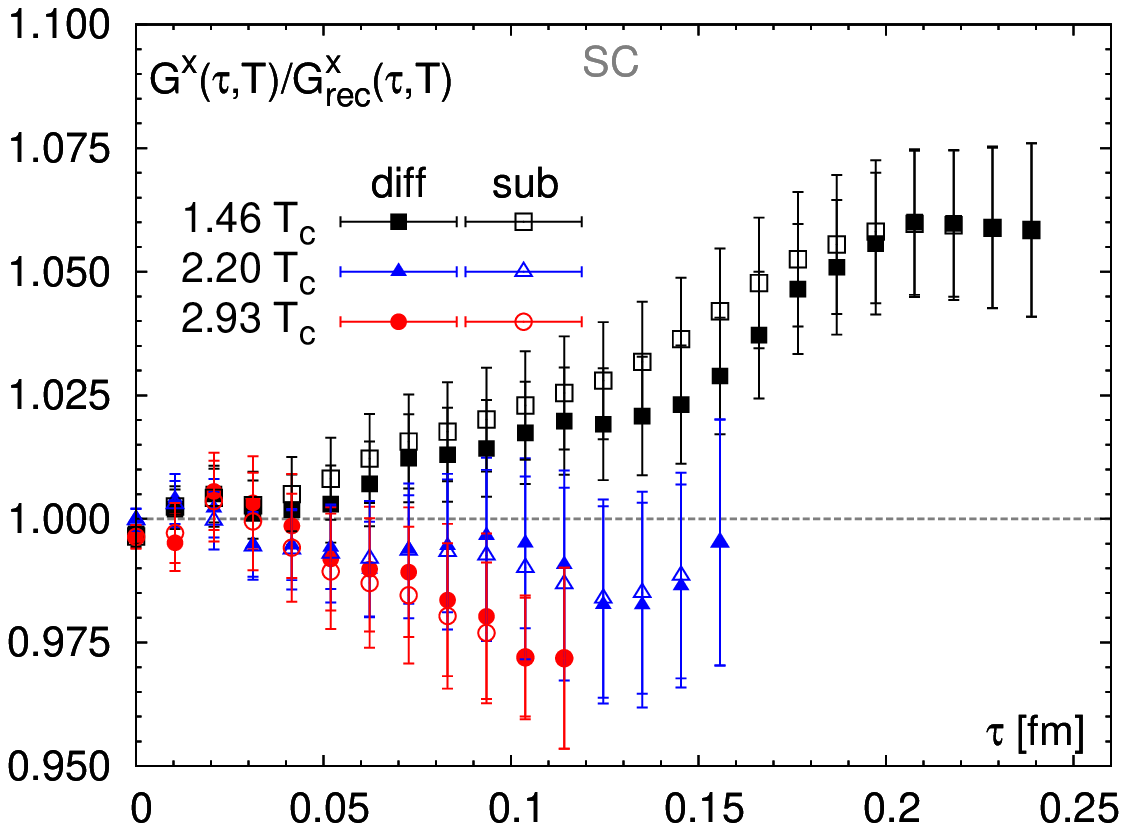}
\caption{Same as Fig.~\ref{fig:GoverGrec_Swave_diff_sub_beta7p793} but for P wave states. The left plot is for the $A_{ii}$ channel and the right one is for the $SC$ channel.} \label{fig:GoverGrec_Pwave_diff_sub_beta7p793}
\end{figure}

In addition to the S wave states we also examined $\RGsub$ and $\RGdiff$ for the P wave states. The corresponding results are shown in Fig.~\ref{fig:GoverGrec_Pwave_diff_sub_beta7p793}, where the left plot is for the $A_{ii}$ channel while the right plot is for the $SC$ channel. The magnitudes of the ratios for both $A_{ii}$ and $SC$ channels are greatly reduced compared to the ratios shown in Fig.~\ref{fig:GoverGrec_Pwave_beta7p793}. This behavior is quite similar to the ratios in the $V_{ii}$ channel and it suggests that the strong rise seen in $G/G_{\rm rec}$ in Fig.~\ref{fig:GoverGrec_Pwave_beta7p793} could be partly due to smeared zero mode contributions. 
However, as mentioned in the case of the vector correlator, also in the $A_{ii}$ and $SC$ channels the bound state contributions get modified in subtracted correlators. We will examine this in more detail in the
next subsection.

\subsection{Difference between $G(\tau ,T)$ and $G_{\rec}(\tau ,T)$}
\label{sec:Gdiff}

The behavior of the ratios $G(\tau ,T)/G_{\rm rec}(\tau ,T)$, $\RGsub$ and $\RGdiff$ provides some insight into the relative importance of different frequency regions for the structure of correlators.
However, the contribution from smeared zero modes and bound states to the correlation function is still difficult to disentangle at this point.
Further information is gained by looking into the differences between measured correlators and reconstructed correlators
\bea
\label{eq:spf_diff}
\Delta G(\tau,T)/T^3= (G(\tau ,T) - G_{\rec}(\tau ,T))/T^3 &=& \int\frac{\md\omega}{2\pi}~\Delta\rho(\omega,T)K(\omega,\tau T)/T^3\\
\label{eq:spf_diff_moments}
&=& \frac{\Delta G^{(0)}}{T^3} +\frac{\Delta G^{(2)}}{T^3}\left(\tau T-1/2\right)^2 + \frac{\Delta G^{(4)}}{T^3}\left(\tau T-1/2\right)^4~+~\cdots,
\eea
where
 \be
 \Delta\rho(\omega,T)= \rho(\omega,T>T_c)-\rho(\omega,0.73T_c)
 \ee and 
\be
\Delta G^{(n)} = G^{(n)} - G^{(n)}_{\rec}.
\label{eq:deltaGn}
\ee
The difference between the measured correlator and the reconstructed correlator provides information on the difference of spectral functions below and above $T_c$. As can be seen from Eq.~(\ref{eq:GoverGrec_expansion}) and Eq.~(\ref{eq:GoverGrec_midpoint-subtracted}), the intercept ($\Delta G^{(0)}$) and curvature of $G(\tau ,T) - G_{\rec}(\tau ,T)$ at large distances ($\Delta G^{(2)}$) are related to the values of $G(\tau ,T)/G_{\rec}(\tau ,T)$ and $G^{\rm sub}(\tau ,T)/G^{\rm sub}_{\rec}(\tau ,T)$ at the largest distance, respectively.  The values of $\Delta G^{(0)}$ and $\Delta G^{(2)}$, obtained by performing a two-parameter quadratic fit to $(G(\tau ,T)-G_{\rec}(\tau ,T))/T^3$, are listed in
Table~\ref{table:moments}.

We first show the differences $(G(\tau ,T) - G_{\rm rec}(\tau ,T))/T^3$ for S wave states in Fig.~\ref{fig:GminusGrec_Swave_beta7p793}.
The first thing to notice is the change in the dependence
of $(G(\tau ,T) - G_{\rm rec}(\tau ,T))/T^3$ on Euclidean time
$\tau T$ as one raises the temperature. In $V_{ii}$ and  $PS$ channels we observe that $(G(\tau ,T) - G_{\rm rec}(\tau ,T))/T^3$
increases with $\tau T$ at all temperatures. This is also reflected in the differences of  second order thermal moments $\Delta G^{(2)}$, which are clearly negative
at the two highest temperatures in the $V_{ii}$ channel and at all temperatures in the $PS$ channel as seen from Table~\ref{table:moments}.
It is obvious that $\Delta G(\tau ,T)$
would decrease with $\tau T$ and $\Delta G^{(2)}$ thus would
be positive, if $\Delta\rho(\omega)>0$ for all $\omega$. 
Increasing differences of correlators $\Delta G(\tau, T)$ and negative
values for $\Delta G^{(2)}$ thus indicate that $\Delta\rho(\omega)$ is
negative in some energy range.  On the other hand,
one clearly cannot draw the reverse conclusion, i.e. we cannot rule
out that $\Delta\rho(\omega)$ is negative in some energy range even
if $\Delta G(\tau ,T)$ decreases with $\tau T$ and $\Delta G^{(2)}$ 
is positive. However, in that case regions with $\Delta\rho(\omega)<0$
need to be compensated by regions with an enhancement in
$\Delta\rho(\omega)>0$. To this extent it is worthwhile to note that
the presence of smeared zero mode contributions above $T_c$, which
do not have a counterpart below $T_c$, will give positive contributions 
to $\Delta\rho(\omega)$, while disappearing bound states will lead to
negative contributions to $\Delta\rho(\omega)$.  We thus conclude that there must be some energy regions in which $\Delta\rho$ is negative in the $V_{ii}$ channel at the two highest temperatures
and in the $PS$ channel at all temperatures we examined. At $1.46~T_c$~the second thermal moment $\Delta G^{(2)}$ is slightly smaller than zero in the $V_{ii}$ channel. It is manifested in the behavior that $G(\tau ,T)- G_{\rec}(\tau ,T)$ is almost flat at large distance and slightly increasing with distance at $1.46~T_c$. However, the increase with $\tau T$ is not statistically significant. If modifications of $\Delta\rho(\omega)$ would arise from the smeared zero mode only, i.e. in  the $\omega\lesssim T$ region, its contribution to correlation function would be either a constant at all distances or decreasing with distances. The former case corresponds to a $\omega\delta(\omega)$ term in $\Delta\rho(\omega)$ and the latter case corresponds to a smeared Breit-Wigner like distribution. Fig.~\ref{fig:GminusGrec_Swave_beta7p793} thus indicates that some modifications in the $\omega > T$ region of the $V_{ii}$ channel are likely to happen already at $1.46~T_c$. 
Combining the above discussion with the information on smeared zero mode
contributions gained from the analysis of the subtracted correlators 
discussed in the previous subsection (and the following section on
spectral functions) we conclude that Fig.~\ref{fig:GminusGrec_Swave_beta7p793} provides strong
evidence for modifications of the spectral functions of S wave states that lead to
$\Delta\rho(\omega)<0$ in some energy range for all $T\ge 1.46 T_c$.
Also note that the differences $(G(\tau ,T) - G_{\rm rec}(\tau ,T))/T^3$ are negative at all temperatures in the $PS$ channel. It is in contrast
to the positive values of $(G(\tau ,T) - G_{\rm rec}(\tau ,T))/T^3$ in the $V_{ii}$ channel. This too suggests the existence of a significant smeared zero mode contribution in the $V_{ii}$ channel and larger thermal modifications in the bound states in the $PS$ channel at $T>T_c$.

  \begin{figure}[htbl]
 \centering
 \includegraphics[width=.45\textwidth]{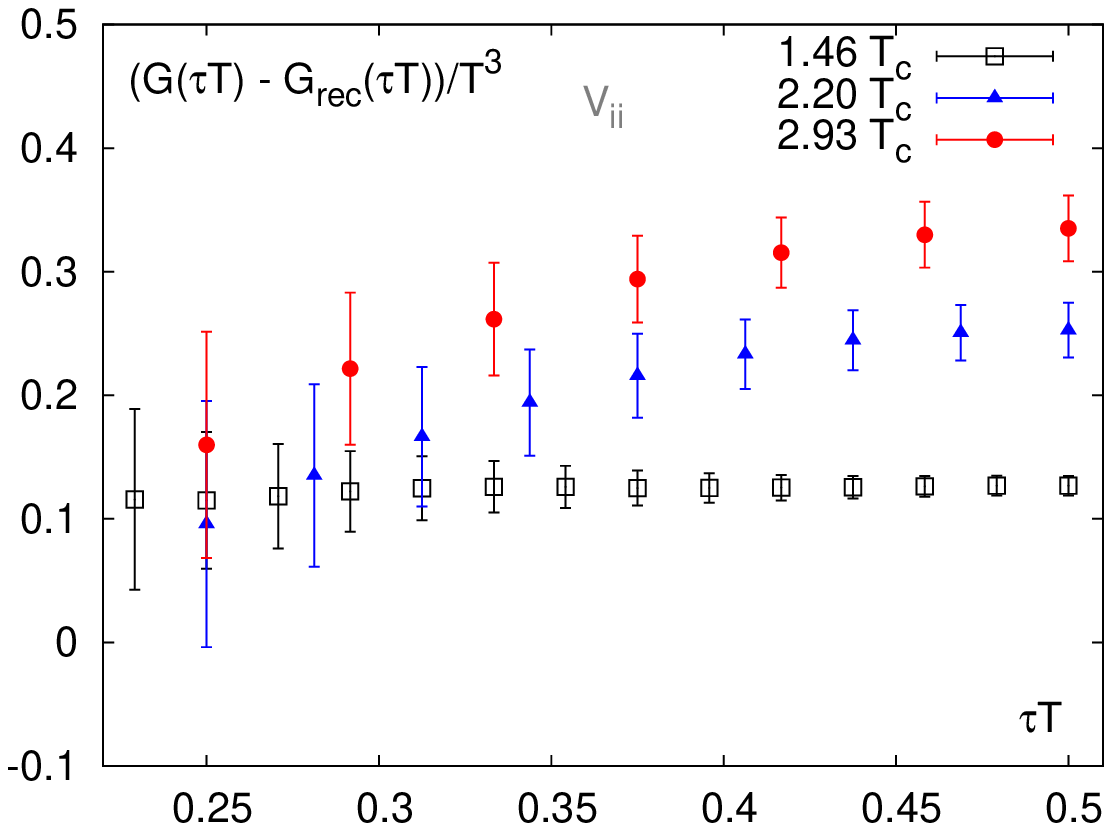}~ \includegraphics[width=.45\textwidth]{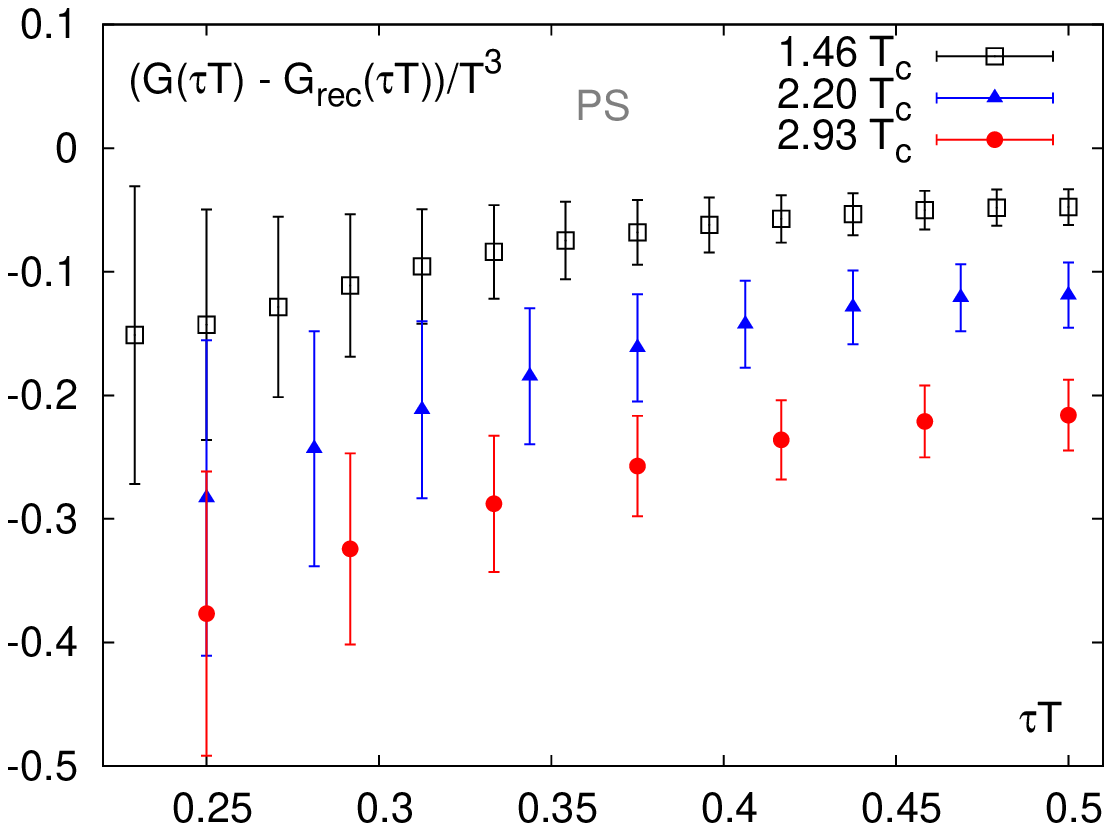}
\caption{$(G(\tau,T)-G_{\rec}(\tau,T))/T^3$ of S wave sates as a function of the Euclidean distance $\tau$ on our finest lattice with $\beta=7.793$ at $T=1.46$, 2.20 and 2.93$~T_c$.  The left plot is for the $V_{ii}$ channel and the right one is for the $PS$ channel.}
 \label{fig:GminusGrec_Swave_beta7p793}
\end{figure}

The results for the differences $(G(\tau ,T) - G_{\rm rec}(\tau ,T))/T^3$ in $A_{ii}$ and $SC$ channels
are given in Fig.~\ref{fig:GminusGrec_Pwave_beta7p793}. At the two highest temperatures $(G(\tau ,T) - G_{\rm rec}(\tau ,T))/T^3$ increases with $\tau T$
in both channels. This indicates that $\Delta \rho<0$ in some energy region in these two channels at the two highest temperatures.
However, the interpretation of the decreasing differences of correlation functions 
in $A_{ii}$ and $SC$ channels, which we observe at $1.46 T_c$,
is a bit more complex. These correlation functions receive positive
contributions to $\Delta\rho(\omega)$ from smeared zero modes and 
positive values for $\Delta \rho(\omega)$ thus will arise at small $\omega/T$,
i.e. for $\omega \lesssim T$. However, if in addition $\Delta\rho(\omega)$
would not change or stay positive also in the region $\omega > T$, the
higher order moments $\Delta G^{(n)}$ would still fulfill an inequality,
$\Delta G^{(0)} >  2\Delta G^{(2)} > 24\Delta G^{(4)}>0 $.
This inequality, however, does not hold in $A_{ii}$ and $SC$ 
channels at $1.46 T_c$ as is evident from Table~\ref{table:moments}. We 
thus conclude that also in the P wave spectral functions
at $1.46~T_c$ thermal modifications in the $\omega >T$ region occur
that lead to negative $\Delta\rho(\omega)$. It is thus plausible
that the P wave states disappear already at $T=1.46 T_c$ (as expected)
but smeared zero mode contributions in these channels are so large
that they still give the dominant contribution to the shape of 
$\Delta G(\tau ,T)$ at this temperature. It is also worthwhile to note that the magnitude of difference, $G(\tau ,T)- G_{\rec}(\tau ,T)$, is smaller at $2.93~T_c$ than at $2.20~T_c$ in the $SC$ channel. This too may be due to a partial cancellation of effects arising from the smeared zero mode contribution and those originating from modifications of the bound states.

  \begin{figure}[htbl]
 \centering
 \includegraphics[width=.45\textwidth]{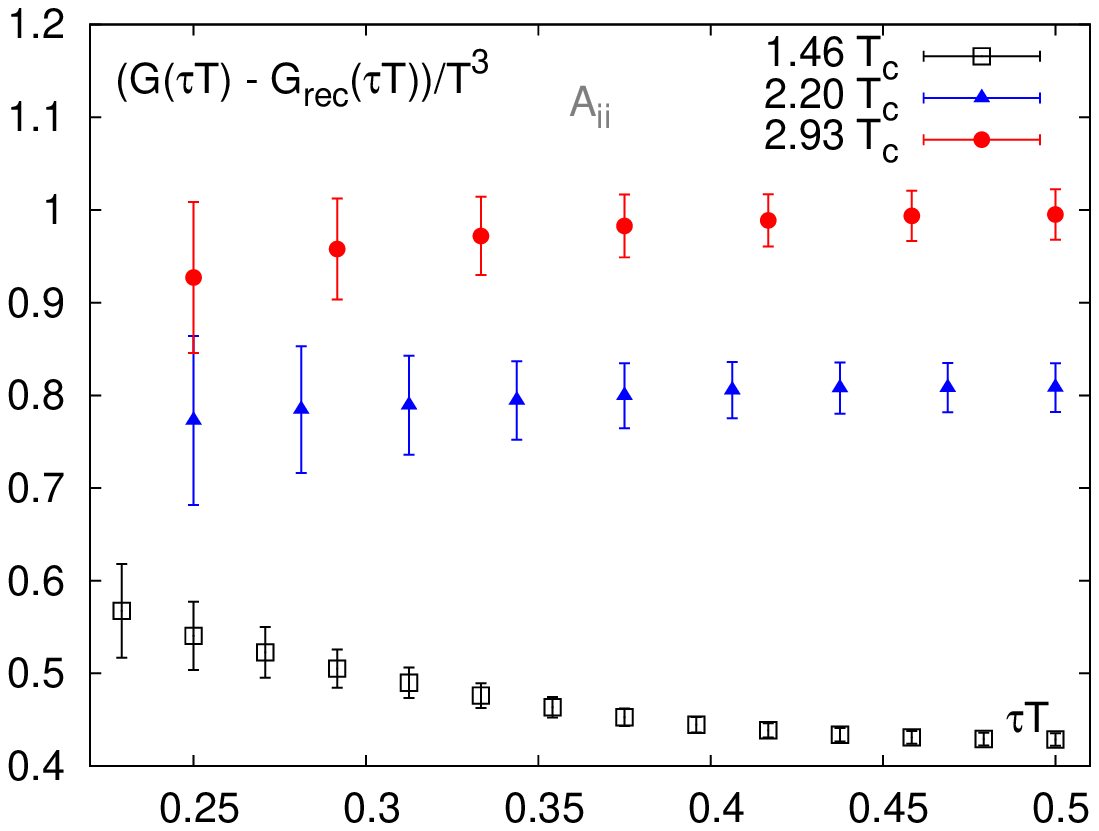}~ \includegraphics[width=.45\textwidth]{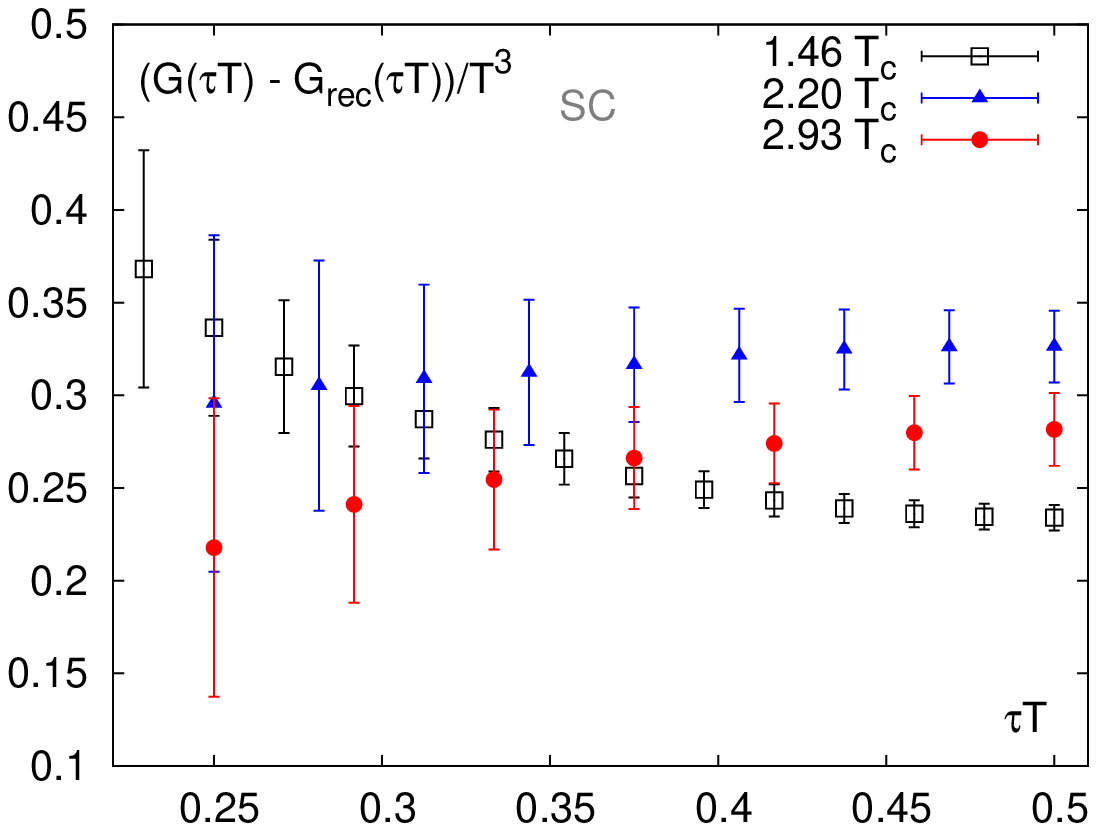}
\caption{Same as Fig.~\ref{fig:GminusGrec_Swave_beta7p793} but for P wave states. The left plot is for the $A_{ii}$ channel and the right one is for the $SC$ channel.}
 \label{fig:GminusGrec_Pwave_beta7p793}
\end{figure}

\subsection{Charm quark diffusion coefficient estimated from correlations functions}
\label{sec:diffusion_corr}

The charm diffusion coefficient is related to the smeared zero mode contribution in the $V_{ii}$ channel.
As the very low frequency structure of the spectral function should manifest itself most strongly at the largest distance of the correlation function, the symmetry point of the correlation function $G(\tau T=1/2)$ should be strongly influenced by the transport contributions.

At $1.46~T_c$, one may assume that the intermediate and high frequency region of the spectral function is similar to that at $0.73~T_c$.
Also based on the fact that there is no zero mode contribution at $0.73~T_c$ in the $V_{ii}$ channel, one could then estimate the charm diffusion coefficient by fitting the value of $G(1/2)-G_{\rec}(1/2)$.
Here we ignore the difference in the  intermediate and high frequency region at $0.73$ and $1.46~T_c$ and only use the ansatz given in Eq.~(\ref{eq:spf_trans}) for the transport peak to fit the value of $G(1/2)-G_{\rec}(1/2)$ at 1.46 $T_c$ . There is only one parameter, i.e. the heavy quark mass $M$,
that needs to be fixed to obtain the charm diffusion coefficient $D$.  We note that the correlation function calculated from Eq.~(\ref{eq:spf_trans}) at $\tau T=1/2$ decreases faster with increasing heavy quark mass  than with
decreasing $D$. Thus there exist a maximum value of quark mass beyond which no solution for $D$ exists. The maximum value of quark mass here is around 1.8 GeV.
As the quark mass extracted from correlation functions is around 1.0 GeV (see Table~\ref{table:QuarkMass}) we vary the charm quark mass from 1.0 GeV to 1.8 GeV. The charm diffusion coefficient $D$ multiplied by $2\pi T$ then ranges from $0.6$ to 3.6, i.e.
\bea
M=1.0~{\rm GeV}, ~~ 2\pi T D \approx 0.6 ,\\
M=1.8 ~{\rm GeV},~~ 2\pi T D \approx 3.6  .
\eea
 If there is no negative contribution from $\Delta\rho(\omega)$ to $G(1/2)-G_{\rec}(1/2)$, 3.6 could be an upper bound on $2\pi T D$ at 1.46 $T_c$. 
 We also performed a fit with a linear form of $b\omega$ describing the very low frequency behavior of the vector spectral function. Fitting to the difference of correlators at the symmetry point $G(1/2)-G_{\rec}(1/2)$ at $1.46~T_c$ gives:
 \be
 2\pi T D = \frac{\pi T}{3\chi_{00}}~b~ \approx 2.
 \ee
 Clearly the estimate of the charm diffusion coefficient is sensitive to the ansatz used for the fits. However, the charm diffusion coefficient estimated from these two different Ans\"atze are compatible.

 When going to higher temperatures at $2.20$ and $2.93~T_c$, the interplay between the change of bound states and diffusion part in the spectral function becomes complicated, thus it is not convincing that one may get a reasonable estimate of the charm diffusion coefficient by using a simple ansatz consisting of only a transport peak.
Nevertheless in the following we will use these current estimates at $1.46~T_c$ as input for the default models that have to be supplied to the MEM analysis. Further details on the choice of default models are given in Appendix~\ref{sec:app}.}

\section{Spectral functions}
\label{sec:spf_MEM}

 In the previous subsections we found that the flatness and the small deviation from unity of the ratios ($G/G_{\rec}(\tau,T)$, $G^{\rm diff}/G_{\rec}^{\rm diff}(\tau ,T)$ and $G^{\rm sub}/G_{\rec}^{\rm sub}(\tau ,T)$) do not necessarily mean that thermal modifications of the ground states are negligible. The more relevant quantities to look at are the sign of $\Delta G^{(2)}$ and the relative strength of $\Delta G^{(4)}$, $\Delta G^{(2)}$ and $\Delta G^{(0)}$.
However, from all these quantities, only a qualitative understanding of the thermal modifications of spectral functions at temperatures from below to above $T_c$ can be
deduced. To really explore the properties of charmonium states at different temperatures, one has to advance to a direct analysis of spectral functions by using the Maximum Entropy Method.
As in the vector channel, the contributions from the bound states and the diffusion part are entangled, it would also be helpful to provide information on the diffusion part of the spectral function estimated in Section~\ref{sec:diffusion_corr}
into the default model in MEM analyses.

In this section we start with a brief introduction to the Maximum Entropy Method in subsection~\ref{sec:MEM_intro}. We then discuss charmonium spectral functions below and above $T_c$ from the Maximum Entropy Method in subsection~\ref{sec:spf_finite_T}. The estimation on the value of the charm quark diffusion coefficient from vector spectral functions at $T>T_c$ will be given in subsection~\ref{sec:diffusion_spf_mem}.

\subsection{Maximum Entropy Method}
\label{sec:MEM_intro}

Inverting Eq.~(\ref{cor_spf_relation}) to extract the spectral function is a typical ill-posed problem. At finite temperature the inversion is more complicated than at $T=0$, 
since the temporal extent is always restricted to the temperature interval, $0 < \tau \leq 1/T$. The spectral functions we want to obtain are continuous while the correlators are calculated 
at a finite set of $N_{\tau}$ Euclidean time points which typically are $\mathcal{O}$(10).  An infinite number of solutions thus exists. The task then is to select the most likely solution which is consistent with additional constraints. Because of the positivity and the normalizability of the spectral function it can be interpreted as a probability function. The guiding principle for the selection thus can be the Bayesian statistical inference, which is the
 basis for the Maximum Entropy Method.

The Maximum Entropy Method (MEM) is a widely used tool for extracting spectral functions from correlation functions. It was introduced to lattice QCD by  Asakawa et al.~\cite{Asakawa01} and has been successfully applied to lattice QCD data at zero
 temperature to extract the parameters of the ground state and excited states of hadrons~\cite{Nakahara99,Yamazaki02, Langfeld02,Fiebig02, Allton02,Sasaki05}. 
 The application to finite temperature lattice QCD has also been explored~\cite{Datta04, Asakawa04, Umeda05, Aarts07, Jakovac07, Aarts07:1,Ding08,Ding09,Ding10, Ding11,Karsch02}.
 Based on the Bayesian theorem, MEM provides a way to select a unique spectral function $\rho(\omega)$ and transfers the problem of specifying a parameterization of $\rho(\omega)$ into the problem of
 specifying a likelihood function and a prior probability. The most probable spectral function $\rho(\omega)$, given lattice data $G$ and prior
 information $H$, can be obtained by maximizing the conditional probability 
  \be
P[\rho|GH]=\mathrm{exp}(\alpha S[\rho] - L[\rho]),
\ee
 where $L[\rho]$ is the standard likelihood function and the Shannon-Jaynes entropy $S[\rho]$ is defined as
\be
S[\rho] = \int_0^{\infty}{\frac{\mathrm{d}\omega}{2\pi}~\left[\rho(\omega) - m(\omega) - \rho(\omega)\mathrm{log}\left(\frac{\rho(\omega)}{m(\omega)}\right)\right]}.
\ee
Here $m(\omega)$ is the default model which introduces the prior information on the spectral function $\rho(\omega)$ as the input, e.g.  $\rho(\omega)$ is positive-definite; $\alpha$ is a real and positive parameter which controls the relative weight of the entropy S and the likelihood function L. The final spectral function is expressed as an integral over $\alpha$:
\be
\rho(\omega) = \int \md\alpha\, \rho_{\alpha}\,P[\alpha|G] \bigg{/}  \int \md\alpha\,P[\alpha|G],
\ee
where $P[\alpha|G]$ is the posterior probability of $\alpha$ given data G and $\rho_{\alpha}$ is the most probable spectral function for a certain $\alpha$.

As pointed out in Ref.~\cite{Aarts07:1}, the integral kernel $K(\tau,\omega)$ diverges at vanishing $\omega$,
\be
K(\tau,\omega) =  \frac{2T}{\omega} + (\frac{1}{6T} - \tau + T\tau^2)\omega + \cO[\omega]^3.
\label{eq:low_w_behaviour_K}
\ee
In order to explore the low frequency behavior of spectral functions, it thus is of advantage to introduce a modified kernel that is free of this divergence and leads to a redefined spectral function.
Since $K(\tau,\omega)$ has the following property,
\be
\sum_{\tau=0}^{N_{\tau}-1}~K(\omega,\tau)=~1/\tanh(\omega/2),
\label{kernel_sum}
\ee
and $\lim_{\omega\rightarrow\infty}\tanh(\omega/2)=1$, we implemented in our analysis the following modified version of kernels and spectral functions~\cite{Engels09,Ding09}
\bea
\label{def:modified_kernel}
\tilde{K}(\tau,\omega) &=& \tanh(\omega/2)\,K(\tau,\omega),\\
\tilde{\rho}(\omega) &=& \coth(\omega/2)\,\rho(\omega).
\eea
The modified kernel $\tilde{K}(\tau,\omega)$ cures the instability of MEM at $\omega\approx 0$ and reproduces the behavior of the original 
kernel in the large $\omega$ region.

Note that we have to specify the default model $m(\omega)$ to extract the spectral function from the correlator data. 
Thus choosing the default model (DM) is an essential part of the MEM analysis. 
Therefore all available prior information needs to be included in the default model, as it can strongly affect the output spectral function if the quality of the correlator data is not sufficient. 
It is  natural to choose a default model which reproduces the behavior of the spectral function in the large $\omega$ region. 
Note that Eq.~(\ref{eq:freespf_continuum_zerop}) describes the propagation of a free quark antiquark pair in the continuum limit. On the lattice, the high frequency  part of the spectral function is strongly distorted due to lattice cutoff effects~\cite{Karsch03,Aarts05}, as seen from Fig.~\ref{fig:spf_lat-cont} in Section~\ref{sec:intro}. Rather than growing as $\omega^2$ the free lattice spectral function vanishes above a maximal frequency. Thus when extracting the spectral function from the correlation function calculated on 
the lattice, it is reasonable to use the free lattice spectral function as the prior information in the MEM analyses. The Breit-Wigner distribution, Eq.~(\ref{eq:spf_trans}), replaces the $\omega\delta(\omega)$ term and  is also added into the default model for the very low frequency region in our MEM analyses at finite temperatures.

 \subsection{Spectral functions at finite temperature}
 \label{sec:spf_finite_T}

In this subsection we will discuss spectral functions obtained from MEM analyses at temperatures below and above $T_c$. We will mainly focus on the results from our finest lattice, i.e. a=0.01fm with $\beta=7.793$. When we analyze  correlation functions using MEM, we fix the number of points in the frequency space to $N_{\omega}=8000$ and the step length $a\Delta\omega$=0.0005, i.e. we fix $a\omega_{max}\approx 4$ or $\omega_{max}\approx 76$ GeV. 
If not mentioned otherwise we use the free lattice spectral functions with quark mass $am\approx0.06$ as part of default models in our MEM analyses, whereby the value  $am\approx0.06$ corresponds to  the value of $m_{\MSbar}(m)$ listed in Table~\ref{table:QuarkMass}.

The lattice spectral function is subject to lattice cutoff effects. These show up in the short distance behavior of correlation function and manifest themselves in the large energy behavior
of spectral functions as shown in Fig.~\ref{fig:spf_lat-cont}. 
To reduce lattice cutoff effects we omit some correlator data points at small distances, i.e. we use $\tilde{\tau}=\tau/a=4,5,6,\cdots,N_{\tau}/2$ in the MEM analysis. 
In addition we need to take into account a default model modification of the large energy part of spectral functions that arises from perturbative corrections to the
free field behavior. When using a free continuum spectral function as ansatz this is usually done by multiplying the large energy part, which is proportional to $\omega^2$, with a suitably
chosen constant. This cannot be done with our ansatz for the default models where we use at large energies the free lattice spectral functions that are cut off at some maximal energy $\omega_{max}$.
Instead we rescale the free lattice spectral function in the default model $DM$ such that the correlator $G_{DM}(\tau,T)$, calculated from the default model, agrees with the lattice data at $\tau/a=4$, i.e. we demand $G_{\rm DM}(\tau/a=4,T)/G(\tau/a=4,T)$=1.   
To suppress the large $\omega$ rise,  in general we plot the spectral function $\rho(\omega)$ divided by $\omega^2$ as a function of $\omega$.
 
 \begin{figure}[htpd]	
  \begin{center}
        \includegraphics[width=.48\textwidth]{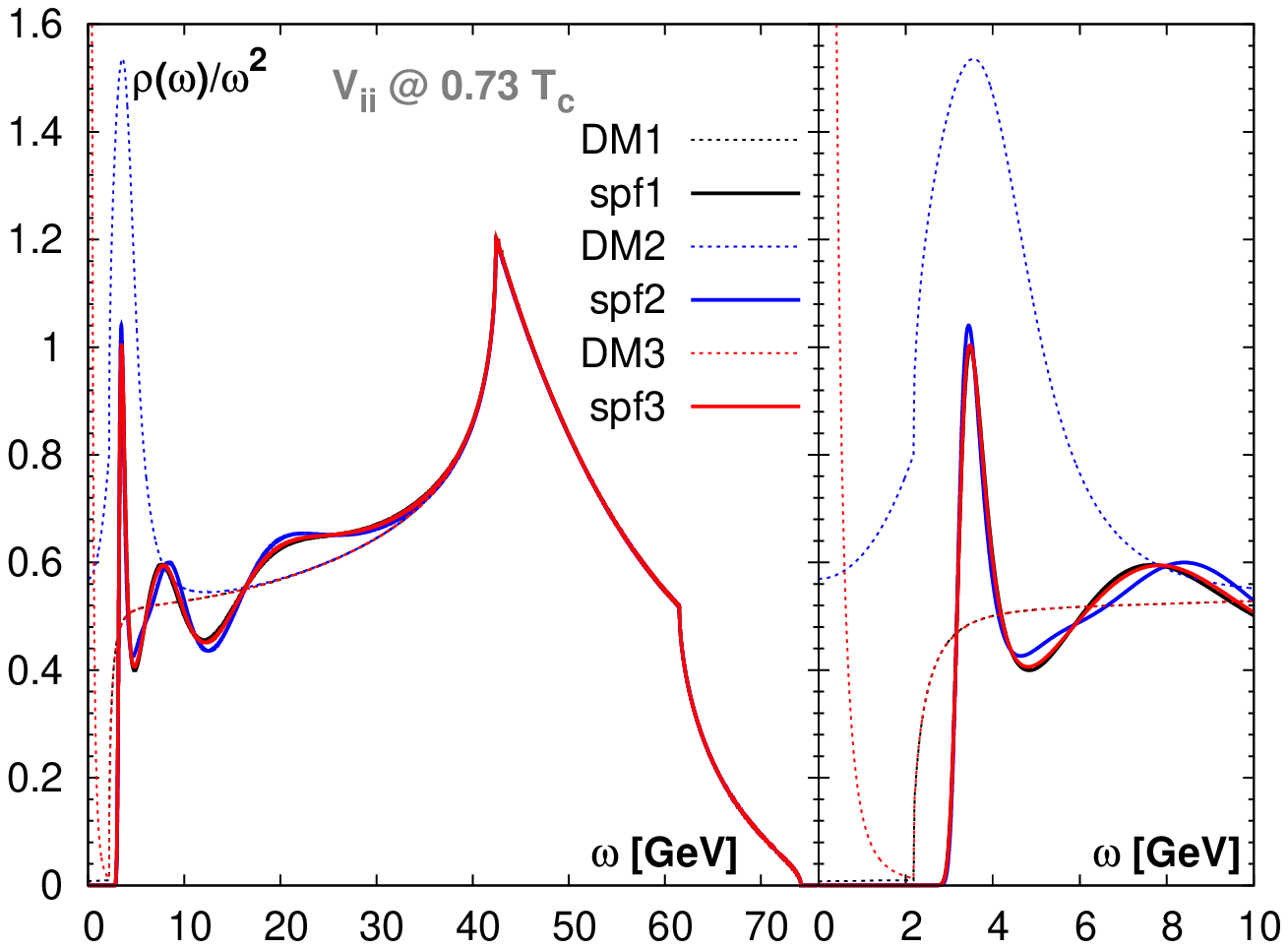}~~~~
    \includegraphics[width=.48\textwidth]{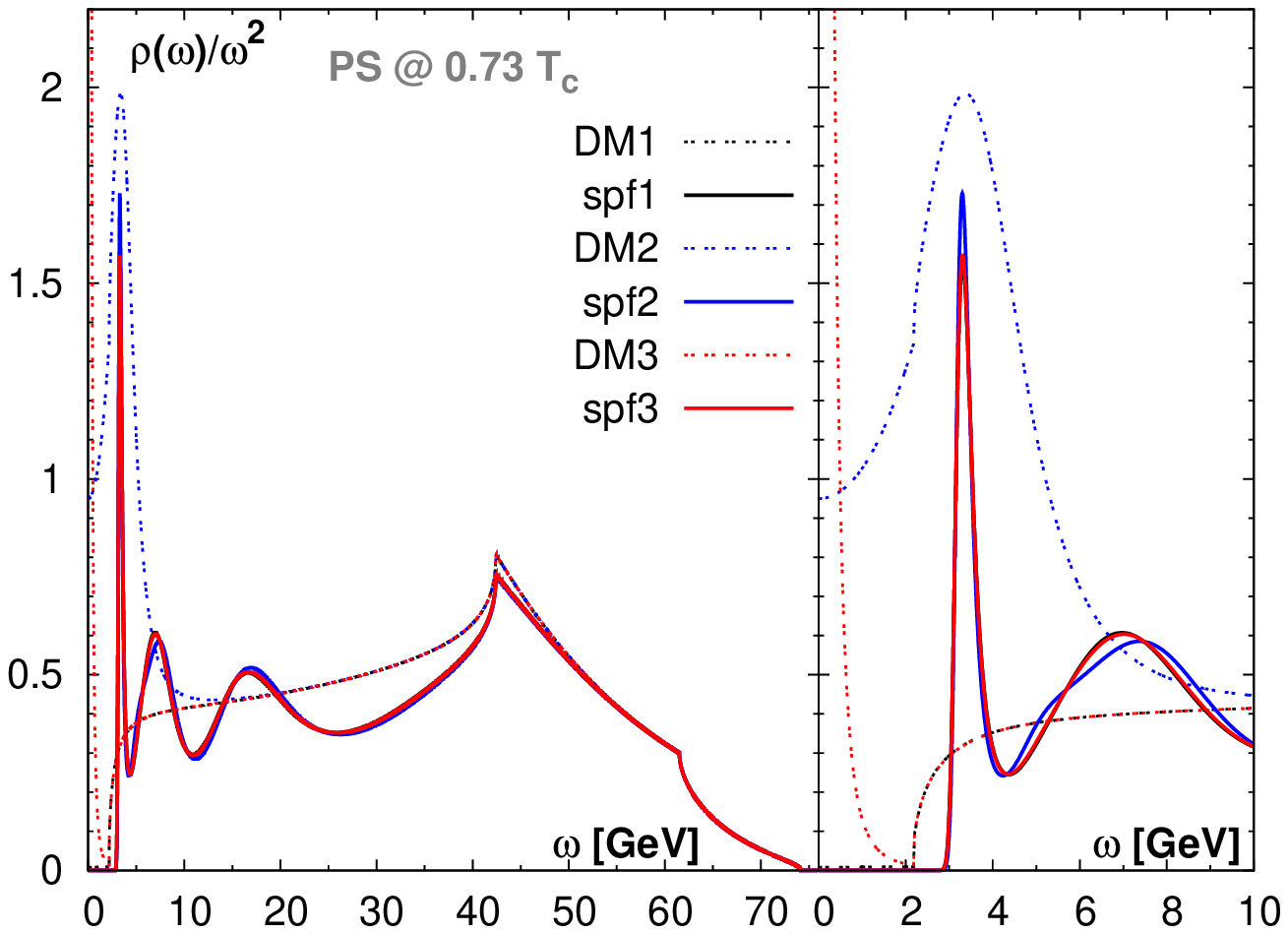}
        \caption{Default model dependences of the output spectral functions in $V_{ii}$ (left) and $PS$ (right) channels at 0.73 $T_c$ on the $128^3\times96$ lattice. The plots in the right panels are blowups of the low frequency region of the left panels. ``DM"s are the input default models while ``spf"s are the corresponding MEM outputs.}
    \label{fig:spf_Swaves_0p75Tc_beta7p793_dm_dependence}
  \end{center}
  \end{figure}

  We first look into the left plot in Fig.~\ref{fig:spf_Swaves_0p75Tc_beta7p793_dm_dependence}, i.e. the $V_{ii}$ channel below $T_c$. We test three different default models, ``DM1" is a rescaled free lattice spectral function, ``DM2" is a rescaled free lattice spectral function supplemented with a resonance peak located in the low frequency region and ``DM3" is a rescaled free lattice spectral function with a transport peak described by e.g. Eq.~(\ref{eq:spf_trans}) in the very low frequency region. In the very high frequency region ($\omega\gtrsim35~$GeV), as seen from the left panel of the left plot in  Fig.~\ref{fig:spf_Swaves_0p75Tc_beta7p793_dm_dependence}, the MEM output just resembles the behavior of the input default models. In the low frequency region, as seen from the right panel of the left plot in Fig.~\ref{fig:spf_Swaves_0p75Tc_beta7p793_dm_dependence},  spectral functions obtained from MEM have a unique form that is different from that of the input default models, i.e. results are independent of the default model and thus might reflect stable features of the spectral function. We found that the default model dependence of the first peak is very weak. Though there are some variations in amplitudes of peaks, the peak location of the first peak is always the same, at $\omega\approx 3.48~$GeV, which in turn is close to the value of the screening mass obtained from the spatial correlator quoted in Table~\ref{table:Meson_Mass}. Thus this peak can be interpreted as the bound state peak of $\Jpsi$. It remains stable and robust in MEM analyses performed with quite different prior information. However, the width of this peak cannot be directly interpreted as the width of $\Jpsi$ due to the limited statistics and small number of data points in the temporal direction. The second and third peak in Fig.~\ref{fig:spf_Swaves_0p75Tc_beta7p793_dm_dependence} could be a mixture of higher excited states or MEM artifacts due to the finite lattice spacing and limited number of correlator points\footnote{We discuss this in more detail in connection with Fig.~\ref{fig:spf_V123_betas_0p75Tc} in Appendix \ref{sec:app}.}. The output spectral function ``spf3", which is obtained from the default model ``DM3",  has no transport peak although such peak is implemented in the default model. We thus conclude that at this temperature there is no (smeared) zero mode contribution in the vector channel, or in other words the charm diffusion coefficient is compatible with zero at $0.73~T_c$. We performed the same analysis in the $PS$ channel (the right plot of Fig.~\ref{fig:spf_Swaves_0p75Tc_beta7p793_dm_dependence}) and find that results in the $PS$ channel are similar to those in the $V_{ii}$ channel. We did not observe a (smeared) zero mode contribution in the $PS$ channel as well. To analyze the modification of bound states in the spectral function with temperature the $PS$ channel is therefore a good candidate.

   \begin{figure}[htpd]
  \begin{center}
    \includegraphics[width=.45\textwidth]{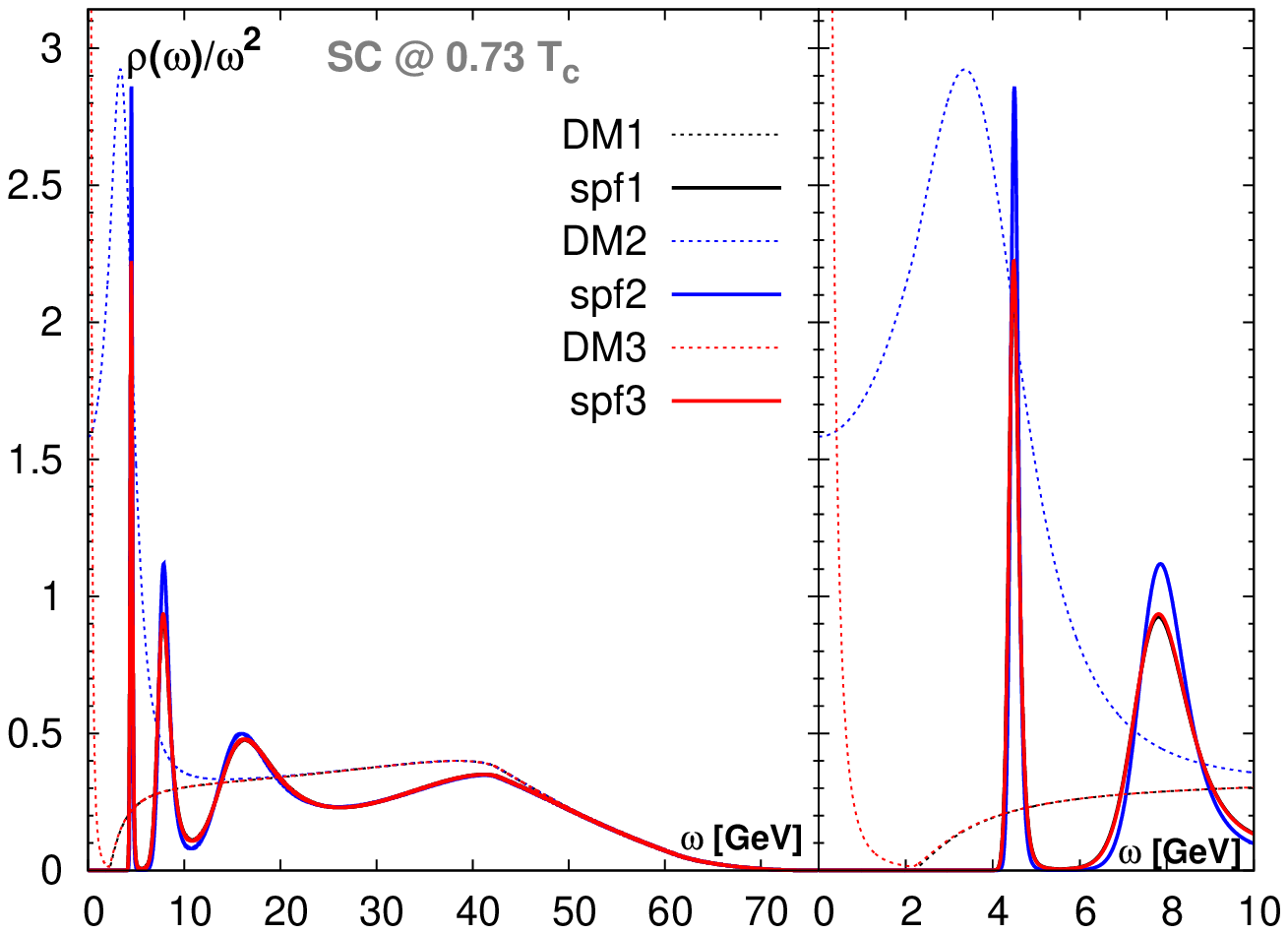}~ \includegraphics[width=.45\textwidth]{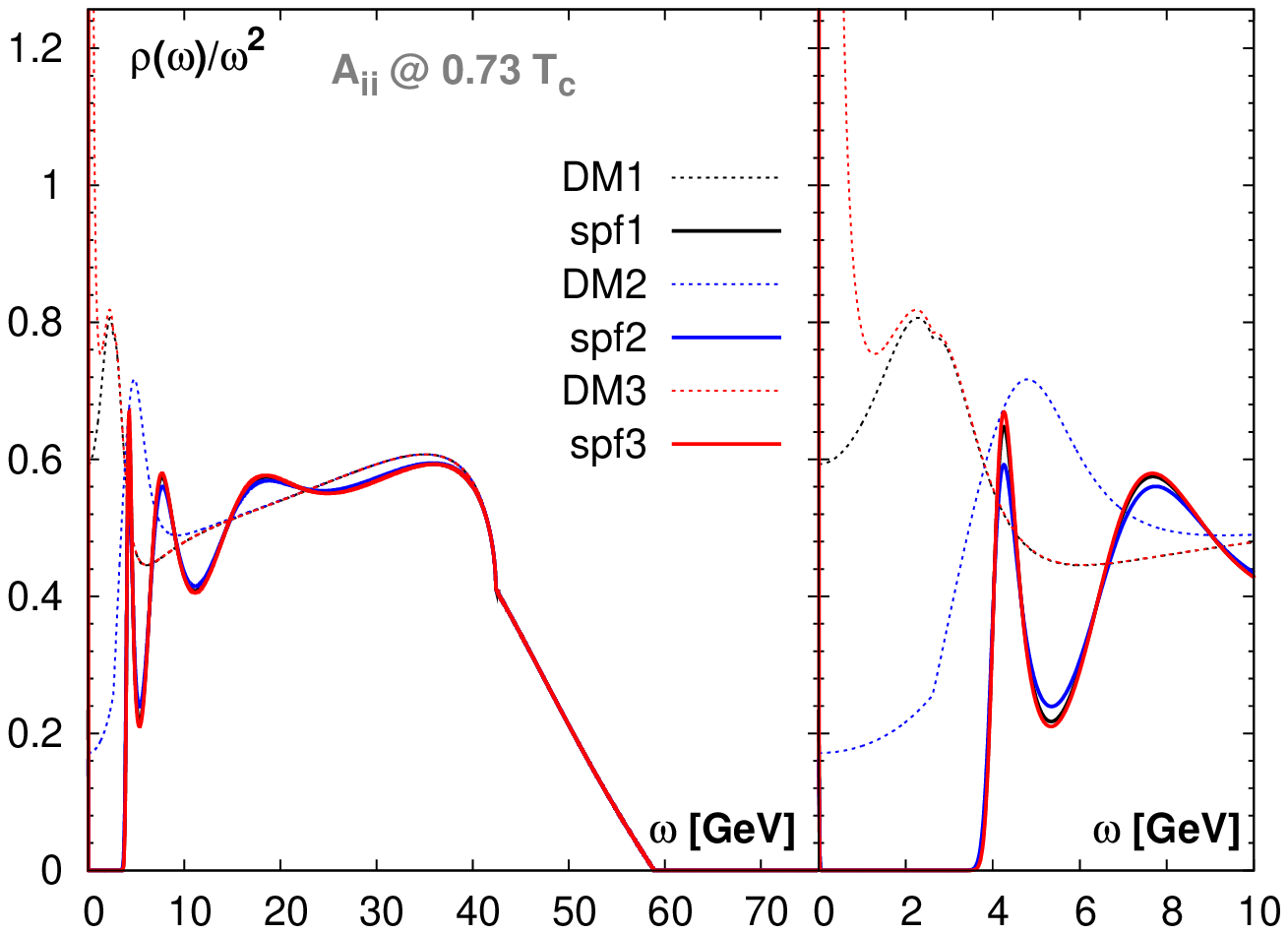}
        \caption{Same as Fig.~\ref{fig:spf_Swaves_0p75Tc_beta7p793_dm_dependence} but for $SC$ (left) and $A_{ii}$ (right) channels.}
           \label{fig:spf_Pwaves_0p75Tc_beta7p793_dm_dependence}
  \end{center}
  \end{figure}

 Spectral functions of P wave states are shown in Fig.~\ref{fig:spf_Pwaves_0p75Tc_beta7p793_dm_dependence}. The default model dependence of the output spectral functions is small in both $SC$ and $A_{ii}$ channels.
 We observed stable P wave ground states.   As in the case of $PS$ and $V_{ii}$ channels, no (smeared) zero mode contribution is observed in the $SC$ channel at $0.73~T_c$. In the $A_{ii}$ channel we always found that some remnant of the input default model is present at  $\omega\approx 0$ for all three spectral functions, although the remnant is not obviously seen in the right plot of Fig.~\ref{fig:spf_Pwaves_0p75Tc_beta7p793_dm_dependence}. Because of the noise level, it is however not clear whether this originates from the zero mode contribution or  insufficient quality of temporal correlator data.


In the following we present the results for charmonium spectral functions above $T_c$. 
In Appendix~\ref{sec:app} we give a detailed analysis of the default model dependence of our results and also quantify the influence of distance windows, the lattice cutoff etc. We found that the outputs from various default models are all compatible with those obtained by using the free lattice spectral functions
with an added transport peak as default models. Consequently all MEM results shown in the following are obtained by using such default models. Since we found some excess of the spectral function in the low frequency region in the $A_{ii}$ channel, we use in that case the massless free lattice spectral function with a Breit-Wigner like peak at low frequencies  as the
default model.

We now focus on the statistical error analysis of the spectral functions. This is done by using the 
Jackknife method.
In the literature statistical errors are often given on the mean of $\rho(\omega)$ over a certain $\omega$ region in the spectral 
function plot~\cite{Datta04, Asakawa04, Aarts07, Jakovac07}. Here we rather perform MEM analyses on Jackknife blocks and calculate the Jackknife error of the amplitude of each point in the spectral function.
MEM cannot reproduce the correct width of the resonance, however, it gives a stable and reliable peak locations of the spectral functions. We thus also estimate the statistical errors of  the peak location of the first peak of the spectral function at $0.73~T_c$ and $1.46~T_c$ (see Table~\ref{tab:peak_locations}).  A signal for the dissociation of charmonium states then is a shift of the peak location and the relative broadening of the peak at different temperatures.

         \begin{figure}[hbpt]
  \begin{center}
    \includegraphics[width=0.45\textwidth]{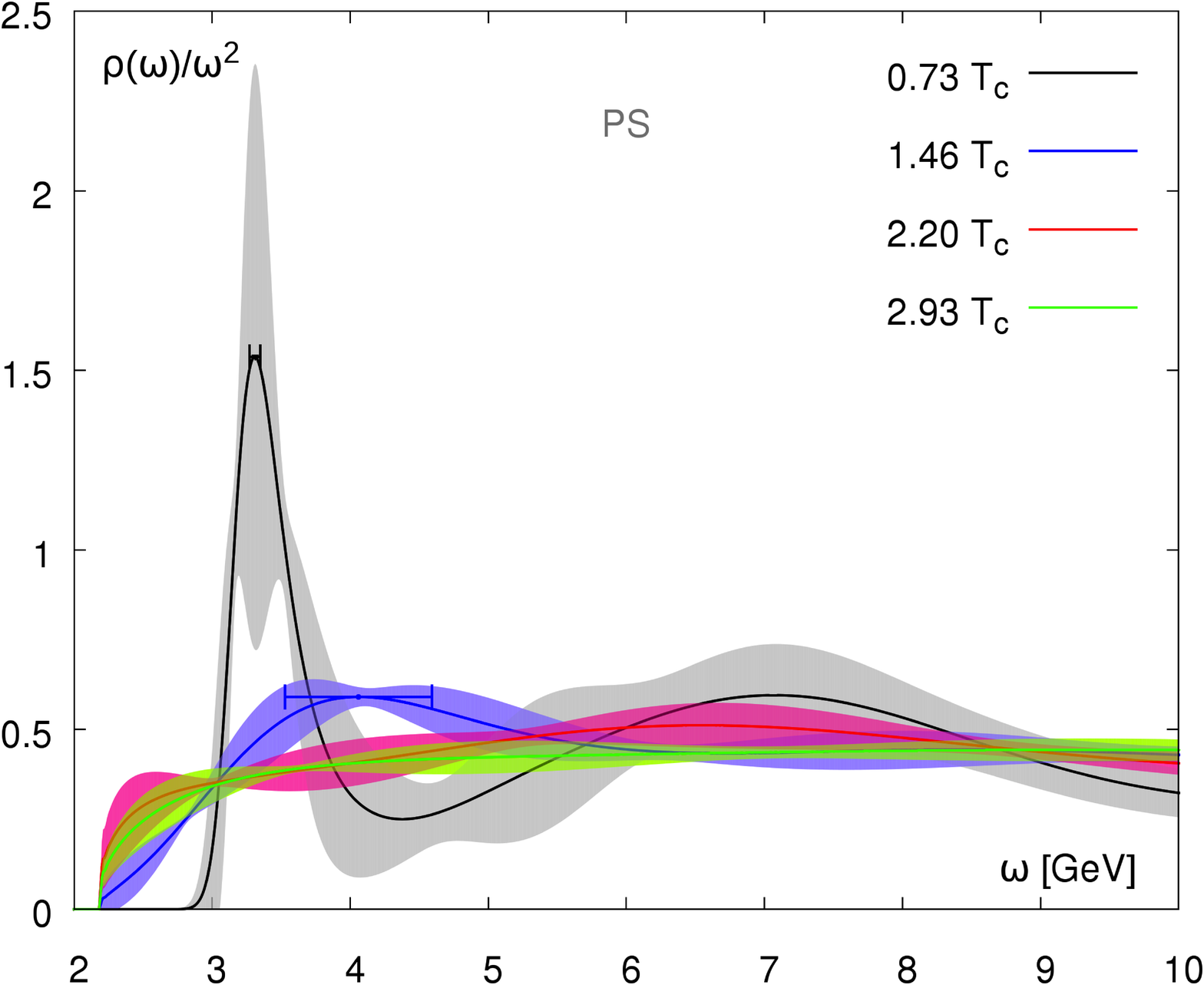}~\includegraphics[width=0.45\textwidth]{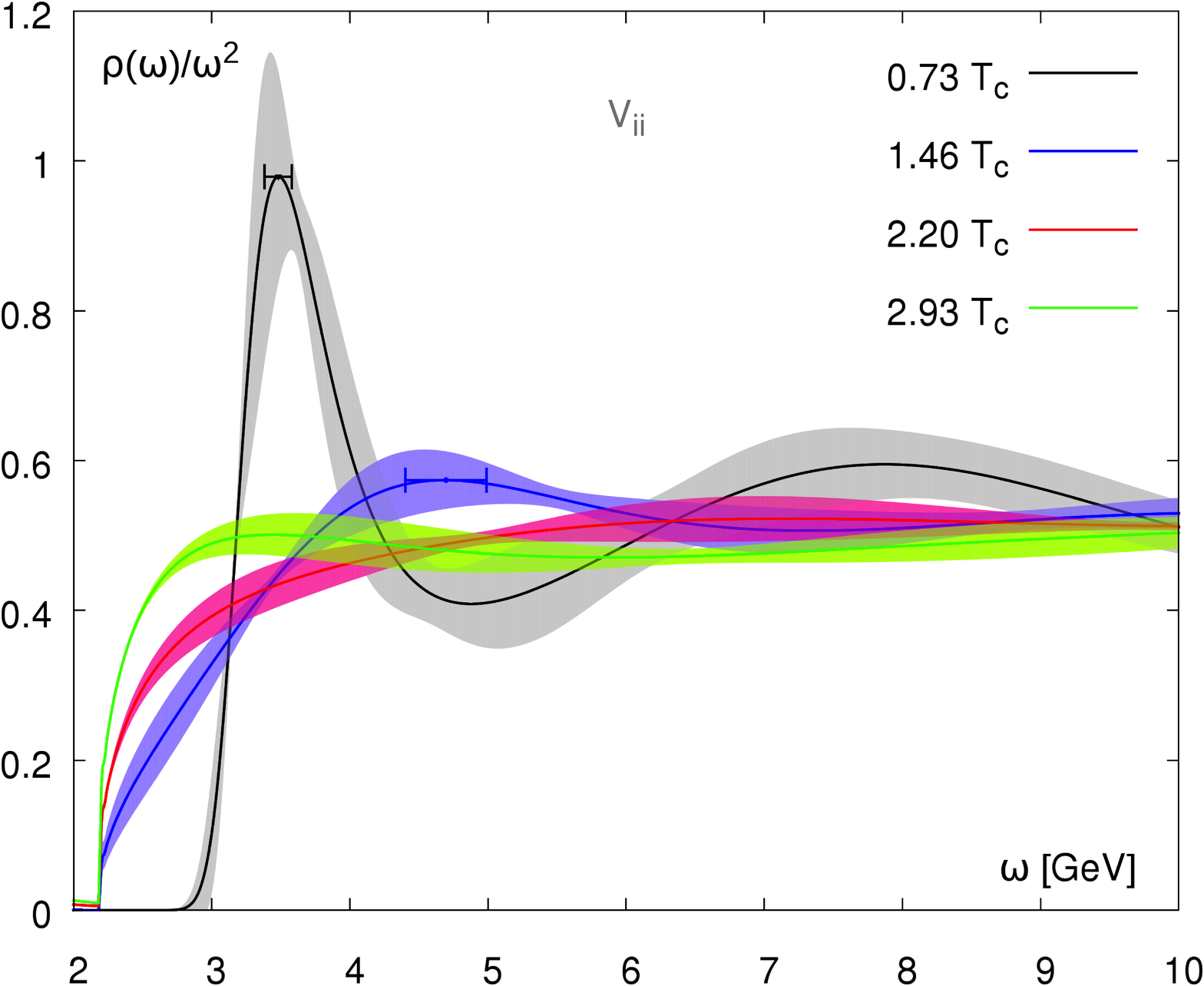}
                    \caption{Statistical uncertainties of output spectral functions in $PS$ (left) and $V_{ii}$ (right)
channels at all available temperatures. The shaded areas are statistical errors of amplitudes of output spectral
functions from Jackknife analyses and the solid lines inside the shaded areas are mean values
of spectral functions. The horizontal error bars at the first peaks of spectral functions at $0.73~T_c$ and $1.46~T_c$ stand for the statistical uncertainties of the peak location obtained from Jackknife analyses. 
          }
                 \label{fig:spf_Swave_statistics_errors}
  \end{center}
\end{figure}

We show the statistical significance of output spectral functions at $\omega\gtrsim 2~$GeV in $PS$ (left) and $V_{ii}$ (right) channels in Fig.~\ref{fig:spf_Swave_statistics_errors}. The shaded areas are statistical uncertainties of amplitudes of output spectral functions from Jackknife analyses and the solid lines inside the shaded areas are mean values
of spectral functions. From the left plot of Fig.~\ref{fig:spf_Swave_statistics_errors} it is apparent that at $0.73~T_c$ the spectral function in the $PS$ channel has large uncertainties in the amplitude at the point which corresponds to the ground state peak location in the mean spectral function. However, even at the lower end of the error band, the amplitude is still larger than the peak amplitudes at the higher temperatures within the errors.
We also show the statistical uncertainties of the first peak location of the spectral function at $0.73~T_c$ and $1.46~T_c$ as horizontal error bars in the left plot of Fig.~\ref{fig:spf_Swave_statistics_errors}.
Unlike the large uncertainties shown in the amplitude of the peak height, the peak location of the ground state peak at $0.73~T_c$ is well determined. A Jackknife analysis yields $m_{\eta_c}=3.31 (4)$ GeV (see Table~\ref{tab:peak_locations}). At $1.46~T_c$ this peak is shifted by about 0.8 GeV to around 4.1 GeV, as is seen from Table~\ref{tab:peak_locations}. At $2.23~T_c$ there is hardly a peak structure that can be identified within the statistical uncertainties. At $2.93~T_c$ the spectral function flattens further. Thus this picture, together with the systematic uncertainties discussed in the Appendix, suggests that $\eta_c$ is melted already at  $1.46~T_c$.

In the right plot of Fig.~\ref{fig:spf_Swave_statistics_errors}, we focus on the resonance part of the spectral function in the $V_{ii}$ channel. One sees that the peak location of the spectral function at $0.73~T_c$ does not have an overlap with the peak location of the spectral function at $1.46~T_c$ and the amplitudes between these two differ a lot (see horizontal error bars and also values in Table~\ref{tab:peak_locations}). At both $2.20~T_c$ and $2.93~T_c$ there is hardly any peak structure. Together with the study of systematic uncertainties discussed in Appendix~\ref{sec:app}, this picture indicates that also $\Jpsi$ is already dissociated at $1.46~T_c$.

     \begin{figure}[hpt]
  \begin{center}
    \includegraphics[width=0.45\textwidth]{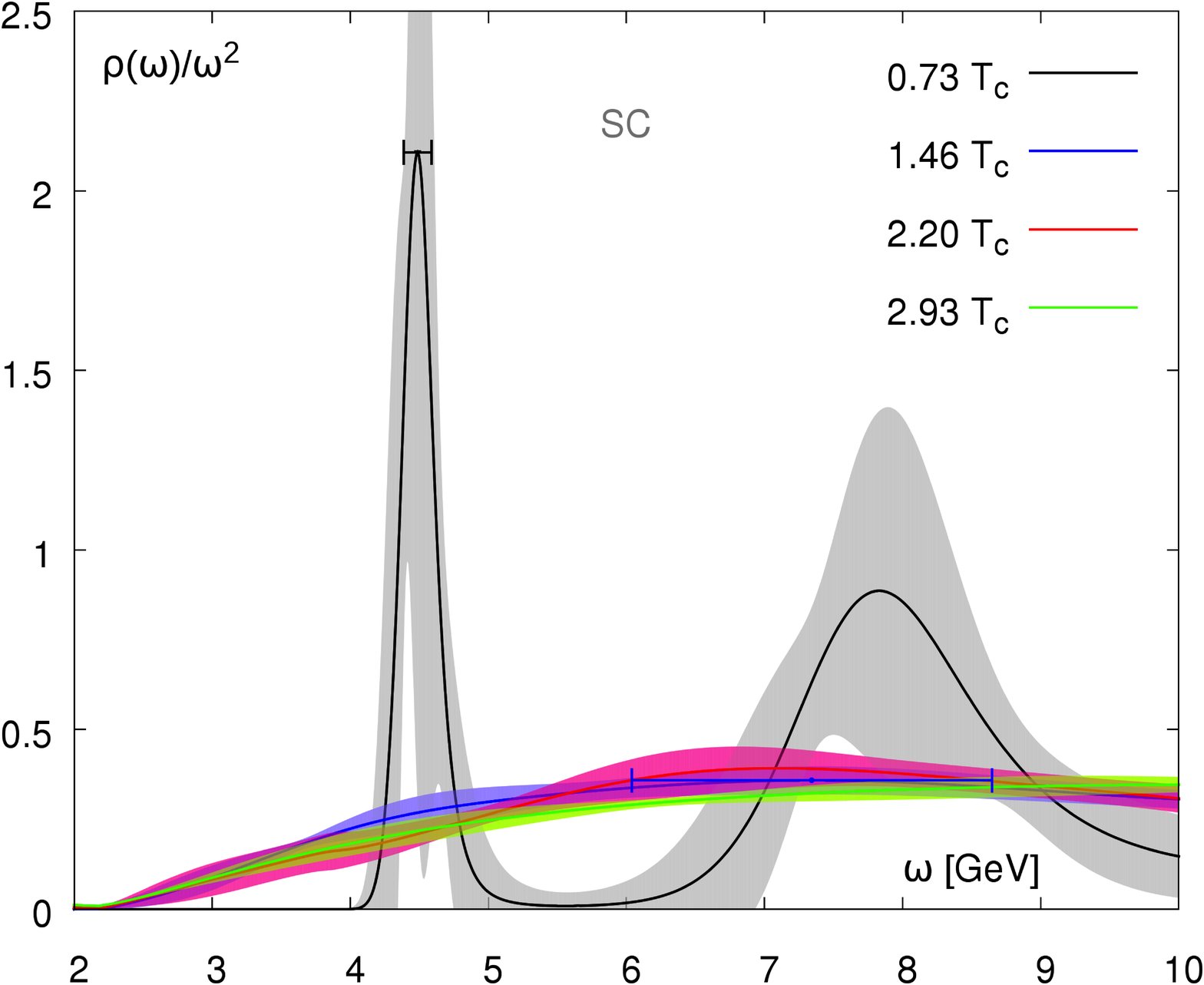}~   \includegraphics[width=0.45\textwidth]{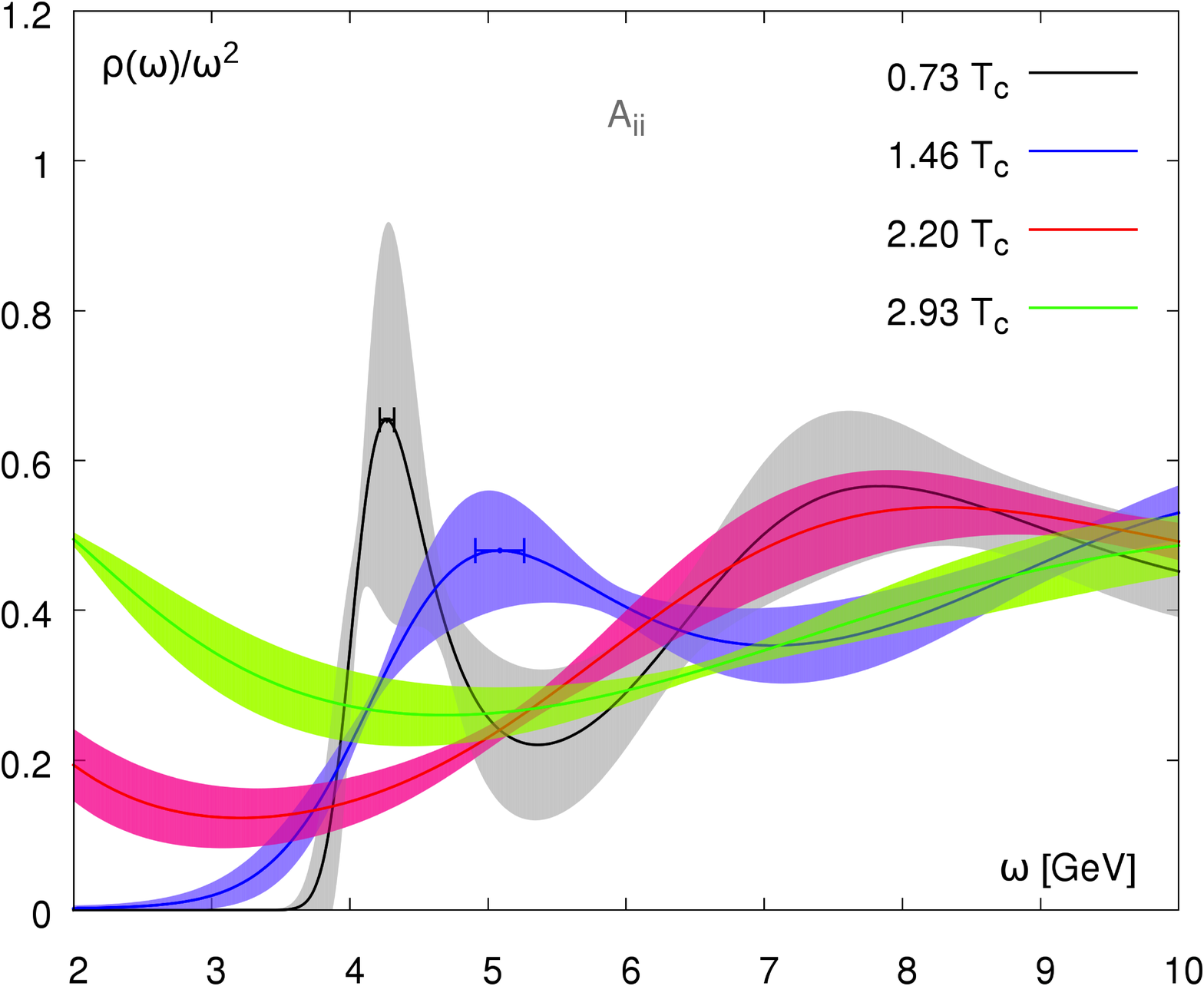}
                    \caption{Same as Fig.~\ref{fig:spf_Swave_statistics_errors}  but for $SC$ (left) and $A_{ii}$ (right) channels.
          }
                 \label{fig:spf_Pwave_statistics_errors}
  \end{center}
\end{figure}

\begin{table}[htbl]
\centering
\begin{tabular}{|c|c|*{5}{c|}}\hline

\backslashbox{{\bf T}}{{\bf channel}}
   &  $PS$ & $V_{ii}$ &  $SC$ & $A_{ii}$ \\ \hline
$0.73T_c$ &  3.31(4)& 3.48(9) &4.5(1) & 4.26(5)  \\ \hline
$1.46T_c$ & 4.1(5) &4.7(3) & 7(1) &5.1(2)\\ \hline

\end{tabular}
\caption{The locations of the first peaks in different channels obtained from MEM. Errors are estimated from the Jackknife analyses. The numbers for peak locations are in units of GeV.}
\label{tab:peak_locations}
\end{table}

The statistical errors on P wave spectral functions are shown in Fig.~\ref{fig:spf_Pwave_statistics_errors}. Here the results for the $SC$ channel are shown in the left plot. When going to temperatures above $T_c$, the structure of the ground state peak is basically gone and results in a rather flat spectral function. This signals the melting of $\chi_{c0}$ at $T\geq 1.46 T_c$.

The right plot of Fig.~\ref{fig:spf_Pwave_statistics_errors} shows the result for the $A_{ii}$ channel. 
As temperature increases from $0.73~T_c$ to $1.46~T_c$, it becomes apparent that the location of the fist peak is shifted to the larger energy region. The bump seen at $1.46~T_c$ becomes much broader at $2.20~T_c$ and flattens at $2.93~T_c$. An enhancement of the small energy part (2 GeV $\lesssim \omega\lesssim$ 4 GeV) in the spectral functions is also observed at the two highest temperatures. 
This originates from our choice of the default model in the $A_{ii}$ channel where we did not introduce a quark mass threshold in the free lattice spectral function. The systematic uncertainties
arising from the choice of the quark mass cutoff have been discussed in the Appendix in connection with Fig.~\ref{fig:spf_V123_aboveT_am_dep}. The change of structures of spectral functions in the $A_{ii}$ channels suggests that $\chi_{c1}$ is dissociated already at $1.46T_c$.

 \subsection{Charm quark diffusion coefficient estimated from spectral  functions}
 \label{sec:diffusion_spf_mem}

       \begin{figure}[hbpt]
  \begin{center}
  \includegraphics[width=0.45\textwidth]{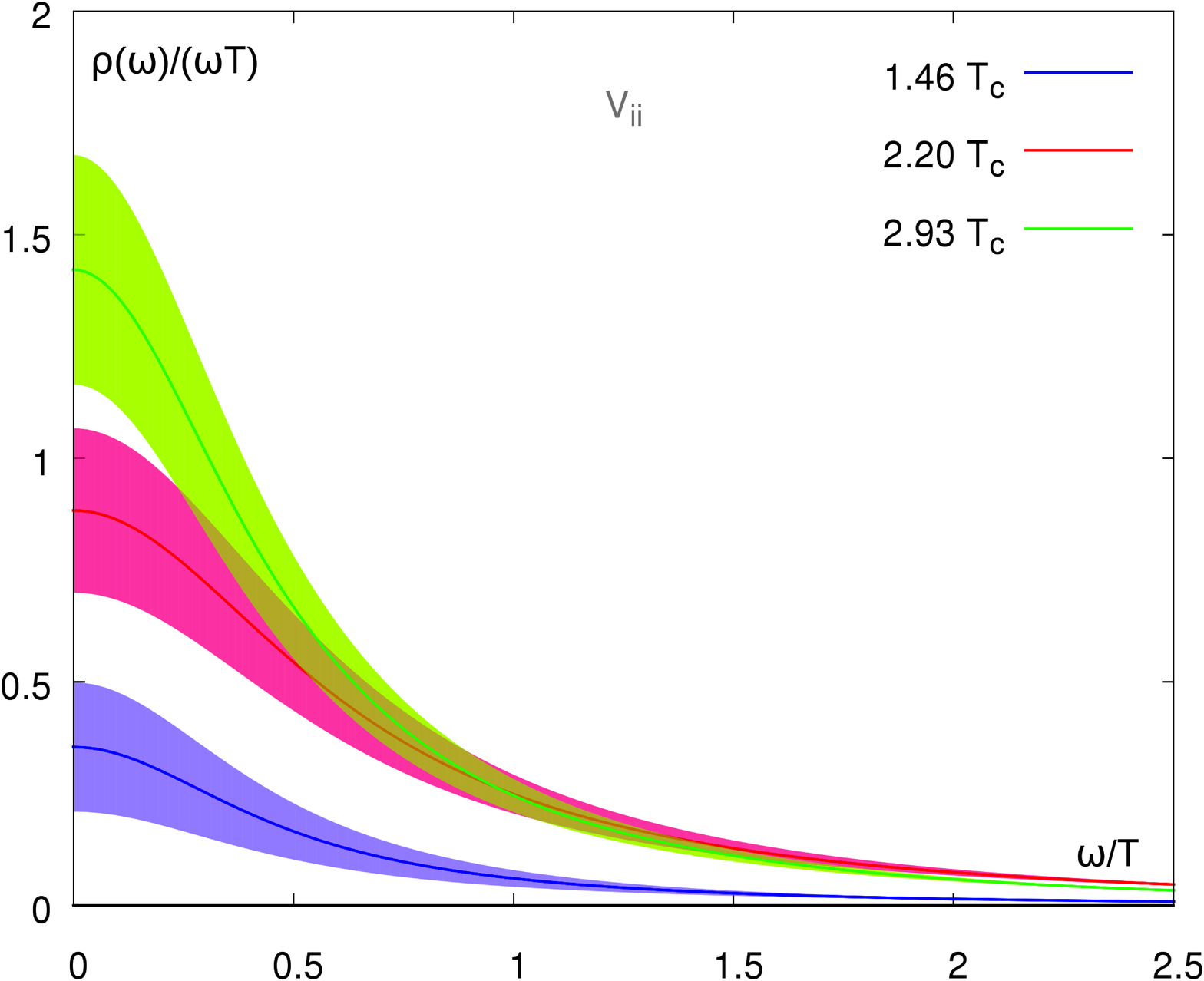}~~ \includegraphics[width=0.45\textwidth,height=0.35\textwidth]{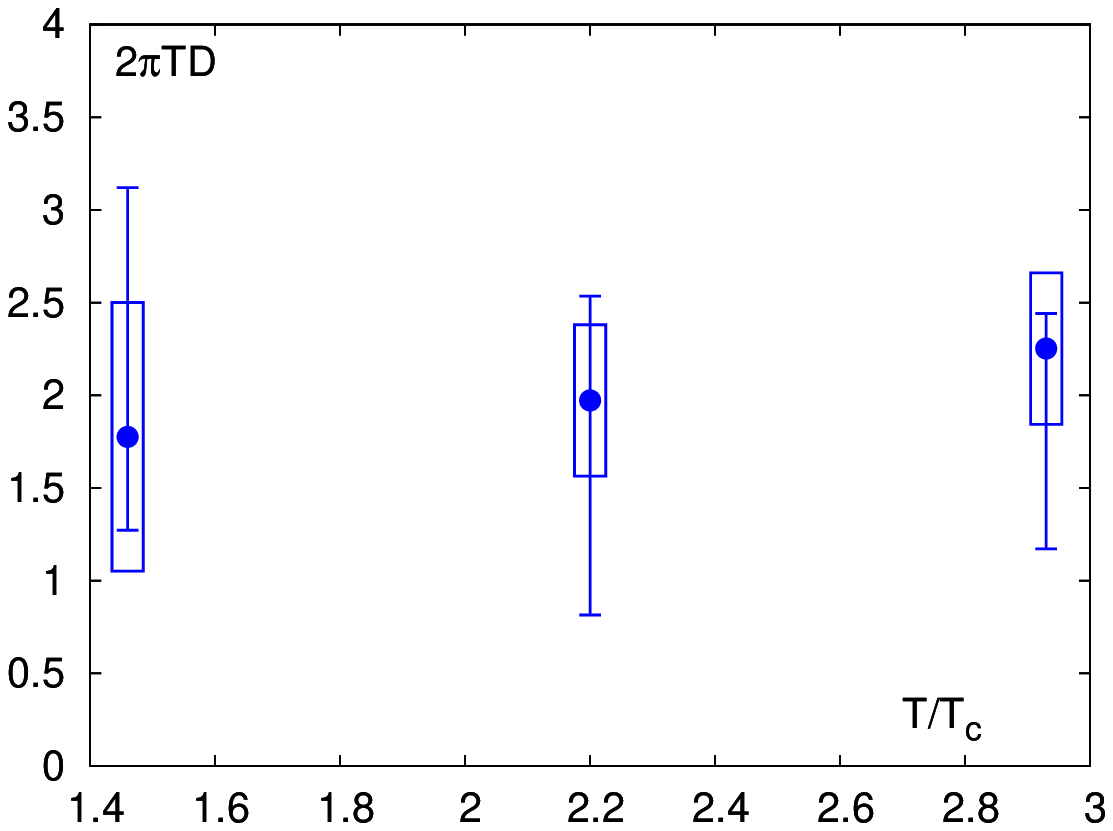}
                \caption{Left: statistical uncertainties of transport peaks at $T>T_c$. Right: the resulting charm diffusion coefficients. The boxes stand for  statistical error estimated from Jackknife method while the bars stand for  systematic uncertainties from MEM analyses. The numbers for charm diffusion coefficients are listed in Table~\ref{table:diffusion}.
}
                 \label{fig:spf_trans_statistics_errors}
  \end{center}
\end{figure}

We now focus on the very low frequency part of the spectral function given in the vector channel, i.e. $\omega/T\leqslant 2.5$ or $\omega\lesssim 1\cdots 2$ GeV at $T/T_c\approx 1.5\cdots 3$.
The statistical uncertainties of the transport peaks observed in the vector channel are shown in the left plot of Fig.~\ref{fig:spf_trans_statistics_errors}. The statistical uncertainties on the amplitude of the peak are relatively small. The charm diffusion coefficient is related to the amplitude of the transport peak at vanishing frequency through the Kubo formula (Eq.~(\ref{eq:HQ_diffusion_formula})). The current estimate for the charm diffusion coefficient $D$ is summarized in the right plot of Fig.~\ref{fig:spf_trans_statistics_errors}. The boxes stand for the statistical uncertainties and the error bars reflect systematic uncertainties obtained from the analyses discussed in the Appendix.
The bound for the systematic uncertainties for charm diffusion coefficients at all temperatures is obtained from the analysis of the default model dependence discussed in the Appendix and is taken from the lowest and highest values in Fig.~\ref{fig:spf_V123_beta7p793_AboveTc_dm_dependence1}, Fig.~\ref{fig:spf_V123_beta7p793_AboveTc_dm_dependence2}, Fig.~\ref{fig:spf_V123_aboveT_Nleft_dep} and Fig.~\ref{fig:spf_V123_aboveT_am_dep}.
The resulting numbers are listed in Table~\ref{table:diffusion}. We find that the mean value of $2\pi T D$ is around two at all three temperatures above $T_c$. As the quark number susceptibility $\chi_{00}$ increases faster with temperature than the amplitude of the transport peak, the mean values of $2\pi T D$ increases only slightly with temperature. The charm diffusion coefficient obtained at $1.46~T_c$ is the most reliable one among the three temperatures above $T_c$ since more prior information is known at this temperature as discussed in Section~\ref{sec:diffusion_corr}. At $2.20~T_c$ and $2.93~T_c$, due to the lack of precise prior information and a fewer number of data points that can be used in the MEM analyses, the uncertainties on the charm diffusion coefficient thus might be underestimated.

\begin{table}[htdp]
\begin{center}
\begin{tabular}{|c|c|c|}
\hline        
$T/T_c$      &     $2\pi$TD           &       $\chi_{00}/T^2$\\
\hline
{1.46}   &   1.8$\pm0.7$(stat.)$^{+1.3}_{-0.5}$(sys.)         &        0.20894(1)  \\
           
                                           \hline
{2.20}    &    2.0$\pm0.4$(stat.)$^{+0.6}_{-1.2}$(sys.)          &        0.46900(2)     \\
                                                                                 \hline
                                                                                 
{2.93}     & 2.3$\pm0.4$(stat.)$^{+0.2}_{-1.1}$(sys.)             &   0.66112(4)     \\

\hline 
\end{tabular} 
\end{center}
\caption{Charm diffusion coefficients and quark number susceptibilities $\chi_{00}$ above $T_c$. The ``stat." stands for statistical errors estimated from Jackknife method and the ``sys." denotes  systematic uncertainties obtained from 
 MEM analyses in Appendix~\ref{sec:app}.}
 \label{table:diffusion}
\end{table}

To close this section, we have studied charmonium spectral functions in different channels at temperatures below and above $T_c$.
The general properties of ground states at $T<T_c$ can be reproduced by the MEM analysis of temporal correlation functions. The extracted peak location of the ground state is close to the physical value and is very reliable. However, the width of the ground state peak cannot be reproduced in the MEM analysis with the current quality of the temporal correlator data. Thus the signature for the dissociation of charmonium states is the shift of the first peak location  and relative broadening of the first peak. Comparing spectral functions below and above $T_c$, our MEM analyses suggest that both the S wave states ($\eta_c$ and $\Jpsi$) and P wave states ($\chi_{c0}$ and $\chi_{c1}$) are melted already at $1.46~T_c$. The charm diffusion coefficient is found to be compatible with zero at $T<T_c$ and around $1/\pi T$ at our available temperatures above $T_c$.

\section{Conclusions}
\label{sec:con}

We have investigated the properties of charmonium states at finite temperature in quenched QCD on large isotropic lattices.
The standard Wilson plaquette action for the gauge field and the nonperturbatively $\cO(a)$ improved clover fermion action for charm quarks were used in the simulation.
In the current study lattices with three different lattice spacings were used to control cutoff effects in the charmonium correlators and spectral functions. Since the use of a temporal extent with a large number of Euclidean time slices is a very important ingredient in the current study, we calculated charmonium correlators on the finest lattices ($a=0.01$fm) with relatively large lattice sizes of $128^{3}\times 96$, $128^3\times48$, $128^3\times32$ and $128^3\times24$ 
at $0.73~T_c$, $1.46~T_c$, $2.20~T_c$ and $2.93~T_c$, respectively.

At $T\lesssim0.73~T_c$ we found stable and reliable ground state peaks of charmonium states from MEM analyses, in which the peak locations are almost the same as the corresponding hadron masses
determined from the large distance behavior of spatial correlation functions at the same temperature. However, the width of the ground state peak still cannot be interpreted as the physical width of hadron states. Thus the dissociation of the ground states in the current study is signaled by the shift of the peak locations and the relative broadening of the width. At $T>T_c$ we first calculated the reconstructed correlation function directly from the correlator data at $T<T_c$. The curvatures of the differences between the measured correlators and the reconstructed correlators indicate that there are obvious thermal modifications to spectral functions at $T \ge1.46~T_c$ in all channels. However, it is hardly possible to distinguish the zero mode contribution and the thermal modification of the bound states in the spectral function from the study at the correlator level alone. We then advanced to the analysis on the spectral functions using the MEM. We utilized an improved integral kernel to avoid the instability of MEM in the very low frequency region. We compared the output spectral functions from lattices with three different lattice spacings and concluded that cutoff effects are small. Results on our finest lattice, which are the most reliable ones, thus should not be affected by severe cutoff effects. Using the correlation functions on the finest lattices, we studied the variation of the output spectral functions using different default models both below and above $T_c$. We checked the systematic uncertainties arising from the number of data points used in the MEM analysis, the lattice cutoff effects present at short distances, and the dependence on the quark mass threshold of the free lattice spectral functions.   Statistical errors of the spectral functions are estimated using a Jackknife analysis. By comparing the spectral functions below and above $T_c$, our analyses suggest that both P wave states ($\chi_{c0}$ and $\chi_{c1}$) and S wave states ($\Jpsi$ and $\eta_c$) are dissociated already at  $1.46~ T_c$.

The determination of dissociation of  $\Jpsi$ and $\eta_c$ already at $1.46~T_c$ is quite different from previous lattice QCD studies~\cite{Datta04, Asakawa04, Umeda05, Aarts07, Jakovac07,Iida06,Ohno11}, which predicted the 1S charmonium states to be dissociated at 
$T\gtrsim 2~T_c$. Since most of the previous results are obtained on anisotropic lattices~\cite{Asakawa04, Umeda05, Aarts07, Jakovac07, Iida06,Ohno11}, lattice cutoff effects may strongly affect the physics deduced from the correlation functions. 
The variational method approach used in Ref.~\cite{Iida06,Ohno11} is a good way to enhance signals of hadron states which are well known to be there, e.g. the ground and excited hadron states at zero temperature,
and will also contribute to a spectral decomposition of thermal correlation functions. However, the crucial question at finite temperature is not whether these states do contribute. Most important is the magnitude of
such a contribution. This does get modified in a variational approach and it is thus difficult to draw firm conclusions on charmonium melting from variational approaches.
Several earlier studies based on the MEM analysis implemented an unimproved integral kernel, which introduces an instability in the MEM algorithm in both low and high frequencies~\cite{Datta04,Asakawa04, Umeda05}. 
Most importantly,
we have doubled the number of points in the temporal direction compared to our previous study on an isotropic lattice~\cite{Datta04}, and  have reduced the spatial lattice spacing by about a factor of 4 or more compared to studies performed on anisotropic lattices~\cite{Asakawa04, Umeda05, Aarts07, Jakovac07,Iida06,Ohno11}. Comparing the number of points in the temporal direction in our study to the studies on anisotropic lattices,  $N_{\tau}$ in our study is about 1.5 times larger than that in Ref.~\cite{Umeda05, Aarts07, Jakovac07,Iida06,Ohno11} and compatible with that in Ref.~\cite{Asakawa04}. Moreover, a very detailed MEM analysis on the default model dependences and the systematic/statistical uncertainties has been performed in the current study. All this supports a better control over systematic effects in our analysis that suggests that both S and P wave states disappear at $T\geqslant 1.46~T_c$. Obviously it does not
necessarily mean that all the charmonium states are dissociated at the same temperature. The magnitude of  thermal effects in the spectral functions is observed to vary in different channels. This may indicate that the charmonium states will dissociate at different temperatures.
The fate of charmonium states at temperatures between $0.73~T_c$ and  $1.46~T_c$, however, remains unknown for us due to the lack of lattice data sets with appropriate temperature values in our study. A lattice QCD study of the screening masses extracted from spatial charmonium correlation functions suggests that $\eta_{c}$ and $\Jpsi$
may survive at $T\lesssim1.5~T_{c}$ in the hot medium~\cite{Mukherjee:2008tr,Karsch:2012na}.
Furthermore, several lattice QCD studies of charmonia suggest that $\chi_{c0}$ and $\chi_{c1}$ melt just above $T_c$~\cite{Datta04, Aarts07, Jakovac07}. The sequential suppression scenario~\cite{Karsch:2005nk,Borghini:2011yq,Strickland:2011mw} thus is not in contradiction to our results and, in fact, appears to be in accordance with the disappearance of bound states in the bottomonium ``spectral function" from the latest experiment
results~\cite{Chatrchyan:2011pe}. Thus  lattice calculations of the temporal correlation function of charmonia at lower temperatures would be interesting and crucial to locate the dissociation temperatures.

For the first time the charm diffusion coefficient has been estimated on the lattice directly from an analysis of spectral functions. The charm diffusion coefficient $D$ is found to be compatible with zero at $T<T_c$ and is about two times larger than the results from AdS/CFT calculations, $1/2\pi T$ at $T>T_c$. However, more efforts are needed to reduce the current uncertainties on the charm diffusion coefficient obtained from lattice QCD calculations especially at higher temperatures. The pQCD results seem to approach to our findings when higher order corrections are included~\cite{CaronHuot:2007gq}. The heavy quark diffusion coefficients obtained from a T-Matrix approach with the internal energy, on the other hand, are close to our results~\cite{He:2012df}. Recently another approach to calculate the heavy quark diffusion on the lattice~\cite{CaronHuot2009uh} has been suggested and an exploratory study has been carried out in Ref.~\cite{Meyer2010tt}. Further progress along this line has been made in Ref.~\cite{Francis11,Banerjee:2011ra}, in which the heavy quark diffusion constant multiplied by $2\pi T$, i.e. $2\pi T D$, is obtained to be
in the range of  (3.5...5) at $T\approx1.5~T_c$. It is quite impressive to see that this estimate for the heavy quark diffusion coefficient is close to our estimate of the charm diffusion coefficient, although totally different approaches have been used. Besides the Maximum Entropy Method, a Fourier method especially designed for addressing the low frequency behavior of  spectral functions has been introduced and has been used to estimate the electrical conductivity~\cite{Burnier:2011jq,Burnier:2012ts}. It would be interesting to implement this method also for the heavy quark sector.

In this study the effects of dynamical quarks are not included. The general picture concerning the properties of charmonium states might not change significantly as concluded from the study of charmonium states in two flavor QCD~\cite{Aarts07}. However, in the medium with sea quarks there exists a $D\bar{D}$ threshold and one might expect charmonium states to dissociate at  lower temperatures. Moreover, it is difficult to predict how the dissociation temperatures are influenced by the change of the pseudo critical temperature $T_c$, as $T_c$ becomes smaller when dynamical quarks are included in the system~\cite{Karsch:2000kv}.  For the charm diffusion coefficient,
one knows that at sufficiently high temperature where perturbative QCD calculations are applicable $2\pi T D$ becomes smaller when sea quarks are included ~\cite{CaronHuot:2007gq}. So far there are no studies that explore
the charm diffusion coefficient in lattice QCD with dynamical quarks included.

\section*{Acknowledgments}
\label{ackn}
This work has been supported in part by the Deutsche Forschungsgemeinschaft under Grant No. GRK 881 and by Contract No. DE-AC02-98CH10886 with the U.S. Department of Energy. 
Numerical simulations have been performed on the BlueGene/P at the New York Center for Computational Sciences (NYCCS) which is supported by the State of New York and 
the BlueGene/P at the John von Neumann Supercomputer Center (NIC) at FZ-J\"ulich, Germany.


\appendix
\section{Uncertainties of the spectral functions}
\label{sec:app}

In the appendix we show the default model dependences and various systematic uncertainties of the charmonium spectral functions. For illustration we will only focus on the $V_{ii}$ channel
as the uncertainties in the other channels are very similar. The basic settings of MEM used here are the same as what we mentioned in Section~\ref{sec:spf_MEM} if without additional description.  As mentioned in Section~\ref{sec:spf_MEM}, MEM cannot reproduce the correct width of the resonance but gives a stable and reliable ground state peak location of the spectral functions, so the signal for the dissociation of charmonium states is the shift of the peak location and relative broadening of the peak at different temperatures. 
 
The appendix is organized as follows: we will first show the default model dependences of output spectral functions in Appendix~\ref{sec:dmdep}. The main point to check is that how the first peak location of the output spectral function changes when resonance peaks with different peak locations are provided in the default model. Then we will study the systematic uncertainties of the output spectral function from MEM in Appendix~\ref{sec:sys}.

\subsection{Default model dependences}
\label{sec:dmdep}

Because of the limited number of correlator data points in the temporal direction, the MEM analysis becomes more difficult at temperatures above $T_c$. As has been done at temperatures below $T_c$, to check the reliability of the output spectral functions from the MEM analysis, the default model dependence test is always the first thing one needs to do. In principle one should put as much physical information into default models as possible. This rule leads to a very straightforward default model dependence test for the spectral functions above $T_c$. That is to fully benefit from the two limits which we already know quite well:  the free lattice spectral function at very high temperature and the spectral function obtained from MEM at a temperature below $T_c$. To put these pieces of information into the default model, one might be able to check to which limit, free or confinement limit, the output spectral function is closer. However, due to the fact that the spectral function at $T<T_c$ has a sharp ground state peak and the quality of correlator data at temperature above $T_c$ is not sufficient, the MEM output basically reproduces the spectral function below $T_c$ with negligible changes at all three available temperatures above $T_c$. Thus in the following default model dependence test we will not use the full information of the spectral function at $T<T_c$ into the default model but rather the information of the peak location of the ground state (``DM2"s in the following analyses). Besides this, we  will also use the free lattice spectral functions with some additional resonance peaks and/or transport peaks. The signature for the dissociation of resonances is then characterized by the shift of resonance peak location and relative broadening of the peak at different temperatures. The default modes used in the current study are 
summarized in Table~\ref{tab:DMs}.

\begin{table}[htbl]
\centering
\begin{tabular}{|c|c|*{2}{c|}}\hline

\backslashbox{{\bf default model}}{{\bf channel}}
  & $V_{ii}$ \\ \hline
DM1 & flspf + BW  \\ \hline
DM2 & flspf + BW + res1 \\ \hline
DM3 & flspf + BW + res2 \\ \hline
DM4 & flspf + BW + res3\\ \hline
\end{tabular}
\caption{The default models investigated in the current section. ``flspf" stands for the free lattice spectral function, ``res" stands for the resonance peak structure according to the relativistic Breit-Wigner distribution, ``res1", ``res2" and ``res3"  are resonance peaks with the peak locations equal, smaller and larger than the corresponding resonance peak locations in each channel at $T<T_c$. ``BW" is a Breit-Wigner like distribution according to Eq.~(\ref{eq:spf_trans}) and may vary in the width and amplitude in different default models.}
\label{tab:DMs}
\end{table}

  \begin{figure}[htpd]
  \begin{center}
    \includegraphics[width=0.75\textwidth]{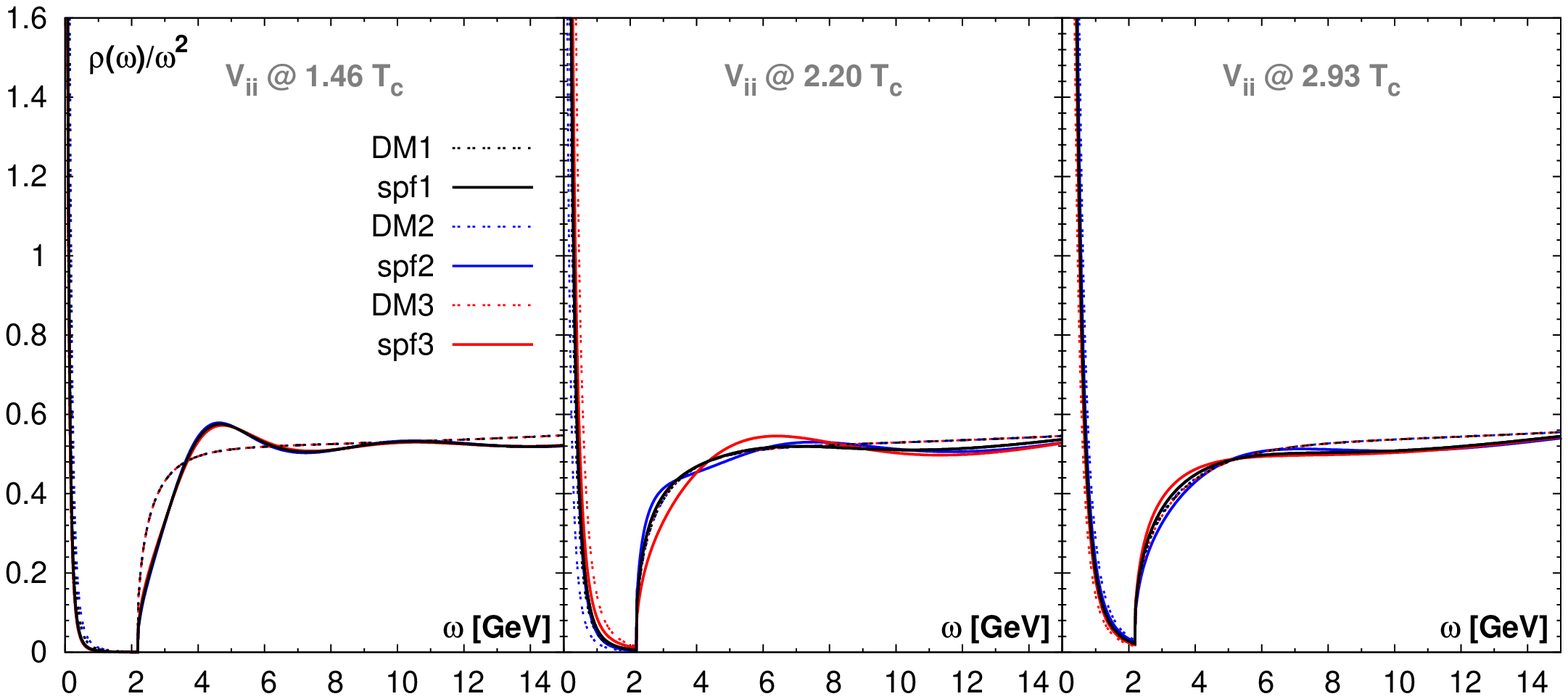}\\
    \vspace{12px}
     \includegraphics[width=0.75\textwidth]{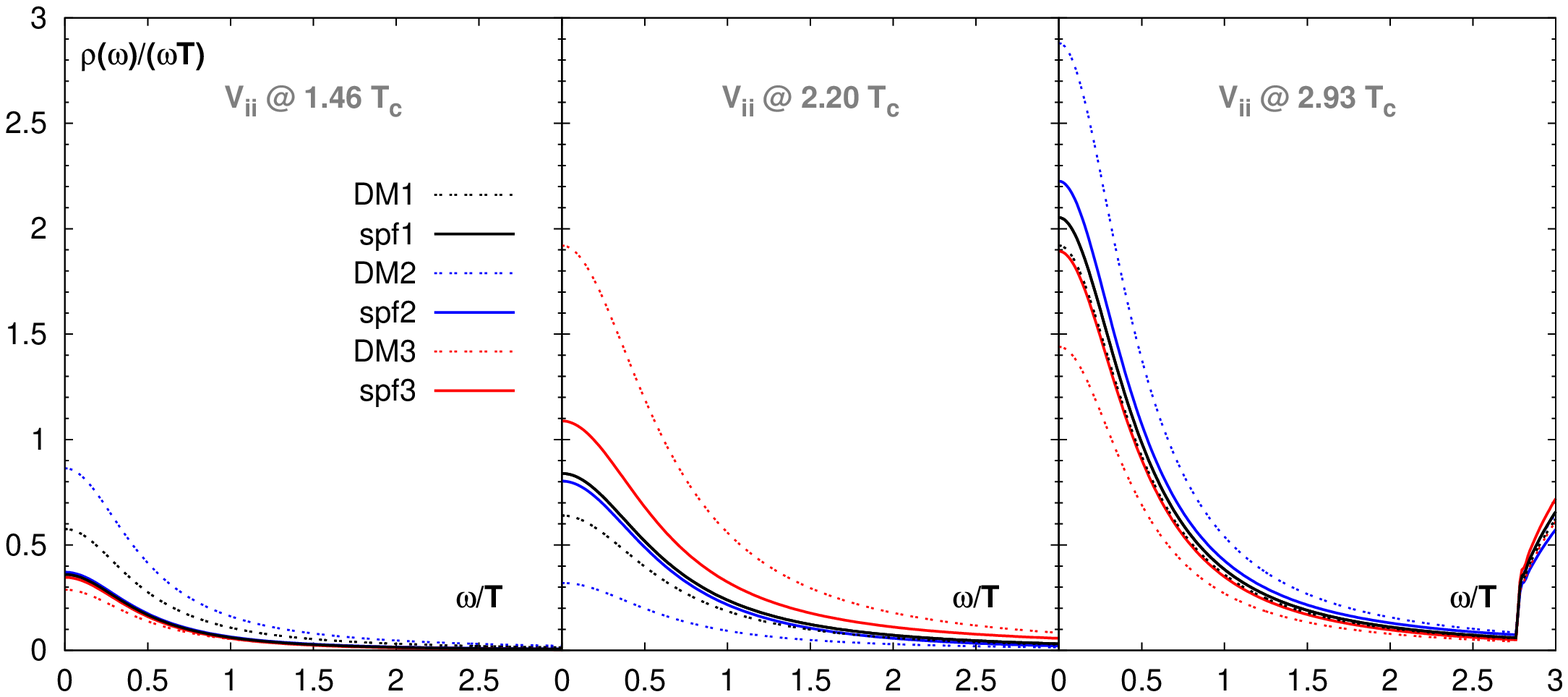}
        \caption{Default model dependences (varying the transport peak) of spectral functions in the $V_{ii}$ channel at temperatures above $T_c$. At each temperature the very large $\omega$ part of the default model is fixed to the behavior of the free lattice spectral function. Upper panel: $\rho(\omega,T)/\omega^2$ as a function of $\omega$, Lower Panel: a blowup of plots in the upper panel in the very low frequency region and plotted as $\rho(\omega,T)/(\omega T)$ versus $\omega/T$. ``DM"s are the input default models while ``spf"s are the corresponding MEM outputs.} 
                   \label{fig:spf_V123_beta7p793_AboveTc_dm_dependence1}
  \end{center}
  \end{figure}
  
Because of the existence of a diffusion contribution to the $V_{ii}$ channel one has to check the dependences both on variations
of the resonance part and diffusion part in the default models. Then for the default model dependence test we first fix the large $\omega$ behavior of the default model by using the rescaled free lattice spectral function and vary the information on the very small $\omega$ part, i.e. the transport peak described in Eq.~(\ref{eq:spf_trans}). The default models used here correspond to ``DM1" ,``DM2" and ``DM3" in Table~\ref{tab:DMs} without resonance parts. We show the result in Fig.~\ref{fig:spf_V123_beta7p793_AboveTc_dm_dependence1}. The upper panel of Fig.~\ref{fig:spf_V123_beta7p793_AboveTc_dm_dependence1} shows $\rho(\omega,T)/\omega^2$ as a function of $\omega$ in the large $\omega$ region while the lower panel of Fig.~\ref{fig:spf_V123_beta7p793_AboveTc_dm_dependence1} shows $\rho(\omega,T)/(\omega T)$ as a function of $\omega/T$ in the very low frequency region. ``DM"s are the input default models while ``spf"s are the corresponding MEM outputs. In particular, the transport parts of ``DM1" in the MEM analysis at $1.46~T_c$ are parameterized in the Breit-Wigner form with $2\pi T D=3.6$ and $M=1.8$ GeV obtained from the fit to the difference between the measured correlator and the reconstructed correlator by using a Breit-Wigner ansatz discussed in Section~\ref{sec:diffusion_corr}. 
Because of the interplay between the contributions from the diffusion and resonance parts 
it is difficult to make an estimate of the charm diffusion constant directly on the correlator level at the two highest temperatures. Here we simply apply the same value of charm diffusion $D$ estimated at $1.46~T_c$ to the default models at  $2.20$ and $2.93~T_c$.  Looking at the output spectral functions at each temperature, we find that the variation of the very small $\omega$ part of the default model gives negligible effects to the intermediate $\omega$ part (resonance part) of the output spectral functions. Concerning the 
temperature dependence of the resonance peak, the upper panel of Fig.~\ref{fig:spf_V123_beta7p793_AboveTc_dm_dependence1} shows that,
already at $1.46~T_c$, the ground state peak becomes much broader and its peak location is shifted to larger energies compared to that at $0.73~T_c$ (see Fig.~\ref{fig:spf_Swaves_0p75Tc_beta7p793_dm_dependence}). When going to the higher temperature of $2.20~T_c$ one can hardly see a bump in the interesting $\omega$ region. At our highest temperature available, $2.93~T_c$, we find that the large $\omega$ part more or less resembles the shape of free lattice spectral functions and no peak structure is observed. For the transport peak shown in the lower panel of Fig.~\ref{fig:spf_V123_beta7p793_AboveTc_dm_dependence1}, the prior information of charm diffusion $D$ estimated from $G(1/2)-G_{\rec}(1/2)$ is put into very low frequency part of the default models. MEM shows the sensitivity to the very low frequency part and the output spectral functions differ from the default model in this very small energy region. We observe that the output transport peak has a weak dependence on the input default models at all three temperatures above $T_c$. And the amplitude of the transport peak at vanishing energy increases with temperature.

\begin{figure}[hptd]
  \begin{center}
    \includegraphics[width=0.75\textwidth]{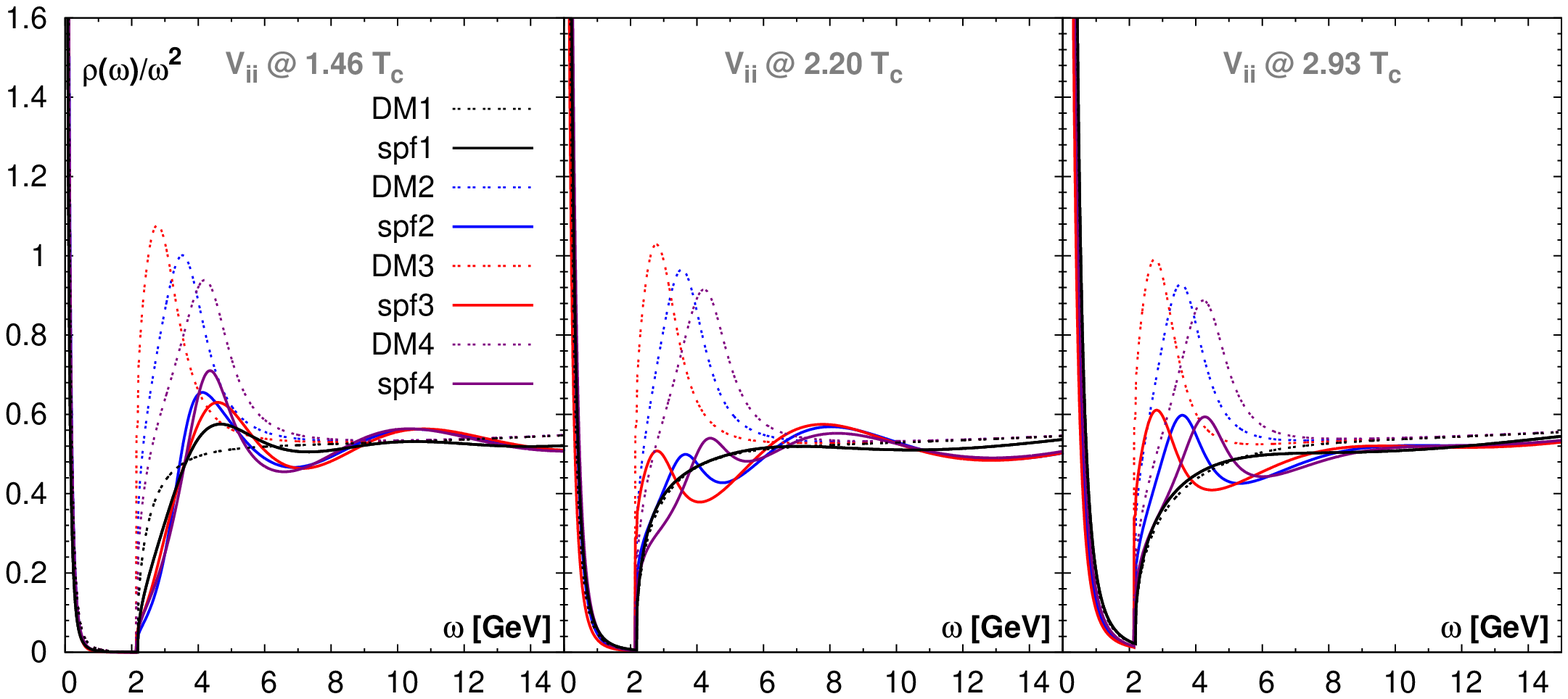}\\
    \vspace{12px}
     \includegraphics[width=0.75\textwidth]{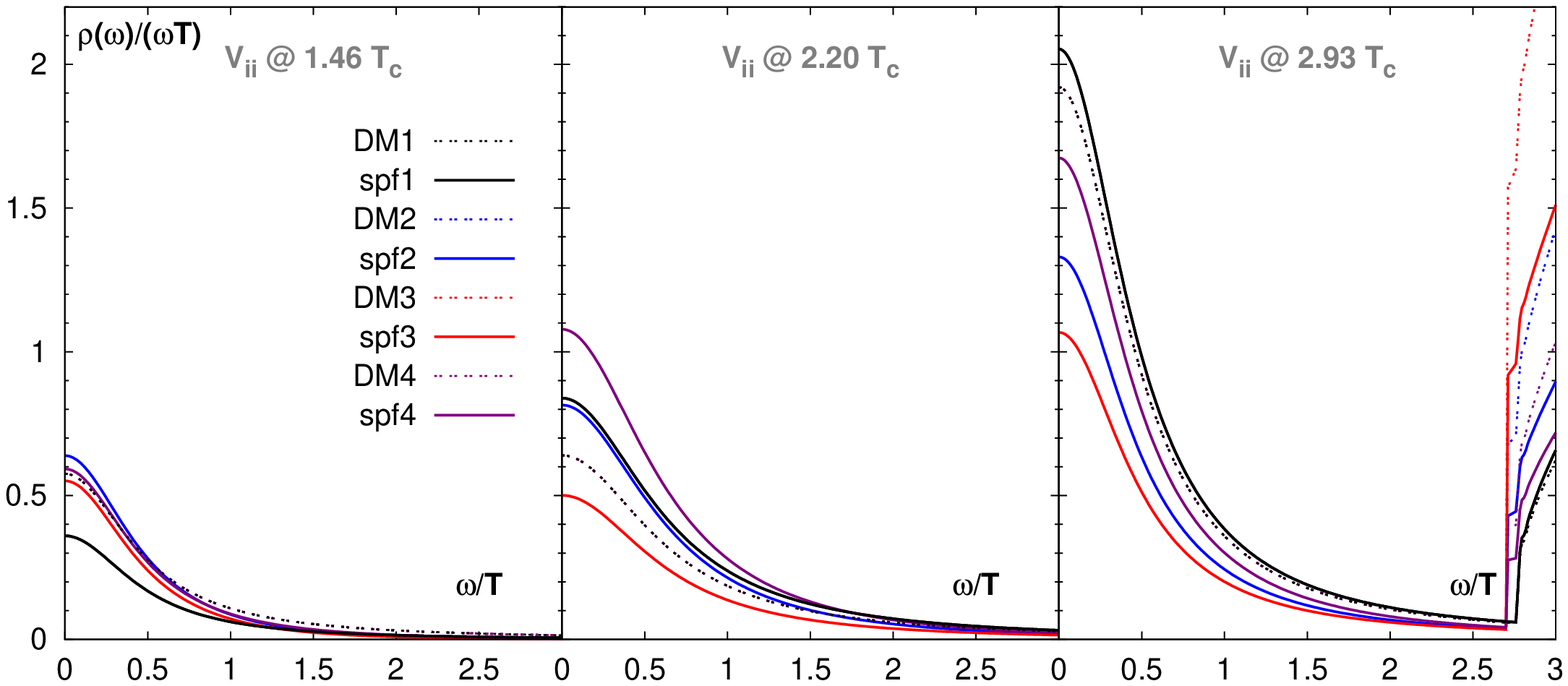}
        \caption{Default model dependences (varying the resonance part) of spectral functions in the $V_{ii}$ channel at temperatures above $T_c$. The transport part of the default model is fixed in each temperature same as that of ``DM1"s in Fig.~\ref{fig:spf_V123_beta7p793_AboveTc_dm_dependence1}. Upper panel: $\rho(\omega,T)/\omega^2$ as a function of $\omega$, Lower Panel: a blowup of plots in the upper panel in the very low frequency region but plotted as $\rho(\omega,T)/(\omega T)$ versus $\omega/T$. ``DM"s are the input default models while ``spf"s are the corresponding MEM outputs.}
                 \label{fig:spf_V123_beta7p793_AboveTc_dm_dependence2}
  \end{center}
  \end{figure}

After studying default model dependences by varying the transport peak in the default model on the output spectral function in the intermediate $\omega$ (resonance peak) region in Fig.~\ref{fig:spf_V123_beta7p793_AboveTc_dm_dependence1}, we now
fix the very low frequency (transport peak) part of the default model and vary the intermediate $\omega$ (resonance part) behavior of the default models. The default models in the very low frequency part are fixed to have 
the same behavior as ``DM1" in Fig.~\ref{fig:spf_V123_beta7p793_AboveTc_dm_dependence1} at each temperature. Again note that the transport part of  ``DM1" is parametrized as $2\pi T D=3.6$ and $M=1.8$ GeV as discussed in Section~\ref{sec:diffusion_corr}. We test four different default models as listed in Table~\ref{tab:DMs}: ``DM1" is a rescaled free spectral function with a transport peak, ``DM2" is a rescaled free spectral function with a transport peak supplemented with a resonance peak whose peak location is the same as that of the spectral function at $0.73~T_c$, ``DM3" and ``DM4" are basically the same as ``DM2" but with a resonance peak whose peak location is smaller and larger than that of the spectral function at $T<T_c$, respectively. We show the default models and their corresponding output spectral functions (``spf"s) divided by $\omega^2$ as functions of $\omega$ in the upper panel of Fig.~\ref{fig:spf_V123_beta7p793_AboveTc_dm_dependence2}. At $1.46~T_c$ there is a minor default model dependence of the output spectral functions, but the trend is similar: the peak location is shifted to a location larger than the peak location of the spectral function at $0.73~T_c$ (peak location shown in ``DM2") and the width becomes larger. At $2.20~T_c$ the default model dependence is a little stronger. This may be due to the smaller number of data points in the temporal direction and lower statistics. However, outputs from MEM still have unique differences from input default models and they all have a trend to resemble the shape of the free spectral function. At $2.93~T_c$ we have only 9 points in the analysis and together with the issue of the transport peak, the default model dependence is considerably stronger than that in the analysis at the other temperatures. Based on the results from 1.46 and 2.20 $T_c$ we do not expect the peak location of the resonance peak at $2.93~T_c$ shifts to smaller energies compared to the case at $0.73~T_c$ and would rather expect that the spectral function at this temperature is much closer to the spectral function in the noninteracting case. In the lower panel of Fig.~\ref{fig:spf_V123_beta7p793_AboveTc_dm_dependence2} we enlarge the very low frequency part of the upper panel in Fig.~\ref{fig:spf_V123_beta7p793_AboveTc_dm_dependence2} and show $\rho(\omega)/(\omega T)$ as a function of $\omega/T$. Unlike the case in the lower panel of Fig.~\ref{fig:spf_V123_beta7p793_AboveTc_dm_dependence1}, the change of the default model in the intermediate $\omega$ part (resonance part) has a relatively large effect on the output in the very low frequency region. It could be mainly due to the compensation of the very low frequency part to the changes of corresponding resonance parts.  Without the quantitative description of the transport peak we can observe a trend that the amplitude of the transport peak becomes larger with increasing temperature.

In a short summary, in this subsection we have checked the reliability of the output spectral function in the $V_{ii}$ channel from the MEM analysis by varying the resonance part and the transport part of input default models.
At $1.46~T_c$, the default model dependence of the resonance part is relatively weak and the resonance peak observed at $T<T_c$ generally shifts to high frequency region and becomes much broader. At higher temperatures, the default model dependences of the resonance part becomes stronger due to the insufficient qualify of the temporal data, however, the general trend is that no clear peak structures are found and the spectral function gets closer to the noninteracting case. In the very low frequency region, we supplied the transport peak parameterized by the estimation in Section~\ref{sec:diffusion_corr}. MEM showed certain sensitivity to the transport peak. The transport peak in the $V_{ii}$ channel has weak dependence in the default models when only the transport peak in the default model is changed while it has relatively large default model dependence when only the resonance part in the default model is changed. The general trend is that the amplitude of the transport peak is increasing with the increasing temperature.

\subsection{Systematic uncertainties}
\label{sec:sys}

In this subsection we explore the systematic uncertainties of the spectral function from the MEM analyses. 
This study includes lattice spacing dependencies, a comparison of the spectral function below and above $T_c$ for a same number of data points used to extract the spectral function, lattice cutoff effects at small distances and the dependence on the threshold of the free spectral function used in the default model.

\begin{figure}[htbl]
  \begin{center}
    ~\includegraphics[width=0.5\textwidth]{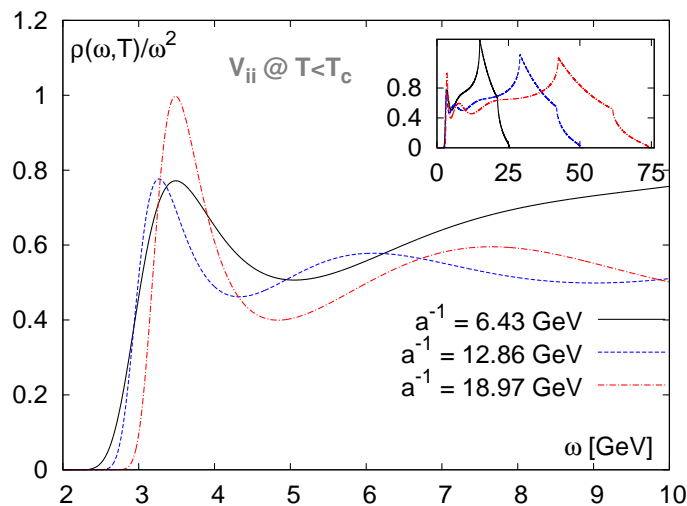}
            \caption{Output spectral functions from MEM in the $V_{ii}$ channel at temperatures below $T_c$ from three different lattices: $128^3\times32$ with $\beta=6.872~(a^{-1}=6.43~$GeV) at 0.74~$T_c$, $128^3\times64$ with $\beta=7.457~(a^{-1}=12.86~$GeV) at $0.74~T_c$ and $128^3\times96$ with $\beta=7.793~(a^{-1}=18.97~$GeV) at $0.73~T_c$. The small plot inside is the output spectral function in the whole energy region.}
                 \label{fig:spf_V123_betas_0p75Tc}
  \end{center}
  \end{figure}

\begin{figure}[htpl]
  \begin{center}
    \includegraphics[width=0.45\textwidth]{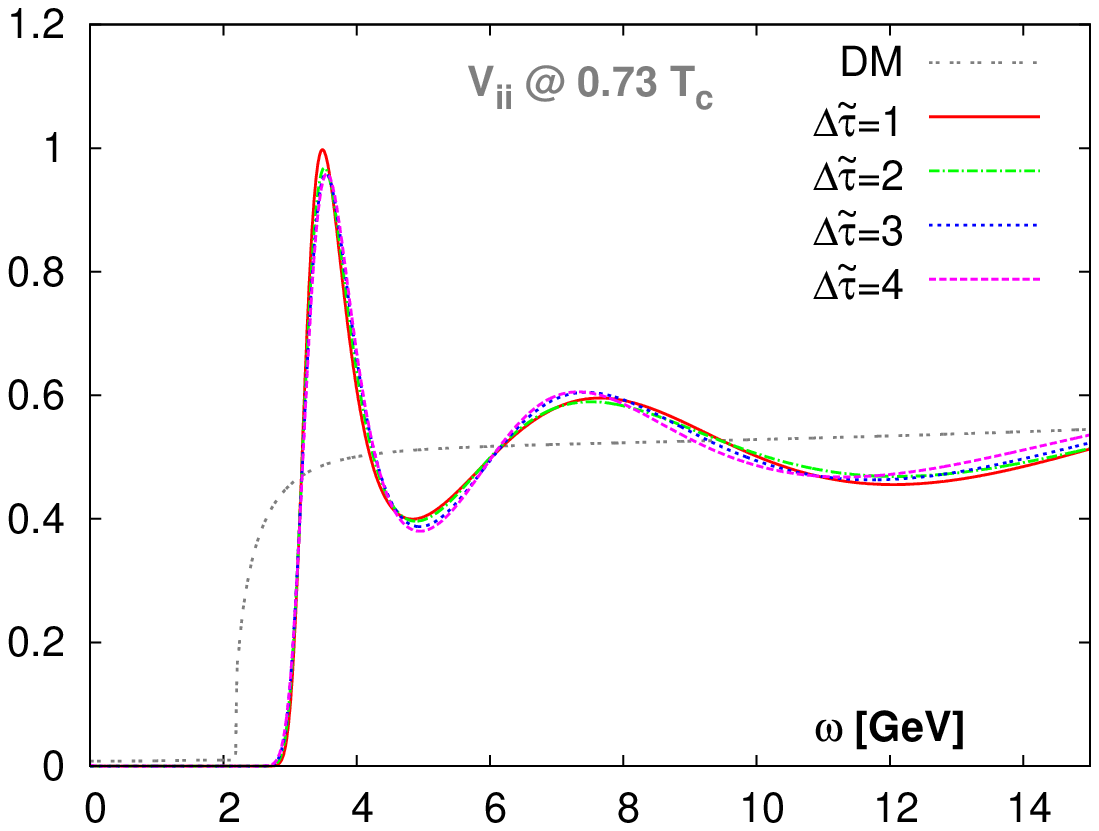}~ \includegraphics[width=0.45\textwidth]{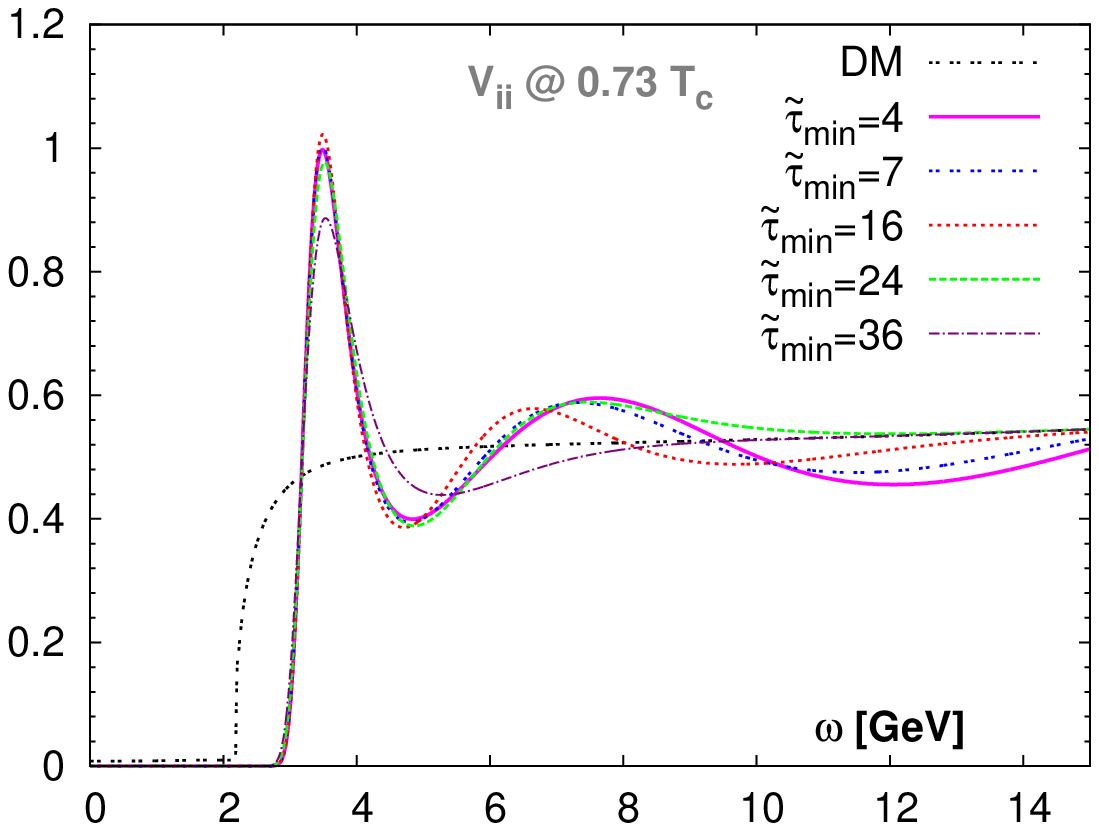}
            \caption{Left: the dependence of the output spectral function on the number of correlator data points used in the MEM analysis at $T=0.73~T_c$. All the points included start at $\tilde{\tau}_{min}=4$. $\Delta\tilde{\tau}$ is the step size between the neighboring data points selected. For instance, $\Delta\tilde{\tau}=4$ means $\tilde{\tau}=4,8,12,\cdots, 48$ are used. Right: the $\tilde{\tau}_{min}$ (number of data points omitted in the short distance) dependence of the output spectral functions at $T=0.73~T_c$.  ``DM" labels the input default model and the other lines are the output spectral functions with different values of $\Delta\tilde{\tau}$.}
                 \label{fig:spf_dtau_taumin_0p75Tc}
  \end{center}
  \end{figure}
  
First we look into the lattice spacing dependence of the output spectral function on our available lattices. We show spectral functions  from the $V_{ii}$ channel at temperatures below $T_c$ in Fig.~\ref{fig:spf_V123_betas_0p75Tc}. The results are obtained from the lattices with $a^{-1}=18.97$~GeV ($\beta=7.793,~128^3\times96$), $a^{-1}=12.86$~GeV ($\beta=7.457,~128^3\times64$) and $a^{-1}=6.43$~GeV ($\beta=6.872,~128^3\times32$). The plot shows the behavior of spectral functions in the low frequency region ($2\le \omega\le 10~$GeV) while the small plot inside shows the behavior in the whole frequency region. One can observe that with smaller lattice spacing the lattice cutoff effects (the cusps in the high frequency region) can be well separated from the physically interesting frequency region. As seen from the low frequency region, the width of the ground state peak becomes narrower with decreasing lattice spacing. We also find that the second peak should be lattice or MEM artifacts since its locations varies a lot from lattice spacings and details in this frequency region can not be resolved on the coarsest lattices used in this study.

  One always has to compare spectral functions at $T>T_c$ to those at $T<T_c$ to study temperature effects. As number of correlator data points at higher temperatures is reduced, we study the dependence of output spectral functions on the number of data points used in the MEM analysis at $T<T_c$, we use the same number of data points below and above $T_c$ to have similar systematic uncertainties and to analyze thermal modifications. Here we restrict the default model to have the behavior of the free lattice spectral function. At $T=0.73~T_c$ we select the data points in the temporal direction as to start at $\tilde{\tau}_{min}=4$ and be separated by a step length of $\Delta\tilde{\tau}$. For instance when $\Delta\tilde{\tau}=2$ we select data points of $\tilde{\tau}=4,6,8,\cdots,48$, in total 13 points. So the number of data points used with $\Delta\tilde{\tau}=2,~3,~4$ at $0.73~T_c$ corresponds to the number of data points used at $1.46~T_c$, $2.20~T_c$ and $2.93~T_c$, respectively. We show the results for the $V_{ii}$ channels in the left plot of Fig.~\ref{fig:spf_dtau_taumin_0p75Tc}. ``DM" labels the input default model and the other lines are the output spectral functions with different values of $\Delta\tilde{\tau}$. We observe negligible dependences on the number of data points used in the interesting frequency region . There are minor changes on the amplitudes of the ground state peak but the ground state peak location always remains the same.

To remove the discretization effects, we normally omit some data points at very small distances. However, it is not very certain how many data points should be omitted or up to what value of $\tilde{\tau}_{min}$ (the shortest time slice $\tilde{\tau}$ used in the MEM analyses) the physics about the bound states is concerned. Thus we check the dependence of the output spectral function on $\tilde{\tau}_{min}$. We vary $\tilde{\tau}_{min}$ to be $4,~7,~16,~24$ and 36 at $0.73~T_c$ to check the effects for the same default model. The default models are fixed in each channel. The results for the $V_{ii}$ channel are shown in the right plot of Fig.~\ref{fig:spf_dtau_taumin_0p75Tc}. ``DM" labels the input default model and the other lines are output spectral functions corresponding different values of $\tilde{\tau}_{min}$. We observe that the large $\omega$ ($\omega\gtrsim 5~$GeV) behavior of the output spectral functions, which is most sensitive to the small distance part of the correlation function, changes with $\tilde{\tau}_{min}$ and in the small $\omega$ region ($\omega\lesssim5$ GeV) the peak location of the ground state peak stays almost unchanged even with $\tilde{\tau}_{min}=36$. Thus the $\tilde{\tau}_{min}$ dependence of the spectral function in the $V_{ii}$ channel in the interesting frequency region is very small at $T < T_c$ on our finest lattice.

         \begin{figure}[htdp]
  \begin{center}
    \includegraphics[width=0.75\textwidth]{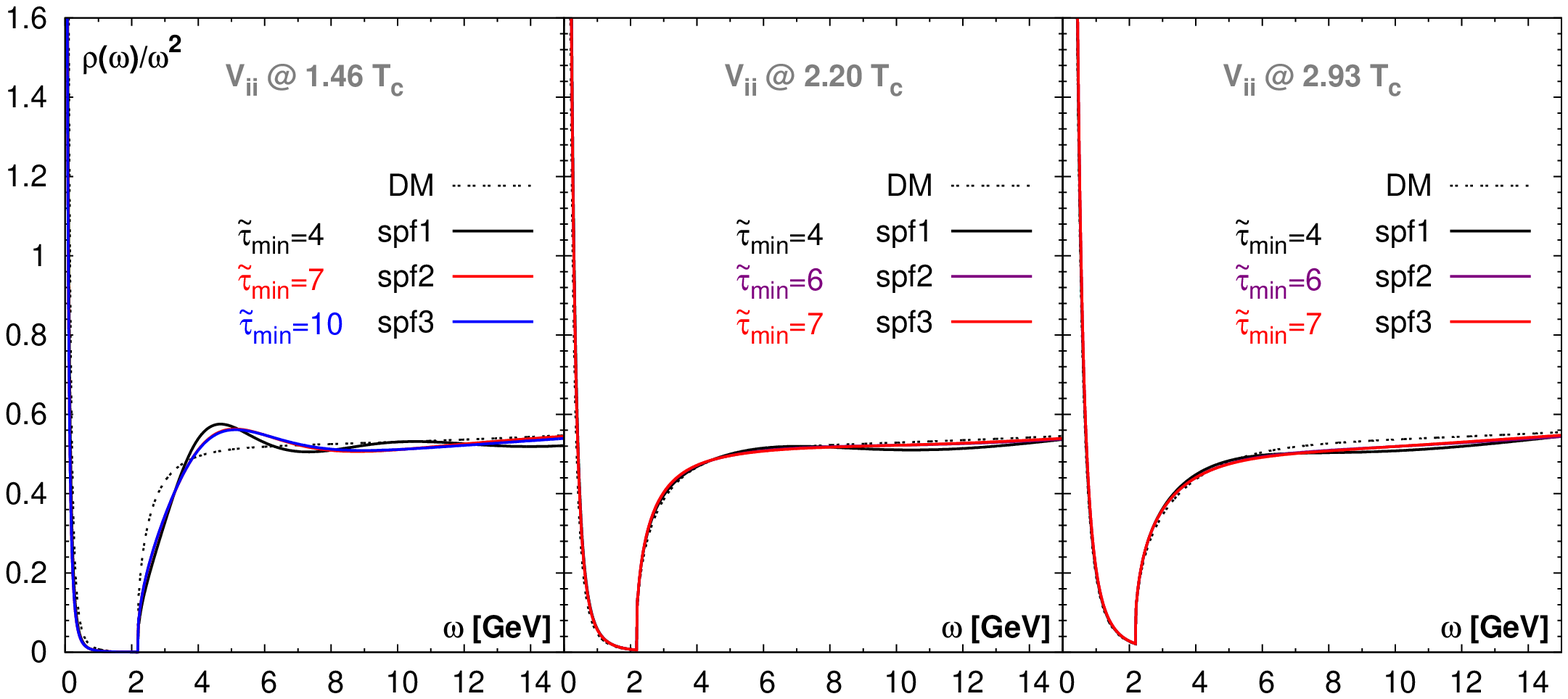}\\
     \vspace{12px}
         \includegraphics[width=0.75\textwidth]{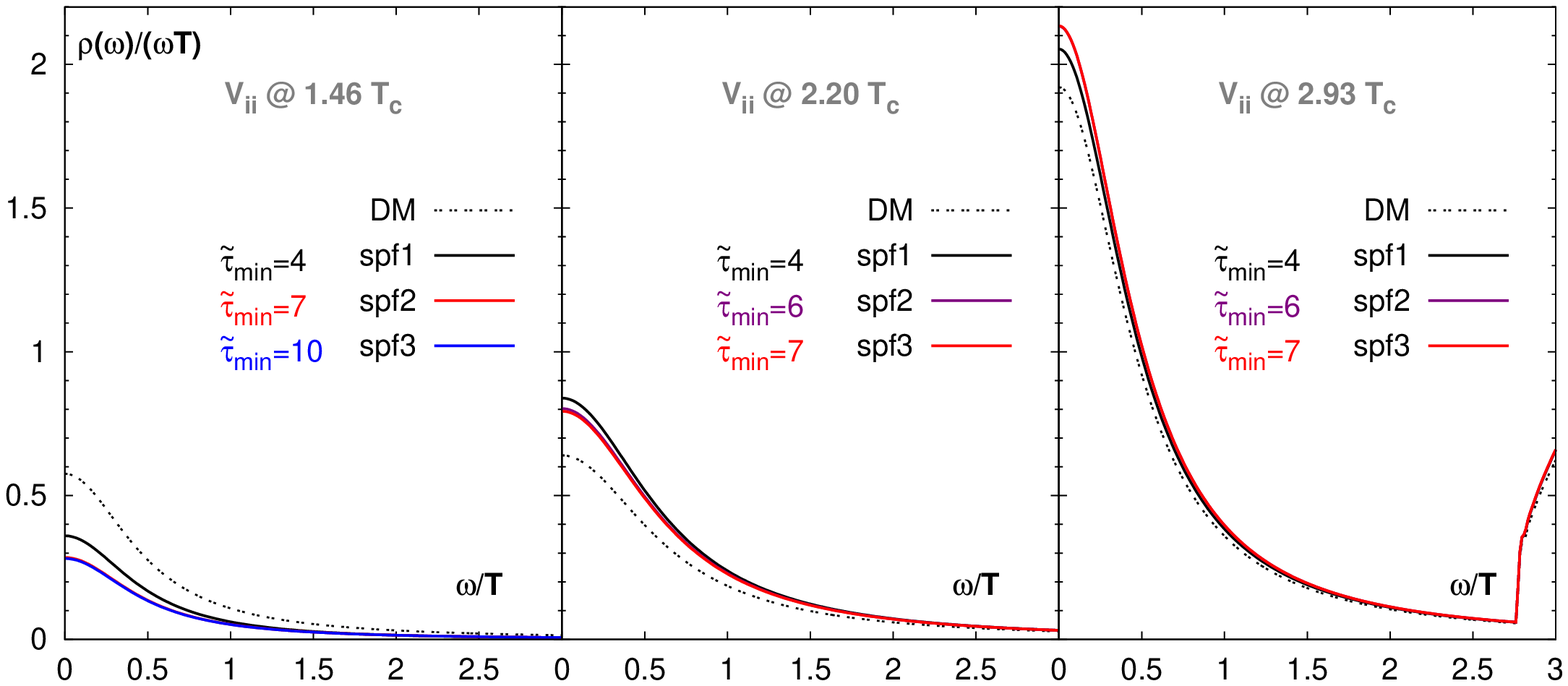}\\
            \caption{The $\tilde{\tau}_{min}$ (number of data points omitted from the short distance) dependence of output spectral functions in the $V_{ii}$ channel at $T>T_c$. The upper plot shows the behavior of $\rho(\omega)/\omega^2$ as a function of $\omega$ while the lower plot shows the transport behavior of $\rho(\omega)/(\omega T)$ as a function of $\omega/T$ which corresponds to the divergent parts in the upper plot at the corresponding temperatures. }
                 \label{fig:spf_V123_aboveT_Nleft_dep}
  \end{center}
   \end{figure}

               \begin{figure}[hdtp]
  \begin{center}
    \includegraphics[width=0.75\textwidth]{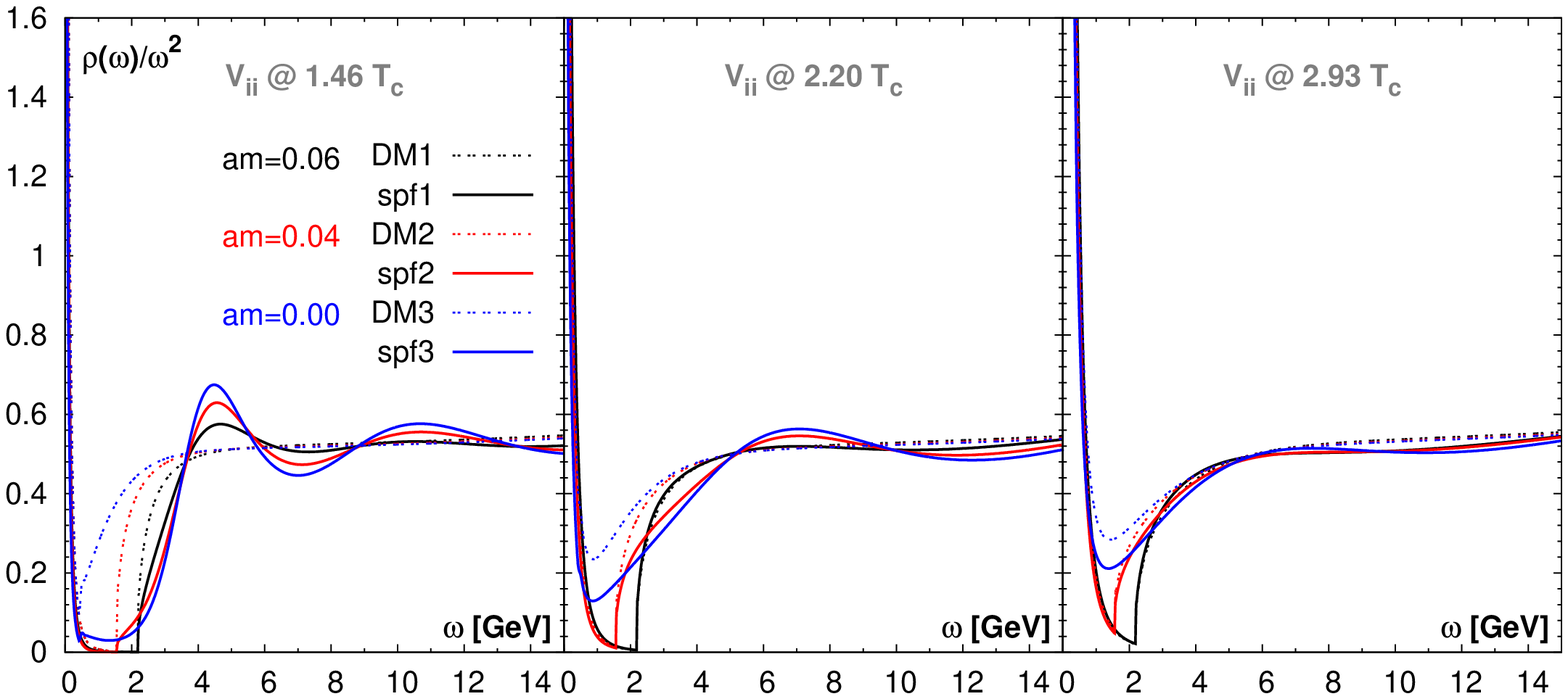}\\
     \vspace{12px}
         \includegraphics[width=0.75\textwidth]{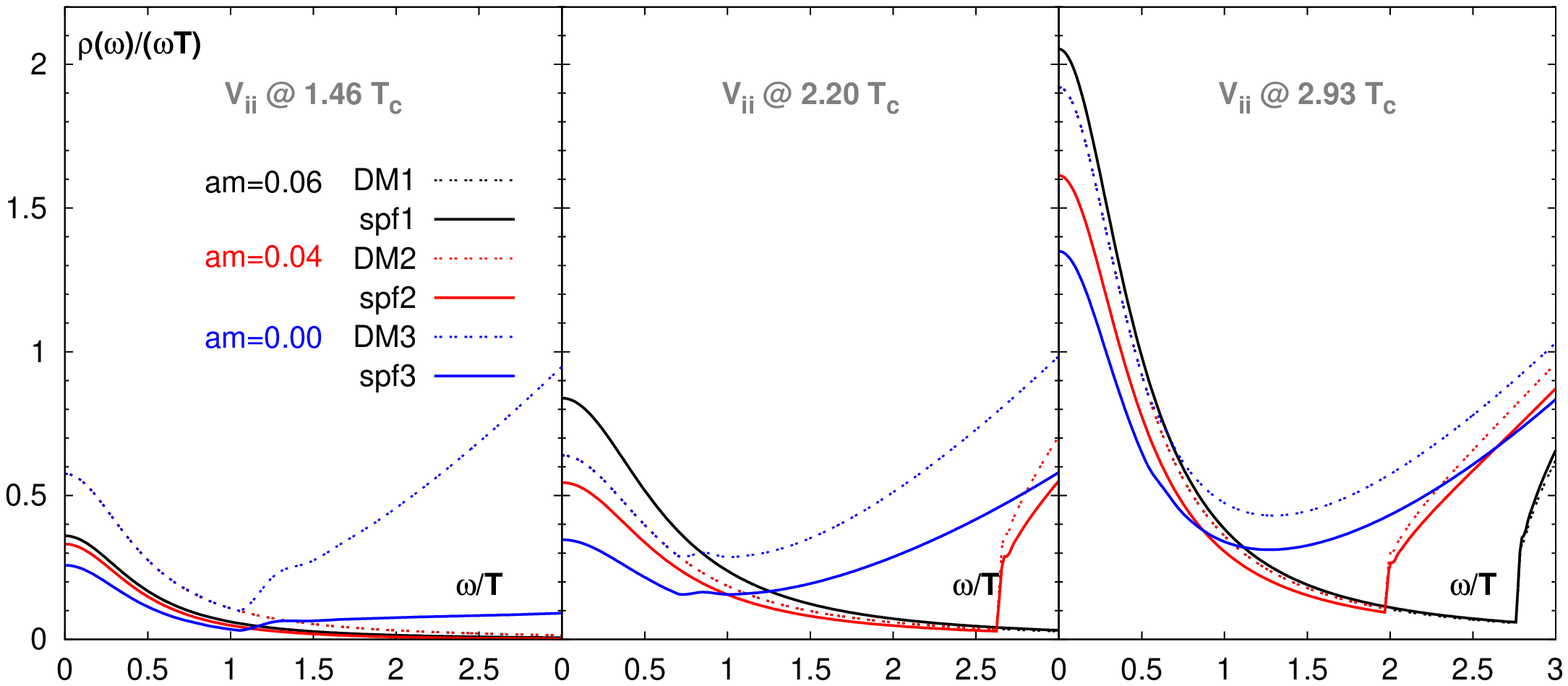}\\
            \caption{The quark mass $am$ dependence of output spectral functions in the $V_{ii}$ channel at $T>T_c$. The upper plot shows the behavior of $\rho(\omega)/\omega^2$ as a function of $\omega$ while the lower plot shows the transport behavior of $\rho(\omega)/(\omega T)$ as function of $\omega/T$ which corresponds to the divergent parts in the upper plot at the corresponding temperatures. }
                 \label{fig:spf_V123_aboveT_am_dep}
  \end{center}
   \end{figure}

  After exploring the uncertainties that can be learned from correlators at $T<T_c$, we then move on to study the uncertainties of output spectral functions at $T>T_c$. Following the spirit we have done at $T<T_c$, we will also check the dependences on the number of data points omitted from the short distances. Besides that we will check the dependences on the threshold of the continuum through the charm quark mass $am$.

  We first show the results for the dependence of the output spectral function on the number of correlator data points omitted at short distances $\tilde{\tau}_{min}$ in Fig.~\ref{fig:spf_V123_aboveT_Nleft_dep}. The upper panel shows $\rho(\omega)/\omega^2$ as a function of $\omega$ at three available temperatures above $T_c$ while the lower panel focuses on the transport behavior of the spectral function in the low frequency region and has $\rho(\omega)/(\omega T)$ as function of $\omega/T$. The default models (``DM") are the same in the whole frequency region at each temperature. ``DM" is provided by a rescaled free lattice spectral function and an additional transport peak. ``DM" is also the same as ``DM1" in Fig.~\ref{fig:spf_V123_beta7p793_AboveTc_dm_dependence1} at each temperature. Note that the transport part of ``DM" is parameterized in the Breit-Wigner form with $2\pi T D=3.6$ and $M=1.8$ GeV. As seen from the upper plot of Fig.~\ref{fig:spf_V123_aboveT_Nleft_dep}, at $1.46~T_c$, from $\tilde{\tau}_{min}=4$ to $\tilde{\tau}_{min}=7$ and 10, the peak location of the ground state peak seems to move a little further to larger energy while at both $2.20~T_c$ and $2.93~T_c$ the output spectral functions show negligible changes due to the variation of $\tilde{\tau}_{min}=4,~6$, and 7. In the lower panel of Fig.~\ref{fig:spf_V123_aboveT_Nleft_dep} the very low frequency behavior of the spectral function is shown. Note that ``DM" in the current frequency region is also fixed at each temperature. At all three temperatures the output transport peaks show minor dependences on $\tilde{\tau}_{min}$, which indicates that the information of the transport peak is mainly enclosed in the large distance part of the correlation function.

  Because of the insensitivity of MEM on the very large $\omega$ behavior of the spectral function, as we observed from, e.g. the left panels of Fig.~\ref{fig:spf_Swaves_0p75Tc_beta7p793_dm_dependence}, the outputs always reproduce the very large $\omega$ behavior of the input default models, which in our case normally is the free lattice spectral function multiplied by a certain constant to reproduce the value of $G(\tilde{\tau}_{min})$. However, we do not really know the exact behavior of the large $\omega$ part as well as the onset point of the continuum. We thus check the effects caused by different quark masses in the default models. The different quark masses $am$ have an effect on the threshold and the structure of the free spectral function.  
  In Fig.~\ref{fig:spf_V123_aboveT_am_dep} we show the dependence of the output spectral function in the $V_{ii}$ channel on the quark mass $am$ at $T>T_c$. The upper panel of Fig.~\ref{fig:spf_V123_aboveT_am_dep} shows the large $\omega$ behavior of the spectral function and the lower panel highlights the transport peak part.  Here we test with free lattice spectral functions having $am=0.06$ (``DM1"), $am=0.04$ (``DM2") and $am=0$ (``DM3") such that $G_{\rm DM}(\tilde{\tau}=4,T)/G(\tilde{\tau}=4,T)=1$. Here $am=0.06$ is the quark mass obtained from the running quark mass on the lattice (see Table~\ref{table:QuarkMass}) and ``DM1"  is the same as ``DM1" in Fig.~\ref{fig:spf_V123_aboveT_Nleft_dep}. The rising side of the ground state peak starts to be nonzero following the trend of the default model already at $1.46~T_c$ and the amplitude of the ground state peak also changes with different values of $am$. However, the location of the first peak remains almost the same and it is much larger than the ground state peak location at $0.73~T_c$. At $2.20~T_c$, the output spectral functions ``spf2" and ``spf3" from the default models ``DM2" and ``DM3" have a small bump structure other than ``spf1". At $2.93~T_c$, ``spf1", ``spf2" and ``spf3" have negligible differences when $\omega\gtrsim 3~$GeV. As seen from the lower panel of Fig.~\ref{fig:spf_V123_aboveT_am_dep}, with decreasing $am$, in general the transport peak's amplitude becomes smaller and its width becomes larger at all the three temperatures. At $1.46~T_c$ the change of the transport peak of ``spf1" is very small, and when going to higher temperatures, $2.20$ and $2.93~T_c$, the deviations become larger, probably as a consequence of the larger differences of ``spf"s in the frequency region of $1 \lesssim\omega\lesssim 7~$GeV.

    In a short summary of this subsection, we have studied the dependence of output spectral functions on the lattice spacing and conclude that our finest lattice gives the most reliable results. To better compare the spectral function below and above $T_c$, we used the same number of data points at below and above $T_c$ in the MEM analysis and observed negligible difference between the cases with and without the same number of data points used. We checked lattice cutoff effects by removing several data points from the short distance and found small dependences of low frequency part of spectral function on the short distance of the correlation function. It supports that spectral functions in the resonance and transport peak region extracted from MEM using our finest lattice really are physical and show no major cutoff dependencies. We also checked the dependences on the threshold of input free lattice spectral function and found minor dependences in the resonance part of the spectral function. So one basically sees that the general picture in the $V_{ii}$ channel is not changed with various different default models and different ways of implementing the correlator data. The transport peak in the $V_{ii}$ channel has a weak dependence on the threshold of the free lattice spectral function at $1.46~T_c$ and  the dependences at $T>1.46T_c$ become somewhat stronger due to the limited available distances.

\end{document}